\documentclass[11pt, a4paper]{article}
\usepackage[top=30truemm,bottom=30truemm,left=20truemm,right=20truemm]{geometry}

\usepackage{amsmath}
\usepackage{amssymb}
\makeatletter
    
    \@addtoreset{equation}{section}
\makeatother

\title{\vbox{%
\baselineskip 14pt
\hfill \hbox{\normalsize KUNS-2545}\\
} \vskip 1.7cm
\Large \bf Anomaly of Tensionless String in Light-cone Gauge
\vskip 0.5cm
}
\author{ \\
 \\
Kenta Murase$^1$\thanks{E-mail: kmurase@gauge.scphys.kyoto-u.ac.jp or kenta1murase2@gmail.com}\\
 \\
{\sl\fontsize{10pt}{0pt}\selectfont $^1$Department of Physics, Kyoto University, Kitashirakawa, Kyoto 606-8502, Japan }\\ }
\date{}
\begin{document}
\maketitle

\begin{abstract}
The classical tensionless string theory has the spacetime conformal symmetry. We expect and require that the quantum tensionless string theory has it too. In the BRST quantization method, the theory has no spacetime conformal anomaly in two dimensions. On the other hand, in the light-cone gauge quantization without the mode expansion, the theory in $D>3$ has the spacetime conformal anomaly in the traceless part of $[\mathcal{J}^{-I}, \mathcal{K}^{J}]$ in some operator order.

In this paper, we consider a tensionless closed bosonic string in the light-cone gauge and investigate the spacetime conformal anomaly in the theory with the mode expansion. The appearance of the spacetime conformal anomaly in the light-cone gauge is different between the case of $D>3$ and the case of $D=3$ and depends on the choice of the operator order. Therefore we must consider dangerous commutators in the spacetime conformal symmetry of $D>3$ and $D=3$ in each operator order separately. Specifically we calculate dangerous commutators, $[\mathcal{J}^{-I},\mathcal{K}^{K}]$ in $D>3$ and $\tilde{\mathcal{K}}^{-}\equiv -i[\mathcal{J}^{-}, \tilde{\mathcal{K}}^{-}]$ and $[\mathcal{J}^{-}, \tilde{\mathcal{K}}^{-}]$ in $D=3$, in two types of the operator order.
\end{abstract}

\newpage

\tableofcontents

\section{Introduction}
Recently the method of dualities, such as the AdS/CFT correspondence, has been used in higher spin gauge theories to understand gauge theories, gravities and M theory: \cite{bib:080, bib:071, bib:078, bib:074, bib:077, bib:052} and \cite{bib:018, bib:054, bib:073, bib:076, bib:075, bib:066, bib:057, bib:053, bib:055}. Because higher spin gauge theories contain the infinite tower of higher spin fields, their relation with the tensionless limit of string theories have been studied \cite{bib:035, bib:061, bib:031, bib:045, bib:059, bib:041}. However, the relations are not clear. One of the reasons is probably the poor understanding of the tensionless string theory, as well as higher spin gauge theories.\cite{bib:072, bib:069, bib:017, bib:060}~ Therefore the understanding of the tensionless string theory is very important to investigate the relation between the tensionless string theory and higher spin gauge theories and to consider dualities on the tensionless string theory.

Classical string theories have the spacetime symmetry as the global symmetry, as well as a point particle \cite{bib:046}. The tensionful string theory\footnote{``Tensionful string" means ``string with tension" in this paper.} has the Poincar\'{e} symmetry and the tensionless string theory has the enhanced symmetry, the spacetime conformal symmetry. As is well known, the quantization of the tensionful string usually causes the anomaly in the Poincar\'{e} symmetry, the Lorentz anomaly. To avoid this anomaly, the spacetime dimensions and the operator ordering constant are determined \cite{bib:001, bib:002, bib:081, bib:025, bib:026, bib:027}. 

Such anomalies and restrictions are known also in the quantum tensionless string theory. There are two kinds of anomalies in the tensionless string theory, the Lorentz anomaly and the spacetime conformal anomaly. The Lorentz anomaly in the tensionless string theory is investigated with various methods. In the BRST quantization method, the absence of the critical dimension is verified in a operator order\cite{bib:006} and the same critical dimension and ordering constant as the tensionful string theory are verified in the normal order \cite{bib:037}. On the other hand, in the light-cone gauge quantization method, the absence of the critical dimension in some operator orders and the same critical dimension in the XP-normal order\footnote{In the XP-normal order, all positive modes of coordinate $X$ and momentum $P$ are to the right of negative ones.} are verified \cite{bib:006, bib:010} as the case of the BRST quantization. 

The spacetime conformal anomaly in the tensionless string theory is also investigated with various methods. In the BRST quantization, the critical dimension of the conformal string theory is two \cite{bib:008}\footnote{The natural BRST quantization of the tensionless string tells only the information of the Lorentz anomaly. Therefore the conformal string, which has the extra coordinates to have the manifest spacetime conformal symmetry, has been considered instead of the tensionless string. Of course, the action of the conformal string is classically equivalent are to that of the tensionless string.}. On the other hand, in the light-cone gauge quantization, the case of two dimensions is not under consideration\footnote{Or non-anomalous trivially or topological.} because of no stringy dynamical variable. Furthermore it is verified by the calculation without the mode expansion that the dangerous commutator, $[\mathcal{J}^{-I}, \mathcal{K}^{J}]$, in the light-cone gauge of higher dimensions causes the anomaly in the Reference order, called R-order for short\footnote{In the R-order, all $P$-modes are to the right of $X$-modes. } \cite{bib:012, bib:047, bib:013}.\\
Similarly, tensionless string theories with the supersymmetry\cite{bib:015, bib:016, bib:048, bib:030} and theories of a higher dimensional object\cite{bib:009, bib:011} are studied.

There is only one transverse direction in three dimensional light-cone coordinate. Therefore the Lorentz anomaly vanishes for the string theory in the light-cone gauge of $D=3$, in addition to the usual critical dimension, $D=26$ or $D=10$. Such an avoidance of the anomaly in three dimensions cannot be obtained with the BRST formalism. The difference between two quantization methods is strange and interesting. Of course, there is no necessity for the coincidence of results in two methods, and we just expect it.\\
Recently the specialty of three dimensional light-cone gauge has gathered attention and then the mass spectra of some 3-dim. tensionful string theories in the light-cone gauge have been investigated in detail \cite{bib:003}. Because there is not any restriction except for $D=3$ in the case of the bosonic tensionful string, the operator ordering constant is undetermined. Such an ambiguity is removed for a tensionful superstring \cite{bib:003, bib:004, bib:005, bib:020}. Similarly, the ambiguity in three dimensional theories may be removed by the requirement of some larger symmetry. \\

In this paper, we consider a tensionless closed bosonic string in the light-cone gauge and investigate whether there is the anomaly of the spacetime conformal symmetry in the theory. Because the appearance of the anomaly in the spacetime conformal symmetry is different between the case of $D>3$ and the case of $D=3$ and depends on the choice of the operator order, we must consider dangerous commutators in the spacetime conformal symmetry of $D>3$ and $D=3$ in each operator order separately. The dangerous commutators which we must calculate at least are $[\mathcal{J}^{-I},\mathcal{K}^{K}]$ in $D>3$ and $\tilde{\mathcal{K}}^{-}\equiv -i[\mathcal{J}^{-}, \tilde{\mathcal{K}}^{-}]$ and $[\mathcal{J}^{-}, \tilde{\mathcal{K}}^{-}]$ in $D=3$.

The main products of our study in this paper are the avoidance of the spacetime conformal anomaly in one type of the operator orders, the Hermitian R-order\footnote{The Hermitian R-order is defined as the Hermitian version of the R-order. It will be explained in section 3.} (and the Weyl order) and the explicit calculation of the spacetime conformal anomaly in another type of the operator orders, the XP-normal order. A part of the first product is based on \cite{bib:999}.\\

The organization of this paper is below.
\par 
In section 2 we review the classical theory of a tensionless string in the light-cone gauge. We give the action for the tensionless string and the classical generators of the spacetime conformal symmetry in the light-cone gauge. 

In section 3 the above tensionless string theory in the light-cone gauge is quantized. Then some candidates for the string ground state and the operator order are shown. One type of the operator orders is the Hermitian R-order (and the Weyl order). Another type of them is the XP-normal order.

In section 4 we consider the structures of commutators in some examples for the operator order. Then we find the structures of dangerous commutators to use them in the following sections.

In section 5 we consider dangerous commutators of the spacetime conformal symmetry in the Hermitian R-order. Although we need some regularization in the Hermitian R-order, we calculate $[\mathcal{J}^{-I}_{\scriptscriptstyle{HR}},\mathcal{K}^{K}_{\scriptscriptstyle{HR}}]$ in $D>3$ and $\tilde{\mathcal{K}}^{-}_{\scriptscriptstyle{HR}}\equiv -i[\mathcal{J}^{-}_{\scriptscriptstyle{HR}}, \tilde{\mathcal{K}}^{-}_{\scriptscriptstyle{HR}}]$ and $[\mathcal{J}^{-}_{\scriptscriptstyle{HR}}, \tilde{\mathcal{K}}^{-}_{\scriptscriptstyle{HR}}]$ in $D=3$ without the explicit regularization to verify that the anomaly irrelevant to the choice of the regularization does not exist. The concrete calculation with the cut-off regularization is shown in appendix.

In section 6 we consider dangerous commutators of the spacetime conformal symmetry in the XP-normal order. We calculate $[\mathcal{J}^{-I}_{\scriptscriptstyle{XP}},\mathcal{K}^{K}_{\scriptscriptstyle{XP}}]$ in $D>3$ and $\tilde{\mathcal{K}}^{-}_{\scriptscriptstyle{XP}}\equiv -i[\mathcal{J}^{-}_{\scriptscriptstyle{XP}}, \tilde{\mathcal{K}}^{-}_{\scriptscriptstyle{XP}}]$ and $[\mathcal{J}^{-}_{\scriptscriptstyle{XP}}, \tilde{\mathcal{K}}^{-}_{\scriptscriptstyle{XP}}]$ in $D=3$ explicitly to find the spacetime conformal anomaly in both case of dimensions.

In section 7 we discuss other types of string, super and open to comment on the difference from the tensionless bosonic closed string theory in supersymmetric and open string theories respectively. Then we estimate whether the anomaly in commutators exists or not. Finally, we summarize this paper and give the outlook for future works. 

In appendix we give the following contents: the commutation relations of the spacetime symmetry in $D>3$ and $D=3$, the cut-off regularization as an explicit example and the calculation with the cut-off regularization of dangerous commutators in the Hermitian R-order.

\section{Tensionless String and Global Symmetry}
In this section we review the classical theory of a tensionless closed bosonic string. 

\subsection{Action of tensionless string and generators of its global symmetry}
We firstly change the Nambu-Goto action to the equivalent action on the phase space and then consider the action of a tensionless string in the same way as the point-particle case.\footnote{The action of a massive point particle, $S_{m}=-m\int dt \sqrt{-\dot{x}^{\mu }\dot{x}_{\mu }} $, is equivalent to $S'_{m}=\int dt \left[ \dot{x}^{\mu }p_{\mu } -\frac{1}{2}v\left( p^{2}+m^{2} \right) \right] $. By choosing $m=0$ in $S'_{m}$, we obtain the action of a massless point particle as $S=\int dt  \left[ \dot{x}^{\mu }p_{\mu } -\frac{1}{2}v p^{2} \right] $.}

The Nambu-Goto action for a string with tension $T$ \footnote{In our convention, $T=\frac{1}{\alpha '}$.} is
\begin{eqnarray}
S[\mathbf{X}]=-T\int d\tau \oint \frac{d\sigma }{2\pi } \sqrt{\left( \left( \dot{\mathbf{X}}^{\mu } \mathbf{X}_{\mu }' \right) ^2 -\dot{\mathbf{X}} ^2 (\mathbf{X}')^2 \right) } ~,
\label{eq:2-1}
\end{eqnarray}
where $\tau $ is the time-like coordinate on the world sheet and $\sigma $ is the space-like coordinate with the period of $2\pi $ on the world sheet.
Here the overdot ($~\dot{ }~$) means $\tau $-derivative and the prime ($~{}'~$) means $\sigma $-derivative.
This action is classically equivalent to 
\begin{eqnarray}
S[\mathbf{X},\mathbf{P};V,U] = \int d\tau \oint \frac{d\sigma }{2\pi } \left\{ \dot{\mathbf{X}} ^{\mu } \mathbf{P}_{\mu } -\frac{1}{2} V\left[ \mathbf{P}^2 +(T\mathbf{X}') ^2 \right] -U \mathbf{X}^{\mu \prime }\mathbf{P}_{\mu } \right\} ~,
\label{eq:2-2}
\end{eqnarray}
where $P_{\mu }$ is a conjugate momentum of $X^{\mu }$, and $V$ and $U$ are Lagrange multipliers. In the case of $T \not= 0$, we solve $P_{\mu }$, $U$ and $V$ in order to get the original action in (\ref{eq:2-1}).

We can choose $T=0$ in (\ref{eq:2-2}) in the same way as the case of a point particle.
\begin{eqnarray}
S[\mathbf{X},\mathbf{P};V,U] = \int d\tau \oint \frac{d\sigma }{2\pi } \left\{ \dot{\mathbf{X}} ^{\mu } \mathbf{P}_{\mu } -\frac{1}{2} V \mathbf{P}^2  -U \mathbf{X}^{\mu \prime }\mathbf{P}_{\mu } \right\} ~.
\label{eq:2-3}
\end{eqnarray}
This is one of the action of a tensionless closed string. Because the tensionlessness is clear and we can easily solve the constraints in the light-cone gauge, we use (\ref{eq:2-3}) as the action of a tensionless string in this paper.\footnote{Of course, there are other equivalent actions for a tensionless string. For example, $S'[\mathbf{X},V^{a}] = \int d^{2}\xi ~ V^{a}\partial _{a}\mathbf{X}^{\mu } V^{b}\partial _{b}\mathbf{X}_{\mu } $ where $a=\{ 0, 1 \}$ \cite{bib:046, bib:012, bib:013}. The world sheet diffeomorphism in this action is clear, $\delta \mathbf{X}^{\mu }=\gamma ^{a}\partial _{a}\mathbf{X}^{\mu },~ \delta V^{a} = -V^{b}\partial _{b}\gamma ^{a} +\gamma ^{b}\partial _{b}V^{a}+\frac{1}{2}(\partial _{b}\gamma ^{b})V^{a}$, and this action is a better representation for the covariant treatment such as the BRST quantization.}\\

The action in (\ref{eq:2-3}) has the next gauge symmetry corresponding to the world sheet diffeomorphism.
\begin{eqnarray}
\left\{ \begin{array}{l}
\delta \mathbf{X}^{\mu } = \alpha \mathbf{P}^{\mu } +\beta \mathbf{X}^{\mu \prime }  \\
\delta \mathbf{P}_{\mu } = (\beta \mathbf{P}_{\mu })'  \\
\delta V = \dot{\alpha } +U'\alpha -U\alpha ' +V'\beta -V\beta '  \\
\delta U = \dot{\beta } +U'\beta -U\beta ' .
\end{array}\right. 
\label{eq:2-4}
\end{eqnarray}
And the field equations of $X^{\mu }$, $P_{\mu }$, $V$ and $U$ are
\begin{eqnarray}\begin{split}
& \dot{\mathbf{P}}_{\mu }-(U \mathbf{P}_{\mu })' = 0 ~,\\
& \dot{\mathbf{X}}^{\mu }-U\mathbf{X}^{\mu \prime }-V \mathbf{P}^{\mu } = 0 ~,
\end{split}\label{eq:2-5}
\end{eqnarray}
\begin{eqnarray}\begin{split}
& \mathbf{P}^2 = 0 ~,\\
& \mathbf{X}^{\mu \prime }\mathbf{P}_{\mu } = 0 ~.
\end{split}\label{eq:2-6}
\end{eqnarray}
The equations in (\ref{eq:2-6}) are constraint conditions.\\

The action in (\ref{eq:2-3}) has the spacetime conformal symmetry as the global symmetry. The generators of the spacetime conformal symmetry are shown below.
\begin{eqnarray}\begin{split}
\mathcal{P}_{\mu } &= \oint \frac{d\sigma }{2\pi } \mathbf{P}_{\mu } ~,\\
\mathcal{J}^{\mu \nu } &= \oint \frac{d\sigma }{2\pi } \left[ \mathbf{X}^{\mu }\mathbf{P}^{\nu } -\mathbf{X}^{\nu }\mathbf{P}^{\mu } \right] ~,\\
\mathcal{D} &= \oint \frac{d\sigma }{2\pi } \mathbf{X}^{\mu } \mathbf{P}_{\mu } ~,\\
\mathcal{K}^{\mu } &= \oint \frac{d\sigma }{2\pi } \left[ \mathbf{X}^{\mu }(\mathbf{X}^{\nu } \mathbf{P}_{\nu }) -\frac{1}{2}\mathbf{X}^2 \mathbf{P}^{\mu } \right] ~.
\end{split}\label{eq:2-7}
\end{eqnarray}
$\mathcal{P}_{\mu }$, $\mathcal{J}^{\mu \nu }$, $\mathcal{D}$ and $\mathcal{K}^{\mu }$ are the generators of the translation, the Lorentz transformation, the dilatation and the special conformal transformation respectively.\\
Under the gauge transformation in (\ref{eq:2-4}), the generators of the translation and the Lorentz transformation are invariant and the generators of the dilatation and the special conformal transformation are weakly invariant. Thanks to the gauge invariance of generators, we can use these generators in the light-cone gauge only by changing the spacetime index.\\
Furthermore, from field equations, we find that generators are independent of $\tau $. Therefore all we consider is only the case of $\tau =0$.

\subsection{Generators in light-cone gauge}
In this subsection we fix the gauge symmetry in (\ref{eq:2-4}) and show the representations of generators (\ref{eq:2-7}) in the light-cone gauge. 

\subsubsection{Gauge fixing}
First we define the light-cone coordinate as
\begin{eqnarray}\begin{split}
& X^{\pm } \equiv \frac{1}{\sqrt{2}} (\mathbf{X}^1\pm \mathbf{X}^0) ~,~~ X^{I} \equiv \mathbf{X}^{I} ~,\\
& P_{\pm } \equiv \frac{1}{\sqrt{2}} (\mathbf{P}_1\pm \mathbf{P}_0)=P^{\mp } ~,~~ P_{I} \equiv \mathbf{P}_{I} ~, 
\end{split}\label{eq:2-8}
\end{eqnarray}
where $I$ is the index of transverse directions, $I=2, \dots ,D-1$. And we define the Dot product in light-cone coordinate as $V\cdot U \equiv \sum _{I=2}^{D-1}V^{I}U^{I}$.\\
The light-cone gauge is 
\begin{eqnarray}
X^{+} =\tau ~,~~ P_{-}=p_{-}(\tau ) \not= 0 .
\label{eq:2-9}
\end{eqnarray}
This gauge choice fixes most of the gauge freedoms such that $\alpha =0$ and $\beta =\beta _0(\tau )$. The residual gauge symmetry by $\beta _0(\tau )$ corresponding to $\sigma $-shift is fixed by $u= \oint \frac{d \sigma }{2 \pi } U = 0$ and gives the only one constraint.

By using field equations\footnote{Field equations of $X$ and $P$ solve the $\tau $-dependence of fields. From here, we assume that the $\tau $-dependence of fields is solved and then we consider only the case of $\tau =0$. } and mode expansions,\footnote{
Mode expansions of $X^{I}$ and $P_{I}$ are $X^{I}(\sigma ) = \sum _{n} X^{I}_{n} e^{in\sigma },~ X^{I}_{0}=x^{I}$ and $P_{I}(\sigma ) = \sum _{n} P_{I,n} e^{in\sigma },~ P_{I,0}=p_{I}$.}~ we can solve the constraints in (\ref{eq:2-6}) to obtain the explicit representations of $X^{-}$ and $P_{+}$ as follows:
\begin{eqnarray}\begin{split}
P_{+}(\sigma ) &= -\frac{1}{2p_{-}} \sum _{n} L_{n} e^{in\sigma } ~,\\
X^{-}(\sigma ) &= x^{-} - \frac{1}{p_{-}} \sum _{n\not= 0} \frac{i}{n} M_{n}(\tau ) e^{in\sigma } ~,
\end{split}\label{eq:2-10}
\end{eqnarray}
where $x^{-}$ is the zero mode of $X^{-}(\sigma )$ and 
\begin{eqnarray}\begin{split}
& L_{n} \equiv \sum _{m} P_{m}\cdot P_{n-m} ~,~~ (L_{n})^{*} = L_{-n} ~,\\
& M_{n} \equiv -i \sum _{m} m X_{m} \cdot P_{n-m} ~,~~ (M_{n})^{*} = M_{-n} ~.
\end{split}\label{eq:2-11}
\end{eqnarray}
From zero mode, $L_{0} = 2p_{-}H = p\cdot p +\mathcal{M}^2$, we obtain the mass square operator as
\begin{eqnarray}
\mathcal{M}^2 = \sum _{n \not= 0} P_{n}\cdot P_{-n} ~.
\label{eq:2-12}
\end{eqnarray}
In our convention of $\tau $, the motion of the center of the tensionless string is restricted by $\dot{x}^{\mu }\dot{x}_{\mu } = -\frac{\mathcal{M}^2}{p_{-}^2}$.\footnote{Each point of the tensionless string moves on the light-cone: $\dot{X}^{\mu }(\tau , \sigma )\dot{X}_{\mu }(\tau , \sigma ) = \frac{1}{p_{-}^2}\left( 2P_{+}(\sigma )p_{-} +P(\sigma )\cdot P(\sigma ) \right) =0$.}\\
Furthermore the residual constraint which corresponds to the gauge choice, $u=0$, is written as
\begin{eqnarray}
M_{0} \equiv -\oint \frac{d\sigma }{2\pi } \bar{X}'\cdot P = -i \sum _{n} nX_{n}\cdot P_{-n} = 0 ~.
\label{eq:2-13}
\end{eqnarray}
$L_{n}$ and $M_{m}$ in (\ref{eq:2-11}) and (\ref{eq:2-13}) satisfy the 2D Galilean conformal algebra (GCA):\footnote{The detail of 2D GCA is referred to \cite{bib:007, bib:029, bib:014}.}
\begin{eqnarray}
[L_{n},L_{m}]=0 ~,~ [L_{n},M_{m}]=(n-m)L_{n+m} ~,~ [M_{n},M_{m}]=(n-m)M_{n+m} ~.
\label{eq:2-14}
\end{eqnarray}

\subsubsection{Classical generators in light-cone gauge}
We give the representations for generators of the spacetime conformal symmetry in the light-cone gauge.

The generators of the translation in the light-cone gauge are 
\begin{eqnarray}
\mathcal{P}_{-} = p_{-} ~,~ \mathcal{P}_{I} = p_{I} ~,~ \mathcal{P}_{+} = p_{+} = -\frac{L_{0}}{2p_{-}}
\label{eq:2-15}
\end{eqnarray}

The generators of Lorentz transformation in the light-cone gauge are
\begin{eqnarray}\begin{split}
& \mathcal{J}^{+I} = -x^{I}p_{-} ~,~ \mathcal{J}^{+-} = -x^{-}p_{-} ~,~ \mathcal{J}^{IJ} = \sum _{n} (X^{I}_{n}P^{J}_{-n} - X^{J}_{n}P^{I}_{-n}) \\
& \mathcal{J}^{-I} = x^{-}p^{I} -\frac{i}{p_{-}}\sum _{n \not= 0} \frac{1}{n}M_{n}P^{I}_{-n} +\frac{1}{2p_{-}}\sum _{n}X^{I}_{n}L_{-n} .
\end{split}\label{eq:2-16}
\end{eqnarray}
In three dimensions, $\mathcal{J}^{IJ}$ does not exist and we can write Lorentz generators as vector, $\mathcal{J}^{\mu }\equiv \frac{1}{2}\epsilon ^{\mu \nu \rho }\mathcal{J}_{\nu \rho }$.\footnote{$\epsilon ^{+-2}=-\epsilon _{+-2}=1$.}
\begin{eqnarray}
\mathcal{J}^{+}=\mathcal{J}^{+2} ~,~ \mathcal{J}=-\mathcal{J}^{+-} ~,~ \mathcal{J}^{-}=-\mathcal{J}^{-2} ,
\label{eq:2-17}
\end{eqnarray}
where we omit the index of transverse direction.

The generator of the dilatation in the light-cone gauge is  
\begin{eqnarray}
\mathcal{D} = x^{-}p_{-} + \sum _{n} X_{n}\cdot P_{-n} .
\label{eq:2-18}
\end{eqnarray}

The generator of special conformal transformation in the light-cone gauge are  
\begin{eqnarray}\begin{split}
\mathcal{K}^{+} =& -\frac{1}{2}\sum _{n}X_{n}\cdot X_{-n} p_{-} \\
\mathcal{K}^{I} =& x^{I}x^{-}p_{-} +i\sum _{m \not= 0}\frac{1}{m}X^{I}_{m}M_{-m} +\sum _{n}\sum _{m} \left[ X^{I}_{n}X^{J}_{m}P_{J,-n-m} -\frac{1}{2} X^{J}_{n}X^{J}_{m}P^{I}_{-n-m} \right] \\
\mathcal{K}^{-} =& x^{-}x^{-}p_{-} +x^{-}\sum _{n}X_{n}P_{-n} +\frac{1}{p_{-}}\sum _{m \not= 0}\frac{1}{m^2}M_{m}M_{-m} \\
& -\frac{i}{p_{-}}\sum _{n}\sum _{m \not= 0} \frac{1}{m}X^{I}M_{m}P_{I,-n-m} +\frac{1}{4p_{-}}\sum _{n}\sum _{m} X^{I}_{n}X^{I}_{m}L_{-n-m} .
\end{split}\label{eq:2-19}
\end{eqnarray}
In three dimensions, $\mathcal{K}^{I}$ becomes simple a little.
\begin{eqnarray}
\mathcal{K} = \mathcal{K}^{I=2}= xx^{-}p_{-} +i\sum _{m \not= 0}\frac{1}{m}X_{m}M_{-m} +\frac{1}{2}\sum _{n}\sum _{m} X_{n}X_{m}P_{-n-m} ,
\label{eq:2-20}
\end{eqnarray}
where we omit the index of transverse direction.\\

In quantum theory, the representation of the generators in the spacetime conformal symmetry become complicated, according to the choice of the operator order. Furthermore quantum effect terms in $\mathcal{K}^{-}$ can become anomalous.

\section{Operator Order and String Ground State}
In this section we consider quantize the tensionless string theory in the last section and then consider some candidates for the operator order and the string ground state. 

\subsection{Quantization of tensionless string in light-cone gauge}
The dynamical variables of the tensionless string theory in the light-cone gauge in the last section are $X^{I}(\sigma )$ and $x^{-}$, and their conjugate momenta, $P_{I}(\sigma )$ and $p_{-}$. 
The theory is quantized by the following commutation relations: 
\begin{eqnarray}
[x^{-},p_{-}]=i ~,~ [X^{I}(\sigma ), P_{J}(\sigma ')]=2\pi i \delta ^{I}_{J} \delta (\sigma -\sigma ') ~,
\label{eq:3-1}
\end{eqnarray}
and otherwise zero. For Fourier modes, the second commutation relation is written as
\begin{eqnarray}
[x^{I},p_{J}]=i\delta ^{I}_{J} ~,~ [X^{I}_{n},P_{J,m}]=i\delta ^{I}_{J}\delta _{n+m,0} ,
\label{eq:3-2}
\end{eqnarray}
where the next delta function for the periodic function was used: 
\begin{eqnarray}
\delta (\sigma ) = \frac{1}{2\pi }\sum _{n} e^{i n \sigma } .
\label{eq:3-3}
\end{eqnarray}

\subsection{Some candidates for operator order and string ground state}
The choice of the operator order is very important in the quantum theory because it determines (or greatly affects) the physical result of the theory. As usual tensionful string theories, the vacuum for the zero-mode part is the eigenstate of momenta, $|p_{I},p_{-}\rangle $. Here, in the quantum tensionless string theory, we consider some candidates for the operator order of non-zero modes and the string ground states corresponding to them. 

In determining the operator order and the string ground state for string theory, the mass square operator, $\mathcal{M}^2$, is very important key.\footnote{The constraint condition or the constraint operator such as $M_{0}$ in (\ref{eq:2-13}) is less important than $\mathcal{M}^2$ because it is enough that there is at least one state which satisfy the constraint condition.}~ If we choose an operator order, the string ground state corresponding to the order must be the eigenstate corresponding to a minimum eigenvalue of $\mathcal{M}^2$ in the order. Or, if we choose a string ground state, the operator order corresponding to the state must order $\mathcal{M}^2$ such that the state is the eigenstate corresponding to a minimum eigenvalue of it.\\

In the case of a usual string theory with non-zero tension, $T \not= 0$, we compose the left-moving and right-moving modes from $X$-mode and $P$-mode,
\begin{eqnarray}
\alpha ^{I}_{\scriptscriptstyle{(T)} n} = -i \sqrt{\frac{T}{2}} n X^{I}_{n} +\frac{1}{\sqrt{2T}}P_{I,n} ~,~ \tilde{\alpha }^{I}_{\scriptscriptstyle{(T)} n} = -i \sqrt{\frac{T}{2}} n X^{I}_{-n} +\frac{1}{\sqrt{2T}}P_{I,-n} 
\label{eq:3-4}
\end{eqnarray}
and then consider the normal order of them. Because the mass square operator in the normal order is the set of the harmonic oscillator, the string ground state is annihilated by both positive modes.\\

In the case of the tensionless string theory, we cannot make the combination such as $\alpha ^{I}_{\scriptscriptstyle{(T)} n}$ or $\tilde{\alpha }^{I}_{\scriptscriptstyle{(T)} n}$, because the tensionless string theory has no dimensionful parameter such as tension $T$. However operator orders which are different from the normal order of left- and right-moving modes have been studied \cite{bib:013, bib:010}, e.g. the {\it Reference} order (R-order)\footnote{In the R-order, all $P$-modes are to the right of $X$-modes.} and the ``XP-normal order."\footnote{In the XP-normal order, all $X$- and $P$-positive modes are to the right of negative ones.}~ We consider these two orders and others in detail below.

\subsubsection*{R-order and similar orders}
The R-order is obtained by referring to the case of a usual tensionful string in the following way. For the sufficiently small $T$, the left- and right-moving modes in (\ref{eq:3-4}) are 
\begin{eqnarray}
\alpha ^{I}_{\scriptscriptstyle{(T)} n} \sim \frac{1}{\sqrt{2T}}P^{I}_{n} ~,~ \tilde{\alpha }^{I}_{\scriptscriptstyle{(T)} n} \sim \frac{1}{\sqrt{2T}}P^{I}_{-n} .
\label{eq:3-5}
\end{eqnarray}
By collecting both positive modes, we find that the natural candidates for the string ground state in the case of the tensionless string theory is the state annihilated by $P_{I,n}$ for $n\not= 0$, $|0\rangle _{\scriptscriptstyle{P}}$:
\begin{eqnarray}
P_{I,n} |0\rangle _{\scriptscriptstyle{P}} = 0 ~~ \mbox{for } n \not= 0.
\label{eq:3-6}
\end{eqnarray}
Because the mass square operator in the tensionless string theory is written only with $P$-mode as seen in (\ref{eq:2-12}), one of the operator orders in which $|0\rangle _{\scriptscriptstyle{P}}$ becomes the eigenstate corresponding to a minimum\footnote{The eigenvalue of the R-ordered $\mathcal{M}^2$ is positive or zero from its structure and it is clear that $|0\rangle _{\scriptscriptstyle{P}}$ is the zero-mass eigenstate of the R-ordered $\mathcal{M}^2$.} eigenvalue of $\mathcal{M}^2$ is the R-order, in which all $P$-modes are to the right of $X$-modes.\\
Of course, there are other operator orders in which $|0\rangle _{\scriptscriptstyle{P}}$ becomes the eigenstate corresponding to a minimum eigenvalue of $\mathcal{M}^2$. For example, the Weyl order and the Hermitian R-order. In the Weyl order, an operator is totally symmetrized. In the Hermitian R-order, an operator $\mathcal{O}$ is ordered into the average of the R-ordered operator $\mathcal{O}_{\scriptscriptstyle{R}}$ and its hermitian conjugate: $\mathcal{O}_{\scriptscriptstyle{HR}}\equiv \frac{1}{2}\left( \mathcal{O}_{\scriptscriptstyle{R}}+ (\mathcal{O}_{\scriptscriptstyle{R}})^{\dagger } \right)$. For a generator or a commutator of lower degree than 4, the Hermitian R-ordered one equals the Weyl ordered one. However, for a generator or a commutator of 4th degree or higher degree such as $\mathcal{K}^{-}$ and $[\mathcal{J}^{-I}, \mathcal{K}^{-}]$, both ordered ones are different generally. Such differences affect the investigation of the anomaly in the spacetime conformal symmetry.\\
$\mathcal{M}^2$ in the R-order, the Hermitian R-order and the Weyl order are the same as each other and $M_{0}$ in these orders are also the same because of $\sum _{n}n=0$. Therefore the physical mass spectrum is equivalent to each other. Furthermore, because there is no difference between the R-order case, the Weyl order case and the Hermitian R-order case in $[\mathcal{J}^{-I}, \mathcal{J}^{-J}]$ as seen in the next section, the check of the Lorentz anomaly in these operator orders is done similarly.

There is a problem in the calculation in the R-order. Because most of generators in the R-order are not hermitian, we have to {\it hermitianize} them. The generators of the translation which consist only of $P$-modes do not need the {\it hermitianization}, but other generators need usually. We define the generator in the R-order as follow:
\begin{eqnarray}
\mathcal{G} = \frac{1}{2}\bigl( \mathcal{G}_{\scriptscriptstyle{R}}+(\mathcal{G}_{\scriptscriptstyle{R}})^{\dagger } \bigr) = \mathcal{G}_{\scriptscriptstyle{R}} -ig_{\scriptscriptstyle{R}} +\cdots ,
\label{eq:3-7}
\end{eqnarray}
where $\mathcal{G}_{\scriptscriptstyle{R}}$, $g_{\scriptscriptstyle{R}}$ and $\cdots $ are R-ordered operators. For some generators in the R-order, $g_{R}$ and $\cdots $ contain divergent terms. Therefore we need some regularization.\\
As seen in appendix, the hermitian version of operators in the R-order has non-hermitian divergent terms and non-hermitian dropping terms. These terms break the hermitianity of operators in the limit of the regulator. Therefore, in this paper, we use the Hermitian R-order as the operator order corresponding to $|0\rangle _{\scriptscriptstyle{P}}$ to calculate commutators in the spacetime symmetry.\footnote{The calculation in the Weyl order is very similar to the case of the Hermitian R-order but more difficult.}

\subsubsection*{XP-normal order}
The mass square operator, $\mathcal{M}^2$, in the tensionless string theory consists only of $P$-mode as seen in (\ref{eq:2-12}). Because $P$-modes commute mutually, we can always move all positive $P$-modes to the right of negative ones without any extra term. Therefore the state annihilated by positive modes of $X$ and $P$ is the eigenstate corresponding to zero eigenvalue of $\mathcal{M}^2$. We represent it as $|0\rangle _{\scriptscriptstyle{XP}}$, which satisfy 
\begin{eqnarray}
X^{I}_{n} |0\rangle _{\scriptscriptstyle{XP}} = P_{I,n} |0\rangle _{\scriptscriptstyle{XP}} =0 ~~ \mbox{for } n>0 .
\label{eq:3-13}
\end{eqnarray}
The operator order corresponding to $|0\rangle _{\scriptscriptstyle{XP}}$ is the XP-normal order, in which all positive modes of $X$ and $P$ are to the right of negative ones \cite{bib:010}.\\
In the XP-normal order, it is known that mass eigenstates except for the string ground state $|0\rangle _{\scriptscriptstyle{XP}}$ have a problem in their norms.\footnote{In the case of $D>3$\cite{bib:010}, the avoidance of the Lorentz anomaly requires $M_{0}=2$. This condition means that physical stets must be at level 2. The mass eigenstates at level 2 are $P^{I}_{-2}|0\rangle _{\scriptscriptstyle{XP}}$, $P^{I}_{-1}P^{J}_{-1}|0\rangle _{\scriptscriptstyle{XP}}$ and $(X^{I}_{-1}P^{J}_{-1}-X^{J}_{-1}P^{I}_{-1})|0\rangle _{\scriptscriptstyle{XP}}$. Their norms are zero or negative. Other states at level 2 are $(X^{I}_{-1}P^{J}_{-1}+X^{J}_{-1}P^{I}_{-1})|0\rangle _{\scriptscriptstyle{XP}}$, $X^{I}_{-1}X^{J}_{-1}|0\rangle _{\scriptscriptstyle{XP}}$ and $X^{I}_{-2}|0\rangle _{\scriptscriptstyle{XP}}$, but they are not mass eigenstate. In the case of $D=3$, the Lorentz anomaly does not exist trivially. Then the various level is possible. However, the discussion of the norm is similar.}~ In section 6, we will find that physical states are restricted to the string ground state in terms of the spacetime conformal anomaly. 

\subsubsection*{Other operator orders}
For other candidates of the operator order in the tensionless string theory, we may consider the orders in which linear-combinational modes of $X$ and $P$ are normally ordered in the same way as the case of the usual tensionful string theory. However, such operator orders have problems in the positivity of the mass square operator and the norm of mass eigenstates \cite{bib:010}. Therefore we don't study them in this paper.

\subsection{Operator orders used in the following sections}
We collect the operator orders and the string ground states used in the following sections below. 
\begin{table}[htb]\begin{center}
Operator Orders used in the Following Sections
\begin{tabular}{|c|c|c|}
\hline
Order name & Ordering rule & ground state \\
\hline
\hline
R-order & all $P$-modes are to the right of $X$-modes & $|0\rangle _{\scriptscriptstyle{P}}$ \\ \hline
Hermitian R-order${}^{\ast }$ & the {\it hermitian average} of the R-order & $|0\rangle _{\scriptscriptstyle{P}}$ \\ \hline
XP-normal order & all positive modes of $X$ \& $P$ are to the right of negatives & $|0\rangle _{\scriptscriptstyle{XP}}$ \\ \hline
\hline
\end{tabular}
\end{center}
$\ast $ : Weyl order is similar.
\end{table}

\section{Structure of Commutators}
Some commutators in the spacetime conformal symmetry contain terms of higher degree with respect to Fourier modes, $X_{n}$ and $P_{n}$. For example, $[\mathcal{J}^{-I},\mathcal{K}^{K}]$ contains operators of degree 4 and $[\mathcal{J}^{-I},\mathcal{K}^{-}]$ contains operators of degree 5. Furthermore, in the (Hermitian) R-order and the Weyl order, we need the regularization because of divergent terms. In the XP-normal order, we will find the central extension terms in some commutators. Therefore the calculation of commutators in the spacetime conformal symmetry is very complicated.

Before we calculate concretely (dangerous) commutators of the spacetime conformal symmetry in the tensionless string theory, we investigate the relation between the choice of operator orders and the structure of commutators. The effect of quantization in the calculation of commutators arises from terms which are exchanged with other operators several times to be ordered correctly. Such quantum effect terms, which is differences from classical results, depend on the choice of the operator order and are in some cases dangerous, even anomalous.

If we know how quantum effects appear in commutators, we can determine the possible structure of commutators of generators to some extent to help the concrete calculation of commutators in the spacetime conformal symmetry. In this section, firstly, we show some examples to investigate the structure of quantum effect terms which arise when a hermitian operator is ordered in a given operator order. Then we give the possible structure of some dangerous commutators in the Hermitian R-order and the XP-normal order.

\subsection{Example for structure of quantum effect terms}
The generators are hermitian operators. The commutator of two hermitian operators is the imaginary unit $i$ times a hermitian operator, anti-hermite.  
Because the hermitian-operator part is usually not ordered, we must order it correctly. At that time, quantum effect terms arise.

In this subsection, by showing some examples, we consider how quantum effect terms arise at the time when we order a hermitian operator in a given operator order. Then we give the possible structure of a dangerous commutator, $[\mathcal{J}^{-I},\mathcal{K}^{K}]$ in $D>3$,\footnote{The commutator corresponding to it in $D=3$ is $[\mathcal{J}^{-},\mathcal{K}]$. This is not dangerous because there is only one transverse direction in $D=3$. This define $\tilde{\mathcal{K}}^{-}$. We will find easily the structure of $\tilde{\mathcal{K}}^{-}$ from that of $[\mathcal{J}^{-I},\mathcal{K}^{K}]$.}~ in the Hermitian R-order and the XP-normal order. Other dangerous commutator are shown in the next subsection.

\subsubsection{Example 1 : Coordinate and Momentum}
First we consider the coordinate $x$ and the momentum $p$ which satisfy the usual commutation relation $[x,p]=i$ as fundamental operators.
We compare some hermitian combinations with Weyl ordered ones. $(A)_{W}$ indicates Weyl ordered version of operator $A$. 
\subsubsection*{Cubic example}
Cubic operators of the product of two $x$ and one $p$ case are $xxp$, $xpx$ and $pxx$. Hermitian combinations of them are $xpx$, $\frac{1}{2}(xxp+pxx)$ and $(xxp)_{W}\equiv \frac{1}{3}(xxp+xpx+pxx)$. 
The first and second combinations cause no extra term when they are Weyl-ordered.
\begin{eqnarray}
xpx = \frac{1}{2}(xxp+pxx) = (xxp)_{W} .
\label{eq:4-1}
\end{eqnarray}
Operators of the product of one $x$ and two $p$ are discussed similarly.
\subsubsection*{Quartic example}
Hermitian combinations of quartic operators which are the product of three $x$ and one $p$ are $\frac{1}{2}(xxxp+pxxx)$, $\frac{1}{2}(xxpx+xpxx)$, $(xxxp)_{W}\equiv \frac{1}{4}(xxxp+xxpx+xpxx+pxxx)$ and so on. The first and second combinations cause no extra term when they are Weyl-ordered:
\begin{eqnarray}
\frac{1}{2}(xxxp+pxxx) = \frac{1}{2}(xxpx+xpxx) = (xxxp)_{W} .
\label{eq:4-2}
\end{eqnarray}
Operators of the product of one $x$ and three $p$ are discussed similarly.

Hermitian combinations of operators which are the product of two $x$ and two $p$ are $xppx$, $pxxp$, $\frac{1}{2}(xpxp+pxpx)$, $\frac{1}{2}(xxpp+ppxx)$, $(xxpp)_{W}\equiv \frac{1}{6}(xxpp+xpxp+xppx+pxxp+pxpx+ppxx)$ and so on. They are Weyl-ordered as
\begin{eqnarray}\begin{split}
xppx = pxxp = (xxpp)_{W} +\frac{1}{2} \\
\frac{1}{2}(xpxp+pxpx) = (xxpp)_{W} \\
\frac{1}{2}(xxpp+ppxx) = (xxpp)_{W} -\frac{1}{2} .
\end{split}\label{eq:4-3}
\end{eqnarray}
In this case, constant quantum effect terms are possible, but quadratic ones are impossible. We find in the last line that there is an quantum difference between the Hermitian R-ordered operator and the Weyl-ordered one. Note that quantum effect terms are in the degree which is 4 degrees lower than the highest degree.\\

Similarly, in the higher degree case, quantum effect terms can appear in terms which is 4 degrees lower than the highest degree.

\subsubsection{Example 2 : Creation and Annihilation Operators}
Next we consider the creation operator $a^{\dagger }$ and the annihilation operator $a$. They are Hermitian conjugate with each other and satisfy the usual commutation relation, $[a,a^{\dagger }]=1$. Here we compare some Hermitian combinations with the normal-ordered ones. 
\subsubsection*{Cubic example}
The example of cubic hermitian operators is below.
\begin{eqnarray}
aa^{\dagger }a +a^{\dagger }aa^{\dagger } = a^{\dagger }aa +a^{\dagger }a^{\dagger }a +a+a^{\dagger } .
\label{eq:4-4}
\end{eqnarray}
The quantum effect terms are linear, which is 2 degrees lower than the highest degree, unlike the case of $x$ and $p$.
The reason is that $a$ and $a^{\dagger }$ are not self Hermitian conjugate and are complex combinations of $x$ and $p$, $a\leftrightarrow \frac{1}{\sqrt{2}}(p-ix),~ a^{\dagger }\leftrightarrow \frac{1}{\sqrt{2}}(p+ix)$.
\subsubsection*{Quartic example}
The examples of quartic operators which are the product of two $a$ and two $a^{\dagger }$ are 
\begin{eqnarray}\begin{split}
a^{\dagger }aa^{\dagger }a =& a^{\dagger }a^{\dagger }aa +a^{\dagger }a ,\\
aa^{\dagger }aa^{\dagger } =& a^{\dagger }a^{\dagger }aa +3a^{\dagger }a +1 ,\\
aaa^{\dagger }a^{\dagger } =& a^{\dagger }a^{\dagger }aa +4a^{\dagger }a +2 ,\\
\frac{1}{2}(aa^{\dagger }a^{\dagger }a+a^{\dagger }aaa^{\dagger }) =& a^{\dagger }a^{\dagger }aa +2a^{\dagger }a .
\end{split}\label{eq:4-5}
\end{eqnarray}
Thus the possible quantum effect terms are quadratic, which is 2 degrees lower than the highest degree, and constant terms, which is 4 degrees lower than the highest degree.\\

Similarly, in the higher degree case, there is the possibility that quantum effect terms appear in the degree which is 2 degrees lower than the highest degree.

\subsubsection{Example 3 : Fourier Modes}
In this subsubsection we consider Fourier modes of coordinate and momentum for a closed string, $X_{n}$ and $P_{n}$. For simplicity, we omit the index of the spacetime. $X_{n}$ and $P_{n}$ satisfy the usual commutation relation and the reality condition, $[X_{n}, P_{m}] =i\delta _{n,-m}$ and $(X_{n})^{\dagger }=X_{-n}$ and $(P_{n})^{\dagger }=P_{-n}$.\\
We consider quartic operators of $X$-modes and $P$-modes as example. Because the ``dangerous" commutators such as $[\mathcal{J}^{-I},\mathcal{J}^{-J}]$ and $[\mathcal{J}^{-I},\mathcal{K}^{K}]$ are quartic\footnote{In the highest degree, the former contain quartic terms of the product of one $X$-mode and three $P$-modes and the latter contain quartic terms of the product of two $X$-modes and two $P$-modes.} and the cases of the higher degree are similarly discussed, it is important and instructive for finding the structure of quantum effect terms in dangerous commutators to consider the quartic operators as examples.\\
Here we concretely consider hermitian combinations of quartic operators which are the product of two $X$-modes and two $P$-modes such that the summation of indexes in each term is zero. We will find below that they can cause a constant quantum effect term as well as quadratic terms.
 
First we order some Hermitian combinations in the Hermitian R-order. For example, 
\begin{eqnarray}\begin{split}
& X_{n}P_{l}X_{m}P_{-n-m-l}+P_{n+m+l}X_{-m}P_{-l}X_{-n} \\
&~~~ = X_{n}X_{m}P_{l}P_{-n-m-l}+P_{n+m+l}P_{-l}X_{-m}X_{-n} -i\delta _{m,-l}(X_{n}P_{-n}-P_{n}X_{-n}) ,\\ 
& X_{n}P_{l}P_{-n-m-l}X_{m}+X_{-m}P_{n+m+l}P_{-l}X_{-n} \\
& ~~~ = X_{n}X_{m}P_{l}P_{-n-m-l}+P_{n+m+l}P_{-l}X_{-m}X_{-n} -i(\delta _{n,-l}+\delta _{m,-l}) (X_{n}P_{-n}-P_{n}X_{-n}) .
\end{split}\label{eq:4-6}
\end{eqnarray}
Thus quantum effect terms appear in the second degree, which is 2 degrees lower than the highest.\\
If there are index-inverted partners, terms of the second degree make commutators, e.g. $[X_{n}, P_{-n}]=i$. In this case, we obtain only the constant quantum effect term, which is 4 degrees lower than the highest. 

More generally, we consider the next operator which has the same structure as generators or commutators in the spacetime symmetry for the tensionless string theory.
\begin{eqnarray}
\sum _{n} \sum _{m} \sum _{l} f_{n,m,l} X_{n}X_{m}P_{l}P_{-n-m-l}
\label{eq:4-7}
\end{eqnarray}
where the summation symbol $\sum _{n}$ indicates $\sum _{n=-\infty }^{\infty }$ and the coefficient has the symmetry which invert the sign of the Fourier mode index, $f_{n,m,l}=f_{-n,-m,-l}$,\footnote{This symmetry corresponds to the world sheet parity symmetry.}~ so called ``mode flipping symmetry." Here we do not use other symmetries for the coefficient such as $f_{n,m,l}=f_{m,n,l}$ and $f_{n,m,l}=f_{n,m,-n-m-l}$.\\
The Hermitian R-order version of (\ref{eq:4-7}) is
{\setlength\arraycolsep{1pt}\begin{eqnarray}
&&\frac{1}{2}\sum _{n} \sum _{m} \sum _{l} f_{n,m,l} ( X_{n}X_{m}P_{l}P_{-n-m-l} + P_{n+m+l}P_{-l}X_{-m}X_{-n} ) \nonumber \\
&&~~~ = \frac{1}{2}\sum _{n} \sum _{m} \sum _{l} f_{n,m,l} ( X_{n}X_{m}P_{l}P_{-n-m-l} + P_{-n-m-l}P_{l}X_{m}X_{n} ) .
\label{eq:4-8}
\end{eqnarray}}
Next we compare other Hermitian combinations with (\ref{eq:4-8}). If the operator order preserves the mode flipping symmetry, a quantum effect term in Hermitian operator of degree 4 is only constatnt. We explicitly show the fact below.

\subsubsection{Operator order with mode flipping symmetry}
We reorder some Hermitian combinations preserving the mode flipping symmetry in the Hermitian R-order and compare them with (\ref{eq:4-8}).
\subsubsection*{v.s. Sandwich order}
For example, the Sandwich ordered-(\ref{eq:4-8}) is reordered in the Hermitian R-order as
\begin{eqnarray}\begin{split}
& \frac{1}{2}\sum _{n} \sum _{m} \sum _{l} f_{n,m,l} ( X_{n}P_{l}P_{-n-m-l}X_{m} + X_{-m}P_{n+m+l}P_{-l}X_{-n} ) \\
& ~~~ = (\mbox{\ref{eq:4-8}}) -\frac{1}{2}i \sum _{n} \sum _{m} (f_{n,m,-n}+f_{n,m,-m}) (X_{n}P_{-n} -P_{n}X_{-n}) \\
& ~~~ = (\mbox{\ref{eq:4-8}}) +\frac{1}{2} \sum _{n} \sum _{m} (f_{n,m,-n}+f_{n,m,-m}) .
\end{split}\label{eq:4-9}
\end{eqnarray}
The quantum effect term is constant, which in 4 degrees lower than the highest degree. 
\subsubsection*{v.s. Weyl order}
The Weyl ordered-(\ref{eq:4-8}) is reordered in the Hermitian R-order as
\begin{eqnarray}\begin{split}
& \sum _{n} \sum _{m} \sum _{l} f_{n,m,l} (X_{n}X_{m}P_{l}P_{-n-m-l})_{W} \\
&~~~ \equiv  \frac{1}{24}\sum _{n} \sum _{m} \sum _{l} f_{n,m,l} ( X_{n}X_{m}P_{l}P_{-n-m-l} + (\mbox{totally symmetric}) ) \\
&~~~ = (\mbox{\ref{eq:4-8}}) +\frac{1}{4} \sum _{n} \sum _{m} (f_{n,m,-n}+f_{n,m,-m}) 
\end{split}\label{eq:4-10}
\end{eqnarray}
and the quantum effect term is constant, which is 4 degrees lower than the highest degree. \\
Thus we find that quantum differences of a hermitian operator between some two orders preserving the mode flipping symmetry is 4 degrees lower than the highest degree. 
\subsubsection*{Structure of $[\mathcal{J}^{-I}_{\scriptscriptstyle{HR}}, \mathcal{K}^{K}_{\scriptscriptstyle{HR}}]$}
From above two example, we can induce the structure of $[\mathcal{J}^{-I},\mathcal{K}^{K}]$ in operator orders preserving the mode flipping symmetry. For example, in the Hermitian R-order,\footnote{The possible structure of commutators in the Weyl order, which is also preserve the mode flipping symmetry, is the same.}~ we find the next structure by seeing the quantum versions of (\ref{eq:2-16}) and (\ref{eq:2-19}). 
\begin{eqnarray}
[\mathcal{J}^{-I}_{\scriptscriptstyle{HR}}, \mathcal{K}^{K}_{\scriptscriptstyle{HR}}] = i(\mbox{ordered classical part}) + i\delta ^{I,K} \frac{C}{p_{-}} ,
\label{eq:4-11}
\end{eqnarray}
where terms with $M_{0}$ at the right end are included in the first part and $C$ is constant. We need some regularization if the constant $C$ diverges. The detail of the regularization is considered in appendix.\\
The structure of $[\mathcal{J}^{-}_{\scriptscriptstyle{HR}}, \mathcal{K}_{\scriptscriptstyle{HR}}]$ in $D=3$ is obtained similarly.

\subsubsection{Operator order without mode flipping symmetry : XP-normal order}
As an example of the operator order which breaks the mode flipping symmetry, we consider the XP-normal order. Here we ignore zero mode for simplicity.\footnote{We assume that the summation of (\ref{eq:4-8}) is restricted over $n\not= 0, m\not= 0, l\not= 0, n+m+l\not= 0$, or that the coefficient $f_{n,m,l}$ is restricted.} Therefore, all summation symbols are unified into the sum over the half range, $\sum _{n>0}$.

Now we reorder (\ref{eq:4-8}) into the XP-normal order.\\ 
We divide the summation into four regions where coefficients equal. The part of $(n,m,l)=(+++),(---)$ with the coefficient $f_{n,m,l}=f_{-n,-m,-l}~(n>0,m>0,l>0)$ in (\ref{eq:4-8}) is written as
\begin{eqnarray}\begin{split}
& \frac{1}{2} (X_{n}X_{m}P_{l}P_{-n-m-l}+h.c.) +\frac{1}{2}(X_{-n}X_{-m}P_{-l}P_{n+m+l}+h.c.) \\
&~ = P_{-n-m-l}P_{l}X_{m}X_{n}+X_{-n}X_{-m}P_{-l}P_{n+m+l} .
\end{split}\label{eq:4-12}
\end{eqnarray}
Note that all operators commute with each other. There is no extra term.\\
The part of $(n,m,l)=(++-),(--+)$ with the coefficient $f_{n,m,-l}=f_{-n,-m,l}~(n>0,m>0,l>0)$ in (\ref{eq:4-8}) is written as
\begin{eqnarray}\begin{split}
& \frac{1}{2}(X_{n}X_{m}P_{-l}P_{-n-m+l}+h.c.) +\frac{1}{2}(X_{-n}X_{-m}P_{l}P_{n+m-l}+h.c.)  \\
&~ = P_{-l}P_{-n-m+l}X_{m}X_{n}+X_{-n}X_{-m}P_{n+m-l}P_{l} \\
&~~~~ -\frac{i}{2}(\delta _{n,l}+\delta _{m,l}) (X_{-n}P_{n}+X_{-m}P_{m}-P_{-n}X_{n}-P_{-m}X_{m}) -(\delta _{n,l}+\delta _{m,l}) .
\end{split}\label{eq:4-13}
\end{eqnarray}
The part of $(n,m,l)=(+-+),(-+-)$ with the coefficient $f_{n,-m,l}=f_{-n,m,-l}~(n>0,m>0,l>0)$ in (\ref{eq:4-8}) is written as
\begin{eqnarray}\begin{split}
& \frac{1}{2}(X_{n}X_{-m}P_{l}P_{-n+m-l}+h.c.) +\frac{1}{2}(X_{-n}X_{m}P_{-l}P_{n-m+l}+h.c.) \\
&~ = X_{-m}P_{-n+m-l}P_{l}X_{n}+X_{-n}P_{-l}P_{n-m+l}X_{m} \\
&~~~~ +\frac{i}{2}\delta _{m,l}(X_{-n}P_{n}+X_{-m}P_{m}-P_{-n}X_{n}-P_{-m}X_{m}) .
\end{split}\label{eq:4-14}
\end{eqnarray}
Finally we exchange $n$ and $m$ in (\ref{eq:4-14}) to obtain the part of $(n,m,l)=(-++),(+--)$ with the coefficient $f_{-n,m,l}=f_{n,-m,-l}~(n>0,m>0,l>0)$ in (\ref{eq:4-8}).

We unify the above results to find that the quantum effect terms in $XP$-normal order are quadratic and constant: 
\begin{eqnarray}\begin{split}
& -\frac{i}{2}\sum _{n>0}\sum _{m>0} (f_{n,m,-n}+f_{n,m,-m}-f_{n,-m,m}-f_{-n,m,n})(X_{-n}P_{n}+X_{-m}P_{m}-P_{-n}X_{n}-P_{-m}X_{m}) \\
& ~~~-\sum _{n>0}\sum _{m>0} (f_{n,m,-n}+f_{n,m-m}) .
\end{split}\label{eq:4-15}
\end{eqnarray}
\subsubsection*{Structure of $[\mathcal{J}^{-I}_{\scriptscriptstyle{XP}}, \mathcal{K}^{K}_{\scriptscriptstyle{XP}}]$}
Thus we can induce the structure of $[\mathcal{J}^{-I},\mathcal{K}^{K}]$ in the XP-normal order. The possible structure in the XP-normal order are\footnote{Note that for simplicity the indexes of modes are omitted and quadratic parts are not written in the correct operator order.}
\begin{eqnarray}\begin{split}
[\mathcal{J}^{-I}_{\scriptscriptstyle{XP}},\mathcal{K}^{K}_{\scriptscriptstyle{XP}}] =& i(\mbox{ordered classical part}) +\delta ^{I,K} \frac{1}{p_{-}}( (X\cdot P) \mbox{-quadratic}) +i\delta ^{I,K} \frac{\tilde{C} }{p_{-}} \\
& +\frac{1}{p_{-}}(X^{I}P^{K}\mbox{-quadratic}) +\frac{1}{p_{-}}(X^{K}P^{I}\mbox{-quadratic}) ~,
\end{split}\label{eq:4-16}
\end{eqnarray}
where terms with $M_{0}$ at the right end are included in the first part and $\tilde{C} $ is constant and the last part arises from the ordering of zero-mode terms. If the third or 4th part exists in the case of $I \not= K$, $[\mathcal{J}^{-I}_{\scriptscriptstyle{XP}}, \mathcal{K}^{K}_{\scriptscriptstyle{XP}}]$ has off-diagonal part and then it becomes anomalous.\\
Similarly, we find the structure of $[\mathcal{J}^{-}_{\scriptscriptstyle{XP}},\mathcal{K}_{\scriptscriptstyle{XP}}]$ in $D=3$. Though, because there is only one transverse direction in $D=3$, $[\mathcal{J}^{-}_{\scriptscriptstyle{XP}}, \mathcal{K}_{\scriptscriptstyle{XP}}]$ is not anomalous.

\subsection{Structure of other dangerous commutators}
Other important dangerous commutators are $[\mathcal{J}^{-I},\mathcal{J}^{-J}]$ in $D>3$ and $[\mathcal{J}^{-},\mathcal{K}^{-}]$ in $D=3$.
$XPPP$-quartic operators appears in the calculation of $[\mathcal{J}^{-I},\mathcal{J}^{-J}]$ and $XXPPP$-quintic operators appears in the halfway of the calculation of $[\mathcal{J}^{-}, \mathcal{K}^{-}]$.

From the consideration in the last two subsubsection, the structure of $[\mathcal{J}^{-I},\mathcal{J}^{-J}]$ in $D>3$ and in the Hermitian R-order\footnote{The possible structure of commutators in the Weyl order is the same.} is
\begin{eqnarray}
[\mathcal{J}^{-I}_{\scriptscriptstyle{HR}}, \mathcal{J}^{-J}_{\scriptscriptstyle{HR}}] = i(\mbox{ordered classical part}) 
\label{eq:4-17}
\end{eqnarray}
where terms with $M_{0}$ at the right end are included in the first part. This has the same structure as the next explicit calculation in the (Hermitian\footnote{Because we do not need the regularization in the calculation of $[\mathcal{J}^{-I}, \mathcal{J}^{-J}]$, the calculation in the R-order gives the same result in the Hermitian R-order. Furthermore there is no constant quantum difference in the calculation of $[\mathcal{J}^{-I}, \mathcal{J}^{-J}]$, which contains $XPPP$-quartic operators in the highest degree. Therefore the calculations in the Hermitian R-order and the Weyl order are identical.}) R-order \cite{bib:010}:
\begin{eqnarray}
[\mathcal{J}^{-I}, \mathcal{J}^{-J}] = -\frac{1}{p_{-}^2}\sum _{n \not= 0} \frac{1}{n} \left( P^{I}_{n}P^{J}_{-n} -P^{J}_{n}P^{I}_{-n} \right) M_{0} ~.
\label{eq:4-18}
\end{eqnarray}
Note that $M_{0}$ is the ordered version of (\ref{eq:2-13}) and $XP$-quadratic.\footnote{The ordering constant absorbs in the right hand side of the constraint: $M_{0}\approx a$.}

Similarly, the structure of $[\mathcal{J}^{-I},\mathcal{J}^{-J}]$ in $D>3$ and in the XP-normal order is\footnote{Note that the indexes of modes are omitted for simplicity.}
\begin{eqnarray}
[\mathcal{J}^{-I}_{\scriptscriptstyle{XP}}, \mathcal{J}^{-J}_{\scriptscriptstyle{XP}}] = i(\mbox{ordered classical part}) +\frac{1}{p_{-}^2}((P^{I}P^{J}-P^{J}P^{I})\mbox{-quadratic})
\label{eq:4-19}
\end{eqnarray}
where terms with $M_{0}$ at the right end are included in the first part. This has the same structure as the next explicit calculation \cite{bib:010}:
\begin{eqnarray}
[\mathcal{J}^{-I}_{\scriptscriptstyle{XP}}, \mathcal{J}^{-J}_{\scriptscriptstyle{XP}}] = \frac{1}{p_{-}^2}\sum _{n =1}^{\infty }\Bigl( P^{I}_{-n}P^{J}_{n} -P^{J}_{-n}P^{I}_{n} \Bigr)  \left[ \Bigl( \frac{D-2}{6}-4 \Bigr) n + \Bigl( 2M_{0}-\frac{D-2}{6} \Bigr) \frac{1}{n} \right] ~.
\label{eq:4-20}
\end{eqnarray}

Furthermore, the structure of $[\mathcal{J}^{-},\mathcal{K}^{-}]$ in $D=3$ and in the Hermitian R-order is
\begin{eqnarray}
[\mathcal{J}^{-}_{\scriptscriptstyle{HR}}, \mathcal{K}^{-}_{\scriptscriptstyle{HR}}] = i(\mbox{ordered classical part}) +i C' \frac{p}{p_{-}^2}
\label{eq:4-21}
\end{eqnarray}
where we included the term with $M_{0}$ at the right end in the first part and $C'$ is constant. From (\ref{eq:4-21}), we find that there is no anomalous term of degree 3. This is calculated concretely in the next section. Because of the algebraic requirement, in the calculation in the next section we must use redefined $\tilde{\mathcal{K}}^{-}$ instead of $\mathcal{K}^{-}$. \\
Similarly, the structure of $[\mathcal{J}^{-},\mathcal{K}^{-}]$ in $D=3$ and in the XP-normal order is\footnote{Note that for simplicity the indexes of modes are omitted and cubic parts are not written in the correct operator order.}
\begin{eqnarray}
[\mathcal{J}^{-}_{\scriptscriptstyle{XP}}, \mathcal{K}^{-}_{\scriptscriptstyle{XP}}] = i(\mbox{ordered classical part}) +\frac{1}{p_{-}^2}(XPP\mbox{-cubic}) +i \tilde{C}' \frac{p}{p_{-}^2}
\label{eq:4-22}
\end{eqnarray}
where we included the term with $M_{0}$ at the right end in the first part and $\tilde{C}'$ is constant. If exists, the second part in the r.h.s. of (\ref{eq:4-22}) becomes an anomalous part of degree 3. This is calculated directly in the next section. Because of the algebraic requirement, in the next section we must use redefined $\tilde{\mathcal{K}}^{-}$ instead of $\mathcal{K}^{-}$. \\

Thus we obtain the structure of dangerous commutators, such as (\ref{eq:4-11}), (\ref{eq:4-16}), (\ref{eq:4-17}), (\ref{eq:4-19}), (\ref{eq:4-21}) and (\ref{eq:4-22}). However we would not find coefficients of the quantum effect terms in dangerous commutators unless we calculate concretely. Therefore, in the following sections, we calculate dangerous commutators in the spacetime conformal symmetry concretely with the help of the information about the structure of commutators obtained in this section.

\section{Calculation of Dangerous Commutators in Hermitian R-order }
In the Poincar\'{e} symmetry, there is no difference between the cases of the pure R-order, the Hermitian R-order and the Weyl order. 
On the other hand, in the spacetime conformal symmetry, because commutators of higher degree exist, there are differences in the quantum effect, which appears in terms of lower degree than the highest. Furthermore, we need some regularization because of inevitable divergent terms. As seen in appendix, the R-order breaks the hermitian property in the cut-off regularization.\footnote{This is seen as some kind of anomaly.}~ Therefore we consider the case of the Hermitian R-order below.

As seen in appendix, commutators which we must calculate for the check of the spacetime conformal symmetry are (at least) $[\mathcal{J}^{-I},\mathcal{K}^{K}]$ in $D>3$ and $\tilde{\mathcal{K}}^{-} \equiv -i[\mathcal{J}^{-},\mathcal{K}]$ and $[\mathcal{J}^{-}, \tilde{\mathcal{K}}^{-}]$ in $D=3$. Other commutators are calculated easily or obtained with the Jacobi identity. In this section we calculate these three commutators in the Hermitian R-order. The case of the Weyl order is similarly discussed. 

\subsection{$[\mathcal{J}^{-I}_{\scriptscriptstyle{HR}},\mathcal{K}^{K}_{\scriptscriptstyle{HR}}]$ in $D>3$ }
In \cite{bib:012, bib:013}, the dangerous commutator $[\mathcal{J}^{-I},\mathcal{K}^{K}]$ is considered with the generic regularization and without Fourier expansion. Then it is verified that the anomaly can appear in the traceless part of $[\mathcal{J}^{-I},\mathcal{K}^{K}]$ with respect to transverse indexes as follows:
\begin{eqnarray}
[\mathcal{J}^{-I},\mathcal{K}^{K}] +i\delta ^{I,K}\tilde{\mathcal{K}}^{-} \equiv \frac{1}{\epsilon p_{-}}L^{IK}_{t.l.} \propto \frac{1}{\epsilon p_{-}}\oint d\sigma [\bar{X}^{K}(\sigma )\bar{P}^{I}(\sigma )+(H.C.)]_{t.l.} ~, 
\label{eq:5-3}
\end{eqnarray}
where $\epsilon $ is regulator\footnote{Some R-ordered generators must be regularized to avoid divergences. In \cite{bib:012, bib:013}, the commutation relation is regularized: $[X^{I}(\sigma ), P_{J}(\sigma )]_{reg } = i\delta _{\epsilon }(0) \sim \frac{i}{\epsilon }$. } and $H.C.$ is the part compensated to preserve the hermitian property. The concrete form of $H.C.$ depends on the operator order. \\

In the case of the Hermitian R-order, however, the possible structure of $[\mathcal{J}^{-I},\mathcal{K}^{K}]$ is (\ref{eq:4-11}). Because there is no traceless part in (\ref{eq:4-11}), we may expect that the tensionless string theory in the Hermitian R-order is anomaly-free. However, we need some regularization because $C$ in (\ref{eq:4-11}) is a divergent constant. Then we must investigate whether the regularization we choose causes problems. If a nice regularization exists, the tensionless string theory in the Hermitian R-order has no spacetime conformal anomaly.

To see what happens in the calculation with a regularization, we consider the regularized version of (\ref{eq:4-11}) firstly. The regularized version of (\ref{eq:4-11}) is
\begin{eqnarray}
[\mathcal{J}^{-I}_{\scriptscriptstyle{HR (\epsilon )}}, \mathcal{K}^{K}_{\scriptscriptstyle{HR (\epsilon )}}] \equiv -i\delta ^{I,K}\tilde{\mathcal{K}}^{-}_{\scriptscriptstyle{HR (\epsilon )}} +i\frac{1}{p_{-}}\delta (O_{\scriptscriptstyle{(\epsilon )}}^{IK})_{t.l.} \approx -i\delta ^{I,K}\Bigl( \mathcal{K}^{-}_{\scriptscriptstyle{HR (\epsilon )}} +\frac{C_{\scriptscriptstyle{(\epsilon )}}}{p_{-}} \Bigr) + i\frac{1}{p_{-}}\delta (O_{\scriptscriptstyle{(\epsilon )}}^{IK})_{t.l.} ~,
\label{eq:5-2}
\end{eqnarray}
where $\epsilon $ is a real regulator and $(\epsilon )$ at the subscript of operators indicates that they are regularized and ``$\approx $" means the dropping of terms with $M_{0}$ at the right end.\footnote{From (\ref{eq:4-17}), the constraint is $M_{0}\approx 0$.}~ The regularized operators become the original ones in the limit of $\epsilon \rightarrow 0$. $C_{\scriptscriptstyle{(\epsilon )}}$ is a regularized constant and $\delta (O_{\scriptscriptstyle{(\epsilon )}}^{IK})_{t.l.}$ is a traceless part consisting of operators of degree 4 and vanishes in the limit of $\epsilon \rightarrow 0$.\\
Because of the algebraic requirement, we must use $\tilde{\mathcal{K}}^{-}_{\scriptscriptstyle{HR (\epsilon )}}$ instead of $\mathcal{K}^{-}_{\scriptscriptstyle{HR (\epsilon )}}$. Although we can use various representations of $\tilde{\mathcal{K}}^{-}_{\scriptscriptstyle{HR (\epsilon )}}$, we use the next one:
\begin{eqnarray}
\tilde{\mathcal{K}}^{-}_{\scriptscriptstyle{HR (\epsilon )}} = \frac{i}{D-2}\sum _{I=1}^{D-2} [\mathcal{J}^{-I}_{\scriptscriptstyle{HR (\epsilon )}}, \mathcal{K}^{I}_{\scriptscriptstyle{HR (\epsilon )}} ] ~.
\label{eq:5-3}
\end{eqnarray}

Because the regularized non-dangerous commutators and the ordered classical part of regularized dangerous commutators do not cause any problem in the limit of the regulator, the rest we have to consider is the quantum effect part of regularized dangerous commutators. The rest of dangerous commutators are $[\mathcal{J}^{-I}_{\scriptscriptstyle{HR (\epsilon )}}, \tilde{\mathcal{K}}^{-}_{\scriptscriptstyle{HR (\epsilon )}}]$ and $[\mathcal{K}^{K}_{\scriptscriptstyle{HR (\epsilon )}}, \tilde{\mathcal{K}}^{-}_{\scriptscriptstyle{HR (\epsilon )}}]$.\footnote{$[\mathcal{J}^{-I}_{\scriptscriptstyle{HR (\epsilon )}}, \mathcal{J}^{-J}_{\scriptscriptstyle{HR (\epsilon )}}]$ becomes (\ref{eq:4-11}) in the limit of $\epsilon \rightarrow 0$. $[\mathcal{K}^{I}_{\scriptscriptstyle{HR (\epsilon )}}, \mathcal{K}^{J}_{\scriptscriptstyle{HR (\epsilon )}}]$ is similar.}~ According to the discussion in section 4, we find the structure of them.
\begin{eqnarray}\begin{split}
[ \mathcal{J}^{-I}_{\scriptscriptstyle{HR (\epsilon )}}, \tilde{\mathcal{K}}^{-}_{\scriptscriptstyle{HR (\epsilon )}} ] &= i(\mbox{ordered classical part}) +i C'_{\scriptscriptstyle{(\epsilon )}} \frac{p^{I}}{p_{-}^2} ~,\\
[ \mathcal{K}^{K}_{\scriptscriptstyle{HR (\epsilon )}}, \tilde{\mathcal{K}}^{-}_{\scriptscriptstyle{HR (\epsilon )}} ] &= i(\mbox{ordered classical part}) +i C''_{\scriptscriptstyle{(\epsilon )}} \frac{x^{K}}{p_{-}} ~,
\end{split}\label{eq:5-4}
\end{eqnarray}
where $C'_{\scriptscriptstyle{(\epsilon )}}$ and $C''_{\scriptscriptstyle{(\epsilon )}}$ are constant. The first classical parts of these should vanish in the limit of $\epsilon \rightarrow 0$. Therefore, if $C'_{\scriptscriptstyle{(\epsilon )}}$ and $C''_{\scriptscriptstyle{(\epsilon )}}$ vanish in the limit of $\epsilon \rightarrow 0$, we can avoid the anomaly of the spacetime conformal symmetry. The check of their vanishings is discussed in the same way as the calculation of $[\mathcal{J}^{-}_{\scriptscriptstyle{HR}}, \tilde{\mathcal{K}}^{-}_{\scriptscriptstyle{HR}}]$ in $D=3$. We refer the discussion in next subsection and appendix. 

\subsection{$\tilde{\mathcal{K}}^{-}_{\scriptscriptstyle{HR}}\equiv -i[\mathcal{J}^{-}_{\scriptscriptstyle{HR}}, \tilde{\mathcal{K}}^{-}_{\scriptscriptstyle{HR}}]$ and $[\mathcal{J}^{-}_{\scriptscriptstyle{HR}}, \tilde{\mathcal{K}}^{-}_{\scriptscriptstyle{HR}}]$ in $D=3$ }
Because the dangerous commutator $[\mathcal{J}^{-}, \mathcal{J}^{-}]$ is trivially zero in three dimensions, the ordering constant in $M_{0}$, $a$, is not determined in three dimensions only by the requirement of the Lorentz symmetry, unlike in $D>3$. Thanks to the freedom of $a$, we obtain the great possibility for the mass spectrum of the tensionless string theory. However, from another point of view, we can say that the theory has an ambiguity. Removing this ambiguity is one of the interesting problem. \\
In this subsection, we calculate the definition of $\tilde{\mathcal{K}}^{-}$ and the dangerous commutator $[\mathcal{J}^{-}, \tilde{\mathcal{K}}^{-}]$ in the Hermitian R--order without any explicit regularization. Then we find the avoidance of the spacetime conformal anomaly and determine the ordering constant such that $a=0$. The calculation with the cut-off regularization is given as example in appendix. 

\subsubsection{Definition of $\tilde{\mathcal{K}}^{-}_{\scriptscriptstyle{HR}}$}
We define the Hermitian R-ordered generator $\tilde{\mathcal{K}}^{-}_{\scriptscriptstyle{HR}}$ as\footnote{Note that $\mathcal{J}^{-}=-\mathcal{J}^{-2}$ in $D=3$.}
\begin{eqnarray}
\tilde{\mathcal{K}}^{-}_{\scriptscriptstyle{HR}} \equiv -i[\mathcal{J}^{-}_{\scriptscriptstyle{HR}}, \mathcal{K}_{\scriptscriptstyle{HR}}] ~.
\label{eq:5-5}
\end{eqnarray}
Because there is only one transverse direction in $D=3$, this commutator is not anomalous. Because of the algebraic requirement, we must use it in commutators of the spacetime conformal symmetry instead of $\mathcal{K}^{-}_{\scriptscriptstyle{HR}}$:
\begin{eqnarray}\begin{split}
\mathcal{K}^{-}_{\scriptscriptstyle{HR}} \equiv & \frac{1}{2} \left( \mathcal{K}^{-}_{\scriptscriptstyle{R}} +h.c. \right) \\
\mathcal{K}^{-}_{\scriptscriptstyle{R}} =& x^{-}x^{-}p_{-} +x^{-}\sum _{n} X_{n}P_{-n} +\frac{1}{4p_{-}}\sum _{n}\sum _{m} X_{n}X_{m}L_{-n-m} \\
& -\frac{i}{p_{-}} \sum _{n}\sum _{m\not= 0}\Bigl( \frac{n}{m^2}+\frac{1}{m} \Bigr) X_{n}M_{m}P_{-n-m} ~, 
\end{split}\label{eq:5-6}
\end{eqnarray}
where $\frac{1}{p_{-}}\sum _{m \not= 0} \frac{1}{m^2} M_{m}M_{-m}$ in the classical $\mathcal{K}^{-}$ has been ordered correctly. 

From the discussion in section 4, the structure of $[\mathcal{J}^{-}_{\scriptscriptstyle{HR}}, \mathcal{K}_{\scriptscriptstyle{HR}}]$ is the same as (\ref{eq:4-11}). Here we consider the structure of the r.h.s. in (\ref{eq:5-5}) in detail. From the coincidence of the spacetime index, we find that the r.h.s. in (\ref{eq:5-5}) consists of the part with $x^{-}$ and the part proportional to $\frac{1}{p_{-}}$. The part with $x^{-}$ is easily calculated. Then we find that it is the same as the part with $x^{-}$ in $\mathcal{K}^{-}_{\scriptscriptstyle{HR}}$.
On the other hand, the part proportional to $\frac{1}{p_{-}}$ consists of $XXPP$-quartic, $XP$-quadratic and a constant. The last two parts are the quantum effect terms, which arise in exchanging operators to get the correctly ordered result. Because $XXPP$-quartic terms are in the highest degree, we can know them by the classical calculation. After a long calculation, we find that they consist of $XXPP$-quartic terms in $\mathcal{K}^{-}_{\scriptscriptstyle{HR}}$ and terms with $M_{0}$ at the right end. As seen in section 4, in the Hermitian R-order, $XP$-quadratic terms make commutators and then are absorbed in the constant part. 

Thus we obtain\footnote{The last term in (\ref{eq:5-7}) is obtained from $[\frac{i}{p_{-}}\sum _{l \not= 0}\frac{1}{l}M_{l}P_{-l},~ i\sum _{m \not= 0}\frac{1}{m}X_{m}M_{-m}] $ in $[\mathcal{J}^{-}, \mathcal{K}]$.}
\begin{eqnarray}
\tilde{\mathcal{K}}^{-}_{\scriptscriptstyle{HR}} = \mathcal{K}^{-}_{\scriptscriptstyle{HR}} +\frac{C}{p_{-}} +\frac{2i}{p_{-}}\sum _{n \not= 0}\frac{1}{n}X_{n}P_{-n} M_{0} ~.
\label{eq:5-7}
\end{eqnarray}
The first term is the original ordered classical generator. The second term is the gap by quantum effect and $C$ is usually divergent constant. The last term,\footnote{$\sum _{n \not= 0}\frac{1}{n}X_{n}P_{-n}$ is anti-hermitian and the same as the Hermitian R-ordered one because it is hermitian and commutes with $M_{0}$.} which has $M_{0}$ at the right end, is obtained also by the classical calculation and becomes quadratic terms under the constraint in $D=3$, $M_{0} \approx a$.

Because there is the contribution from $[X_{n}, P_{-n}]=i$ for non-zero $n$ to $C$, $C$ is a divergent constant even in the Hermitian R-order. Therefore we need some regularization. \\
Now we want to know whether the anomaly unrelated to the choice of the regularization exists in the spacetime conformal symmetry.
Therefore we assume below the existence of a good regularization and then use $\tilde{\mathcal{K}}^{-}_{\scriptscriptstyle{HR}}$, as it is in (\ref{eq:5-7}). The calculation with the cut-off regularization is given in appendix.

\subsubsection{Dangerous commutator : $[\mathcal{J}^{-}_{\scriptscriptstyle{HR}}, \tilde{\mathcal{K}}^{-}_{\scriptscriptstyle{HR}}]$}
Here we calculate the dangerous commutator, $[\mathcal{J}^{-}, \tilde{\mathcal{K}}^{-}]$, which may be anomalous.
Firstly we see its structure. Then we calculate it explicitly. The calculation with the cut-off regularization is given in appendix.

From the number of the spacetime index, $[\mathcal{J}^{-}_{\scriptscriptstyle{HR}}, \tilde{\mathcal{K}}^{-}_{\scriptscriptstyle{HR}}]$ consists of the part proportional to $x^{-}x^{-}$, the part proportional to $\bigl( x^{-}\frac{1}{p_{-}}+\frac{1}{p_{-}}x^{-} \bigr) $ and the part proportional to $\frac{1}{p_{-}^2}$.\footnote{Although we may consider $x^{-}\frac{1}{p_{-}}$ and $\frac{1}{p_{-}}x^{-}$ separately, the hermitian combination is only $\bigl( x^{-}\frac{1}{p_{-}}+\frac{1}{p_{-}}x^{-} \bigr) $. Their differences from the hermitian combination are $\mp \frac{i}{2p_{-}^2}$. Then they are included in the part proportional to $\frac{1}{p_{-}^2}$.}

Furthermore, from the mass dimension, the first part can contain only $x^{-}x^{-}p$. It is easily calculated and then we find its cancellation. The second part can consist of $XPP$-cubic terms and $p$-linear terms. $XPP$-cubic terms in the second part are ordered classical results. On the other hand, $p$-linear terms are forbidden because they are hermitian. The third part consists of $XXPPP$-quintic terms, $XPP$-cubic terms and $p$-linear terms. $XXPPP$-quintic terms in the third part are ordered classical results.  As seen in section 4, $XPP$-cubic terms in the third part make a commutator $[X_{n}, P_{m}]=i\delta _{n,-m}$ in the Hermitian R-order to be contained in $p$-linear term. 

Thus we find the structure of $[\mathcal{J}^{-}_{\scriptscriptstyle{HR}}, \tilde{\mathcal{K}}^{-}_{\scriptscriptstyle{HR}}]$ in the Hermitian R-order.
\begin{eqnarray}\begin{split}
[\mathcal{J}^{-}_{\scriptscriptstyle{HR}},~ \tilde{\mathcal{K}}^{-}_{\scriptscriptstyle{HR}}] =&~ i x^{-}x^{-}p \times (\mbox{const.}) + i \Bigl( x^{-}\frac{1}{p_{-}} + \frac{1}{p_{-}}x^{-} \Bigr) ( \mbox{cubic term}: XPP+h.c.) \\
&~ +i\frac{1}{p_{-}^2} (\mbox{quintic term}: XXPPP+h.c. ) +i C' \frac{p}{p_{-}^2} \\
=&~ i(\mbox{ordered classical part}) +i C' \frac{p}{p_{-}^2} ~,
\end{split}\label{eq:5-8}
\end{eqnarray}
where $C'$ is a constant and we contained terms with $M_{0}$ at the right end in the ordered classical part. The second term in the last line is a quantum effect term. 

Because there is no term with the square of $M_{0}$, the constraint, $M_{0}\approx a$, does not mix the first term and the second term in the last line of (\ref{eq:5-8}). Therefore they must be zero separately.

\subsubsection*{Calculation : Classical part of $[\mathcal{J}^{-}_{\scriptscriptstyle{HR}}, \tilde{\mathcal{K}}^{-}_{\scriptscriptstyle{HR}}]$}
$\mathcal{K}^{-}_{\scriptscriptstyle{HR}}$ and the third term among $\tilde{\mathcal{K}}^{-}_{\scriptscriptstyle{HR}}$ defined in (\ref{eq:5-4}) contribute to the classical part of $[\mathcal{J}^{-}_{\scriptscriptstyle{HR}}, \tilde{\mathcal{K}}^{-}_{\scriptscriptstyle{HR}}]$ in (\ref{eq:5-8}). Because we want the classical information, we can use R-ordered generators.
\begin{eqnarray}\begin{split}
\Bigl[ \mathcal{J}^{-}_{\scriptscriptstyle{HR}},~ \mathcal{K}^{-}_{\scriptscriptstyle{HR}}+\frac{2i}{p_{-}}\sum _{n \not= 0}\frac{1}{n}X_{n}P_{-n} M_{0} \Bigr] & \simeq \Bigl[ \mathcal{J}^{-}_{\scriptscriptstyle{R}},~ \mathcal{K}^{-}_{\scriptscriptstyle{R}} \Bigr] + \Bigl[ \mathcal{J}^{-}_{\scriptscriptstyle{R}},~ \frac{2i}{p_{-}}\sum _{n \not= 0}\frac{1}{n}X_{n}P_{-n} M_{0} \Bigr] \\
& \Rightarrow i(\mbox{ordered classical part}) ~,
\end{split}\label{eq:5-9}
\end{eqnarray}
where $\simeq $ means the extraction of the classical part.\\

First we consider the first commutator in (\ref{eq:5-9}). Because the calculation is long, we separate $\mathcal{J}^{-}_{\scriptscriptstyle{R}}$ into three parts as below to calculate the contribution from each part.
\begin{eqnarray}
\mathcal{J}^{-}_{\scriptscriptstyle{R}} = -x^{-}p -\frac{1}{2p_{-}} \sum _{l} X_{l}L_{-l} +\frac{i}{p_{-}}\sum _{l \not= 0}\frac{1}{l}M_{l}P_{-l}
\label{eq:5-10}
\end{eqnarray}
Commutation relations in (\ref{eq:2-14}) which are satisfied also in the (Hermitian) R-order may help us in the following calculations.\\
The commutator of the first part in (\ref{eq:5-10}) and $\mathcal{K}^{-}_{\scriptscriptstyle{R}}$ is classically
\begin{eqnarray}\begin{split}
\left[ -x^{-}p,~ \mathcal{K}^{-}_{\scriptscriptstyle{R}} \right] \simeq & ~ i \Bigl( \frac{1}{4}\sum _{n}\sum_{m}X_{n}X_{m}L_{-n-m} -i\sum _{n}\sum _{m \not= 0}\frac{n+m}{m^2}X_{n}M_{m}P_{-n-m} \Bigr) \frac{p}{p_{-}^2} \\
& +\frac{i}{2}x^{-}\frac{1}{p_{-}}\sum _{n}X_{n}L_{-n} +x^{-}\frac{1}{p_{-}}\sum _{m \not= 0}\frac{1}{m}M_{m}P_{-m} ~.
\end{split}\label{eq:5-11}
\end{eqnarray}
The commutator of the second part in (\ref{eq:5-10}) and $\mathcal{K}^{-}_{\scriptscriptstyle{R}}$ is classically
\begin{eqnarray}\begin{split}
\Bigl[ -\frac{1}{2p_{-}}\sum _{l}X_{l}L_{-l},~ \mathcal{K}^{-}_{\scriptscriptstyle{R}} \Bigr] \simeq  -\frac{i}{2}x^{-}\frac{1}{p_{-}}\sum _{n}X_{n}L_{-n} &-\frac{i}{4p_{-}^2}\sum _{n}\sum _{m}\sum _{l}X_{n}X_{m}P_{l}L_{-n-m-l} \\
& -\frac{1}{2p_{-}^2}\sum _{n}\sum _{m \not= 0}\frac{1}{m}X_{n}M_{m}L_{-n-m} ~.
\end{split}\label{eq:5-12}
\end{eqnarray}
The commutator of the third part in (\ref{eq:5-10}) and $\mathcal{K}^{-}_{\scriptscriptstyle{R}}$ is classically
\begin{eqnarray}\begin{split}
\Bigl[ \frac{i}{p_{-}}\sum _{l \not= 0}\frac{1}{l}M_{l}P_{-l},~\mathcal{K}^{-}_{\scriptscriptstyle{R}} \Bigr] & \simeq -x^{-}\frac{1}{p_{-}}\sum _{m \not= 0}\frac{1}{m}M_{m}P_{-m} +\frac{i}{4p_{-}^2}\sum _{n}\sum _{m}\sum _{l}X_{n}X_{m}P_{l}L_{-n-m-l} \\
&~~~~ -i\Bigl( \frac{1}{4}\sum _{n}\sum_{m}X_{n}X_{m}L_{-n-m} -i\sum _{n}\sum _{m \not= 0}\frac{n+m}{m^2}X_{n}M_{m}P_{-n-m} \Bigr) \frac{p}{p_{-}^2} \\
&~~~~ +\frac{1}{2p_{-}^2}\sum _{n}\sum _{m \not= 0}\frac{1}{m}X_{n}M_{m}L_{-n-m} +\frac{2}{p_{-}^{2}}\sum _{n}\sum _{m \not= 0}\frac{n}{m^2}X_{n}M_{m}P_{-m}P_{-n} ~.
\end{split}\label{eq:5-13}
\end{eqnarray}
We collect (\ref{eq:5-11})-(\ref{eq:5-13}) to obtain 
\begin{eqnarray}
[\mathcal{J}^{-}_{\scriptscriptstyle{R}},~ \hat{\mathcal{K}}^{-}_{\scriptscriptstyle{R}}] \simeq \frac{2}{p_{-}^{2}}\sum _{n}\sum _{m \not= 0}\frac{n}{m^2}X_{n}M_{m}P_{-m}P_{-n} \simeq \frac{2i}{p_{-}^2}\sum _{m \not= 0}\frac{1}{m^2}M_{m}P_{-m} M_{0} ~,
\label{eq:5-14}
\end{eqnarray}
where we used $M_{0} = -i\sum _{n}nX_{n}P_{-n}$ and then moved it to the right end. 
In the calculation of the quantum part, we must not forget the contribution which arises in moving $M_{0}$ to the right end.\\

Next we consider the commutator of the second term in (\ref{eq:5-9}). Because $M_{0}$ commutes with all generators, we obtain
\begin{eqnarray}\begin{split}
& \Bigl[ \mathcal{J}^{-}_{\scriptscriptstyle{R}},~ \frac{2i}{p_{-}}\sum _{n \not= 0}\frac{1}{n}X_{n}P_{-n} M_{0} \Bigr] \\
& = \Bigl( -\frac{6}{p_{-}^2}\sum _{n \not= 0}\frac{1}{n}X_{n}P_{-n}p +\frac{1}{p_{-}^2}\sum _{n \not= 0}\frac{1}{n}X_{n}L_{-n} +\frac{2i}{p_{-}^2}\sum _{n \not= 0}\frac{1}{n^2}M_{n}P_{-n} \Bigr) M_{0} ~.
\end{split}\label{eq:5-15}
\end{eqnarray}
This contains $M_{0}$ at the right end again.\\

We collect (\ref{eq:5-14}) and (\ref{eq:5-15}) to find that the classical part in (\ref{eq:5-8}) consists only of terms with $M_{0}\approx a$ at the right end. Therefore we determine the ordering constant so that $a=0$.\footnote{Note that the classical part and the quantum effect part must vanish separately.}

\subsubsection*{Calculation : Quantum effect part of $[\mathcal{J}^{-}_{\scriptscriptstyle{HR}}, \tilde{\mathcal{K}}^{-}_{\scriptscriptstyle{HR}}]$}
Next we calculate the quantum effect part of $[\mathcal{J}^{-}_{\scriptscriptstyle{HR}}, \tilde{\mathcal{K}}^{-}_{\scriptscriptstyle{HR}}]$, that is the second part in the last line of (\ref{eq:5-8}). Because the second part contains $p$-zero mode, we can know its cancellation from the commutator of $x$ and $[\mathcal{J}^{-}_{\scriptscriptstyle{HR}}, \tilde{\mathcal{K}}^{-}_{\scriptscriptstyle{HR}}]$. By using the Jacobi identity, we deform it as follows.
\begin{eqnarray}\begin{split}
& \bigl[ x,~ [\mathcal{J}^{-}_{\scriptscriptstyle{HR}},~ \tilde{\mathcal{K}}^{-}_{\scriptscriptstyle{HR}}] \bigr] = \bigl[ [x, \mathcal{J}^{-}_{\scriptscriptstyle{HR}}],~ \tilde{\mathcal{K}}^{-}_{\scriptscriptstyle{HR}} \bigr] +\bigl[ \mathcal{J}^{-}_{\scriptscriptstyle{HR}},~ [x, \tilde{\mathcal{K}}^{-}_{\scriptscriptstyle{HR}}]\bigr] \\
&= \frac{1}{2}\Bigl( \bigl[ [x, \mathcal{J}^{-}_{\scriptscriptstyle{HR}}],~ \mathcal{K}^{-}_{\scriptscriptstyle{R}}\bigr] +\bigl[ \mathcal{J}^{-}_{\scriptscriptstyle{HR}},~ [x, \mathcal{K}^{-}_{\scriptscriptstyle{R}}]\bigr] + h.c. \Bigr) +\biggl[ [x, \mathcal{J}^{-}_{\scriptscriptstyle{HR}}],~ \frac{C}{p_{-}} +\frac{2i}{p_{-}}\sum _{n \not= 0}\frac{1}{n}X_{n}P_{-n} M_{0} \biggr] ~.
\end{split}\label{eq:5-16}
\end{eqnarray}
Below, we calculate each part.

The first commutator in the second line of (\ref{eq:5-16}) is calculated as follows.
\begin{eqnarray}\begin{split}
\left[ [x, \mathcal{J}^{-}_{\scriptscriptstyle{HR}}],~ \mathcal{K}^{-}_{\scriptscriptstyle{R}}\right] =& x^{-}x^{-} +2x^{-}x\frac{p}{p_{-}} +\frac{i}{2}x\frac{p}{p_{-}^2} -\frac{i}{2}x^{-}\frac{1}{p_{-}} +\frac{1}{p_{-}^2} \\
& +x^{-}\mathcal{D}_{\scriptscriptstyle{R}}\frac{1}{p_{-}} +2x\mathcal{D}_{\scriptscriptstyle{R}}\frac{p}{p_{-}^2} -\frac{i}{2}\mathcal{D}_{\scriptscriptstyle{R}}\frac{1}{p_{-}^2} +x\mathcal{J}^{-}_{\scriptscriptstyle{HR}}\frac{1}{p_{-}} -\mathcal{K}_{\scriptscriptstyle{R}}\frac{p}{p_{-}^2} -\mathcal{K}^{-}_{\scriptscriptstyle{R}}\frac{1}{p_{-}} ~,
\end{split}\label{eq:5-17}
\end{eqnarray}
where we used 
\begin{eqnarray}\begin{split}
& \mathcal{J}^{-}_{\scriptscriptstyle{HR}} = \mathcal{J}^{-}_{\scriptscriptstyle{R}} +\frac{i}{2p_{-}}p ~,\\
& [x, \mathcal{J}^{-}_{\scriptscriptstyle{HR}}] = (-i) \Bigl( x^{-} +x\frac{p}{p_{-}} -\frac{i}{2p_{-}} \Bigr) ~.
\end{split}\label{eq:5-18}
\end{eqnarray}
The second commutator in the second line of (\ref{eq:5-16}) is calculated as follows.
\begin{eqnarray}\begin{split}
\left[ \mathcal{J}^{-}_{\scriptscriptstyle{HR}}, [x, \mathcal{K}^{-}_{\scriptscriptstyle{R}}]\right] &= -x^{-}x^{-} -2x^{-}x\frac{p}{p_{-}} -ix\frac{p}{p_{-}^2} +\frac{i}{2}x^{-}\frac{1}{p_{-}} -\frac{1}{p_{-}^2} \\
&~~ -x^{-}\mathcal{D}_{\scriptscriptstyle{R}}\frac{1}{p_{-}} -2x\mathcal{D}_{\scriptscriptstyle{R}}\frac{p}{p_{-}^2} +\frac{i}{2}\mathcal{D}_{\scriptscriptstyle{R}}\frac{1}{p_{-}^2} -x\mathcal{J}^{-}_{\scriptscriptstyle{HR}}\frac{1}{p_{-}} +\mathcal{K}_{\scriptscriptstyle{R}}\frac{p}{p_{-}^2} -i\bigl[ \mathcal{J}^{-}_{\scriptscriptstyle{HR}}, \mathcal{K}_{\scriptscriptstyle{R}} \bigr] \frac{1}{p_{-}} ~,
\end{split}\label{eq:5-19}
\end{eqnarray}
where we used the following relations:
\begin{eqnarray}
[x, \mathcal{K}^{-}_{\scriptscriptstyle{R}}] = i\Bigl( x^{-}x +x\mathcal{D}_{\scriptscriptstyle{R}}\frac{1}{p_{-}} -\mathcal{K}_{\scriptscriptstyle{R}}\frac{1}{p_{-}} \Bigr) ~.
\label{eq:5-20}
\end{eqnarray}
We collect (\ref{eq:5-17}) and (\ref{eq:5-19}) to obtain
\begin{eqnarray}
\left[ [x, \mathcal{J}^{-}_{\scriptscriptstyle{HR}}],~ \mathcal{K}^{-}_{\scriptscriptstyle{R}}\right] + \left[ \mathcal{J}^{-}_{\scriptscriptstyle{HR}}, [x, \mathcal{K}^{-}_{\scriptscriptstyle{R}}]\right] = -\frac{i}{2}x\frac{p}{p_{-}^2} -\mathcal{K}^{-}_{\scriptscriptstyle{R}}\frac{1}{p_{-}} -i\bigl[ \mathcal{J}^{-}_{\scriptscriptstyle{HR}}, \mathcal{K}_{\scriptscriptstyle{R}} \bigr] \frac{1}{p_{-}} ~.
\label{eq:5-21}
\end{eqnarray}
Then we find that the second line of (\ref{eq:5-16}) is as follows:
\begin{eqnarray}\begin{split}
&\frac{1}{2} \left( \left[ [x, \mathcal{J}^{-}_{\scriptscriptstyle{HR}}],~ \mathcal{K}^{-}_{\scriptscriptstyle{R}}\right] + \left[ \mathcal{J}^{-}_{\scriptscriptstyle{HR}}, [x, \mathcal{K}^{-}_{\scriptscriptstyle{R}}]\right]  +h.c. \right) \\
& = \frac{1}{4p_{-}^2} +(\tilde{\mathcal{K}}^{-}_{\scriptscriptstyle{HR}} -\mathcal{K}^{-}_{\scriptscriptstyle{HR}})\frac{1}{p_{-}} -\frac{1}{2}\Bigl[ \frac{1}{p_{-}}, (\mathcal{K}^{-}_{\scriptscriptstyle{R}})^{\dagger } \Bigr] +\frac{i}{2}\Bigl[ \frac{1}{p_{-}}, [\mathcal{J}^{-}_{\scriptscriptstyle{HR}}, \mathcal{K}^{-}_{\scriptscriptstyle{R}}]^{\dagger } \Bigr] ~.
\end{split}\label{eq:5-22}
\end{eqnarray}
The third term in the r.h.s. of (\ref{eq:5-22}) is calculated as
\begin{eqnarray}
-\frac{1}{2}\Bigl[ \frac{1}{p_{-}}, (\mathcal{K}^{-}_{\scriptscriptstyle{R}})^{\dagger } \Bigr] = -\frac{1}{2}\Bigl[ \frac{1}{p_{-}}, (\mathcal{D}_{\scriptscriptstyle{R}})^{\dagger }x^{-} \Bigr] = -\frac{i}{2} \Bigl( \frac{1}{p_{-}}x^{-} +(\mathcal{D}_{\scriptscriptstyle{R}})^{\dagger }\frac{1}{p_{-}} \Bigr)
\label{eq:5-23}
\end{eqnarray}
and the 4th term in the r.h.s. of (\ref{eq:5-22}) is calculated as
\begin{eqnarray}\begin{split}
\frac{i}{2}\Bigl[ \frac{1}{p_{-}}, [\mathcal{J}^{-}_{\scriptscriptstyle{HR}}, \mathcal{K}^{-}_{\scriptscriptstyle{R}}]^{\dagger } \Bigr] &= -\frac{i}{2}\Bigl[ \frac{1}{p_{-}}, [\mathcal{J}^{-}_{\scriptscriptstyle{HR}}, \mathcal{K}^{-}_{\scriptscriptstyle{R}}] \Bigr] ^{\dagger } = \frac{i}{2}\Bigl( \Bigl[ \mathcal{K}^{-}_{\scriptscriptstyle{R}}, \Bigl[ \frac{1}{p_{-}}, \mathcal{J}^{-}_{\scriptscriptstyle{HR}} \Bigr] \Bigr] +\Bigl[ \mathcal{J}^{-}_{\scriptscriptstyle{HR}}, \Bigl[ \mathcal{K}^{-}_{\scriptscriptstyle{R}}, \frac{1}{p_{-}} \Bigr] \Bigr] \Bigr) ^{\dagger } \\
&= \frac{i}{2}\Bigl( \Bigl[ \mathcal{K}^{-}_{\scriptscriptstyle{R}}, -i\frac{p}{p_{-}^2} \Bigr] +\Bigl[ \mathcal{J}^{-}_{\scriptscriptstyle{HR}}, -i\frac{x}{p_{-}} \Bigr] \Bigr) ^{\dagger } = \frac{i}{2} \Bigl( \frac{1}{p_{-}}x^{-} +(\mathcal{D}_{\scriptscriptstyle{R}})^{\dagger }\frac{1}{p_{-}} \Bigr) -\frac{1}{4p_{-}^2} ~.
\end{split}\label{eq:5-24}
\end{eqnarray}
Thus we get 
\begin{eqnarray}\begin{split}
\frac{1}{2} \left( \left[ [x, \mathcal{J}^{-}_{\scriptscriptstyle{HR}}],~ \mathcal{K}^{-}_{\scriptscriptstyle{R}}\right] + \left[ \mathcal{J}^{-}_{\scriptscriptstyle{HR}}, [x, \mathcal{K}^{-}_{\scriptscriptstyle{R}}]\right]  +h.c. \right) = (\tilde{\mathcal{K}}^{-}_{\scriptscriptstyle{HR}} -\mathcal{K}^{-}_{\scriptscriptstyle{HR}})\frac{1}{p_{-}} = \frac{C}{p_{-}^2} +\frac{2i}{p_{-}^2}\sum _{n \not= 0}\frac{1}{n}X_{n}P_{-n} M_{0} ~.
\end{split}\label{eq:5-25}
\end{eqnarray}
\par 
The last commutator in the second line of (\ref{eq:5-16}) is calculated as
\begin{eqnarray}
\biggl[ [x, \mathcal{J}^{-}],~ \frac{C}{p_{-}} +\frac{2i}{p_{-}}\sum _{n \not= 0}\frac{1}{n}X_{n}P_{-n} M_{0} \biggr] = -\frac{C}{p_{-}^2} -\frac{2i}{p_{-}^2}\sum _{n \not= 0}\frac{1}{n}X_{n}P_{-n} M_{0} ~.
\label{eq:5-26}
\end{eqnarray}
\par
We collect (\ref{eq:5-25}) and (\ref{eq:5-26}) to get
\begin{eqnarray}
\left[ x,~ [\mathcal{J}^{-}_{\scriptscriptstyle{HR}}, \hat{\mathcal{K}}^{-}_{\scriptscriptstyle{HR}}] \right] = 0 ~.
\label{eq:5-27}
\end{eqnarray}
Thus we find that $p$-zero mode in $[\mathcal{J}^{-}_{\scriptscriptstyle{HR}}, \hat{\mathcal{K}}^{-}_{\scriptscriptstyle{HR}}]$, including terms of higher degree with $p$, is canceled. Other commutators are calculated easily or obtained with the Jacobi identity or similarly discussed. Therefore there is no anomalous term in $[\mathcal{J}^{-}_{\scriptscriptstyle{HR}},~ \hat{\mathcal{K}}^{-}_{\scriptscriptstyle{HR}}]$. \\ 

In this subsection, in order to know whether the anomaly irrelevant to an explicit regularization exists, we have calculated the dangerous commutator without the regularization. In appendix, we will calculate the dangerous commutator explicitly with the cut-off regularization and without using the Jacobi identity. There we will calculate regularized-$C$ in (\ref{eq:5-7}) concretely and then verify the cancellation of $C'$ in (\ref{eq:5-8}).

\section{Calculation of Dangerous Commutators in XP-normal Order}
In this section we investigate three commutators which we must calculate for the check of the spacetime conformal symmetry, $[\mathcal{J}^{-I},\mathcal{K}^{K}]$ in $D>3$ and $\tilde{\mathcal{K}}^{-} \equiv -i[\mathcal{J}^{-},\mathcal{K}]$ and $[\mathcal{J}^{-}, \tilde{\mathcal{K}}^{-}]$ in $D=3$, in the XP-normal order.

\subsection{$[\mathcal{J}^{-I}_{\scriptscriptstyle{XP}},\mathcal{K}^{K}_{\scriptscriptstyle{XP}}]$ in $D>3$}
Because the off-diagonal part in the second line of (\ref{eq:4-16}) can exist in the XP-normal order of $D>3$, $[\mathcal{J}^{-I}_{\scriptscriptstyle{XP}}, \mathcal{K}^{K}_{\scriptscriptstyle{XP}}]$ may be anomalous. However, because these terms possibly vanish by a profound reason, we have to calculate the dangerous commutator $[\mathcal{J}^{-I}_{\scriptscriptstyle{XP}}, \mathcal{K}^{K}_{\scriptscriptstyle{XP}}]$ explicitly to verify the concrete coefficient of these terms.\\
Before the explicit calculation, we investigate the expected result of $[\mathcal{J}^{-I}_{\scriptscriptstyle{XP}}, \mathcal{K}^{K}_{\scriptscriptstyle{XP}}]$ in a little more detail.

From (\ref{eq:4-20}), in XP-normal order, the requirement of the Lorentz symmetry determines the critical dimension and the ordering constant in the constraint such that $D=26$ and $M_{0}\approx a=2$. By these conditions and the Jacobi identity, we obtain
\begin{eqnarray}
[\mathcal{J}^{-I}_{\scriptscriptstyle{XP}}, \mathcal{K}^{K}_{\scriptscriptstyle{XP}}] \approx [\mathcal{J}^{-K}_{\scriptscriptstyle{XP}}, \mathcal{K}^{I}_{\scriptscriptstyle{XP}}]
\label{eq:6-1}
\end{eqnarray}
where $\approx $ means the dropping of terms which vanish under $D=26$ and $M_{0}=2$. Thus we expect that $[\mathcal{J}^{-I}_{\scriptscriptstyle{XP}}, \mathcal{K}^{K}_{\scriptscriptstyle{XP}}]$ consists of the symmetric part in exchanging of $I$ and $K$ and the part which vanishes under $D=26$ and $M_{0}=2$.

\subsubsection{Preparation for explicit calculation}
Here we give the explicit representation of important generators and some useful commutation relations.

First, we give the representation of generators which we need for the calculation, $\mathcal{J}^{-I}_{\scriptscriptstyle{XP}}$ and $\mathcal{K}^{K}_{\scriptscriptstyle{XP}}$. In XP-normal ordered generators, non-zero modes are normal-ordered but zero mode are not. Therefore we decouple zero mode. $\mathcal{J}^{-I}_{\scriptscriptstyle{XP}}$ is written as
\begin{eqnarray}\begin{split}
\mathcal{J}^{-I}_{\scriptscriptstyle{XP}} =& x^{-}p^{I} +\left( \frac{1}{2p_{-}}x^{I}p\cdot p -\frac{i}{2p_{-}}p^{I} \right) +\frac{1}{2p_{-}}x^{I}\bar{L}_{0} \\
& +\frac{p_{J}}{p_{-}}\sum _{l=1}^{\infty } \bigl[ (X^{I}_{-l}P^{J}_{l}+P^{J}_{-l}X^{I}_{l}) -(X^{J}_{-l}P^{I}_{l}+P^{I}_{-l}X^{J}_{l}) \bigr] -\frac{\bar{\Lambda }^{I}_{\scriptscriptstyle{XP}}}{p_{-}} ,
\end{split}\label{eq:6-2}
\end{eqnarray}
where barred operators are defined by removing zero mode from unbarred ones and we used 
\begin{eqnarray}
\bar{\Lambda }^{I}_{\scriptscriptstyle{XP}} \equiv \sum _{l=1}^{\infty } \left[ -\frac{1}{2}(X^{I}_{-l}\bar{L}_{l}+\bar{L}_{-l}X^{I}_{l}) +\frac{i}{l}(P^{I}_{-l}\bar{M}_{l}-\bar{M}_{-l}P^{I}_{l}) \right] .
\label{eq:6-3}
\end{eqnarray}
Here we note that commutation relations of $\bar{L}_{n}$ or $\bar{M}_{n}$ is the same as the relations of unbarred ones because unbarred ones does not contain any $x$ zero mode.\\
$\mathcal{K}^{K}_{\scriptscriptstyle{XP}}$ is 
\begin{eqnarray}\begin{split}
\mathcal{K}^{K}_{\scriptscriptstyle{XP}} &= x^{K}x^{-}p_{-} +x^{K}(x\cdot p) -\frac{1}{2}(x\cdot x)p^{K} -i\frac{D-1}{2}x^{K}\\
&~ +x^{K}\sum _{n=1}^{\infty }(X_{-n}\cdot P_{n}+h.c.) -p^{K}\sum _{n=1}^{\infty }X_{-n}\cdot X_{n} +x_{L}\sum _{n=1}^{\infty }\Bigl( (X^{K}_{-n}P^{L}_{n}-X^{L}_{-n}P^{K}_{n}) +h.c. \Bigr) \\
&~ +\sum _{n=1}^{\infty }\sum _{m=1}^{\infty } \Bigl[ \bigl( X^{K}_{-n}(X_{-m}\cdot P_{n+m}) +X^{K}_{-n}(P_{-m}\cdot X_{n+m}) +X^{K}_{-n-m}(P_{m}\cdot X_{n}) \bigr) +h.c. \Bigr] \\
&~ -\frac{1}{2} \sum _{n=1}^{\infty }\sum _{m=1}^{\infty } \Bigl[ \bigl( (X_{-n}\cdot X_{-m})P^{K}_{n+m} +2(X_{-n-m}\cdot X_{m})P^{K}_{n} \bigr) +h.c. \Bigr] \\
&~ -i\sum _{n=1}^{\infty } \frac{1}{n} (X^{K}_{-n}\bar{M}_{n} -\bar{M}_{-n}X^{K}_{n}) ,
\end{split}\label{eq:6-4}
\end{eqnarray}
where $h.c.$ indicates hermitian conjugate of the adjacent part. Because we do not need $\mathcal{K}^{-}$ in the calculation in the next subsubsection, we omit the representation of $\mathcal{K}^{-}$ in $D>3$.

Next we give some useful commutation relations below. For $n>0$,
\begin{eqnarray}\begin{split}
\bigl[ \bar{\Lambda }^{I}_{\scriptscriptstyle{XP}}, X^{J}_{n} \bigr] =& \delta ^{I,J}\frac{1}{n}\bar{M}_{n} +i\sum _{l=1}^{n-1}\Bigl( P^{J}_{n-l}X^{I}_{l}+\frac{n-l}{l}X^{J}_{n-l}P^{I}_{l} \Bigr) \\
& +i\sum _{l=1}^{\infty } \Bigl( X^{I}_{-l}P^{J}_{n+l}+P^{J}_{-l}X^{I}_{n+l}-\frac{n+l}{l}P^{I}_{-l}X^{J}_{n+l}-\frac{l}{n+l}X^{J}_{-l}P^{I}_{n+l} \Bigr) ~,\\
\bigl[ \bar{\Lambda }^{I}_{\scriptscriptstyle{XP}}, P^{J}_{n} \bigr] =& -\delta ^{I,J}\frac{i}{2}\bar{L}_{n} +in\sum _{l=1}^{n-1}\frac{1}{l} P^{J}_{n-l}P^{I}_{l} +i n \sum _{l=1}^{\infty } \Bigl( \frac{1}{l}P^{I}_{-l}P^{J}_{n+l}-\frac{1}{n+l}P^{J}_{-l}P^{I}_{n+l} \Bigr) ~,\\
\bigl[ \bar{\Lambda }^{I}_{\scriptscriptstyle{XP}}, X^{J}_{-n} \bigr] =& -[\bar{\Lambda }^{I}_{\scriptscriptstyle{XP}}, X^{J}_{n}] ^{\dagger } ~,~ \bigl[ \bar{\Lambda }^{I}_{\scriptscriptstyle{XP}}, P^{J}_{-n} \bigr] = -[\bar{\Lambda }^{I}_{\scriptscriptstyle{XP}}, P^{J}_{n}] ^{\dagger } ~.
\end{split}\label{eq:6-5}
\end{eqnarray}
and 
\begin{eqnarray}\begin{split}
\bigl[ \bar{\Lambda }^{I}_{\scriptscriptstyle{XP}}, \bar{M}_{n} \bigr] =& i n P^{I}_{n} \left( -(n+1) +\biggl[ 2M_{0}\frac{1}{n} +\frac{D-2}{6}\Bigl( n-\frac{1}{n} \Bigr) \biggr] \right) \\
& +\frac{1}{2}n \left[ \sum _{l=1}^{n-1} \bar{L}_{n-l}X^{I}_{l} +\sum _{l=1}^{\infty } (\bar{L}_{-l}X^{I}_{n+l}+X^{I}_{-l}\bar{L}_{n+l}) \right] \\
& +i n \left[ \sum _{l=1}^{n-1} \frac{1}{l}\bar{M}_{n-l}P^{I}_{l} +\sum _{l=1}^{\infty } \Bigl( \frac{1}{n+l} \bar{M}_{-l}P^{I}_{n+l}-\frac{1}{l}P^{I}_{-l}\bar{M}_{n+l} \Bigr) \right] ~,\\
\bigl[ \bar{\Lambda }^{I}_{\scriptscriptstyle{XP}}, \bar{M}_{-n} \bigr] =& -\bigl[ \bar{\Lambda }^{I}_{\scriptscriptstyle{XP}}, \bar{M}_{n} \bigr] ^{\dagger } ~.
\end{split}\label{eq:6-6}
\end{eqnarray}

\subsubsection{Calculation of $[\mathcal{J}^{-I}_{\scriptscriptstyle{XP}}, \mathcal{K}^{K}_{\scriptscriptstyle{XP}}]$ for $I \not= K$ and anomaly}
Here we calculate $[\mathcal{J}^{-I}_{\scriptscriptstyle{XP}}, \mathcal{K}^{K}_{\scriptscriptstyle{XP}}]$ for $I \not= K$ to find the anomaly in the spacetime conformal symmetry explicitly. Because the calculation is too lengthy, we omit the ordered classical part except for terms with $M_{0}$ at the right end and then show only the anomalous part. The contribution to the anomalous part in $[\mathcal{J}^{-I}_{\scriptscriptstyle{XP}}, \mathcal{K}^{K}_{\scriptscriptstyle{XP}}]$ comes from terms with $M_{0}$ at the right end and quantum effect terms, which arise from extra exchanging operators to get the correctly ordered result.

We firstly calculate the contribution from each part of $\mathcal{J}^{-I}_{\scriptscriptstyle{XP}}$ in (\ref{eq:6-2}) and then collect the results. The commutator of the first term in (\ref{eq:6-2}) is
\begin{eqnarray}
[x^{-}p^{I},~ \mathcal{K}^{K}_{\scriptscriptstyle{XP}}] = ix^{-}x^{I}p^{K} -ix^{-}\sum _{n=1}^{\infty } \Bigl[ (X^{I}_{-n}P^{K}_{n}-X^{K}_{-n}P^{I}_{n}) +h.c. \Bigr] \sim 0 ~,
\label{eq:6-7}
\end{eqnarray}
where ``$\sim $" means the equivalence up to non-anomalous part.\\
The commutator of the second part in (\ref{eq:6-2}) is
{\setlength\arraycolsep{1pt}\begin{eqnarray}
&& \Bigl[ \frac{1}{2p_{-}}x^{I}p\cdot p -\frac{i}{2p_{-}}p^{I} ,~ \mathcal{K}^{K}_{\scriptscriptstyle{XP}} \Bigr] \nonumber \\
&=& -ix^{-}x^{I}p^{K} +i\frac{D-3}{2p_{-}}\Bigl[ \bigl( x^{I}(x\cdot p)p^{K} -x^{I}x^{K}(p\cdot p) \bigr) + h.c. \Bigr] \nonumber \\
&& -\frac{i}{p_{-}}x^{I}p^{K}\sum _{n=1}^{\infty }(X_{n}\cdot P_{-n} +h.c. ) -\frac{i}{2p_{-}}(x^{I}p_{J}+p_{J}x^{I}) \sum _{n=1}^{\infty } \Bigl[ (X^{J}_{-n}P^{K}_{n}-X^{K}_{-n}P^{J}_{n}) +h.c. \Bigr] \sim 0 ~.
\label{eq:6-8}
\end{eqnarray}}
And the commutator of the third part in (\ref{eq:6-2}) is
\begin{eqnarray}
\Bigl[ \frac{1}{2p_{-}}x^{I}\bar{L}_{0} ,~ \mathcal{K}^{K}_{\scriptscriptstyle{XP}} \Bigr] = \frac{i}{p_{-}}x^{I}p^{K}\sum _{n=1}^{\infty }(X_{-n}\cdot P_{n}+h.c.) +\frac{i}{p_{-}}x^{I}\bar{\Lambda }^{K}_{\scriptscriptstyle{XP}} \sim 0 ~.
\label{eq:6-9}
\end{eqnarray}
Thus we find that commutators of the first three parts in (\ref{eq:6-2}) do not contribute to the anomalous part.

However, there are contributions from the commutators of 4th and 5th parts in (\ref{eq:6-2}) to the anomalous part. Because terms with odd degree of $x^{I}$ and $p^{I}$ does not contribute to the anomalous part, we find that the contribution from the 4th part to the anomalous part comes from commutators with the first and second lines in (\ref{eq:6-4}). From a short calculation, we find that the commutator of the 4th part in (\ref{eq:6-2}) and the first lines in (\ref{eq:6-4}) does not contribute to the anomalous part. Then we obtain 
{\setlength\arraycolsep{1pt}\begin{eqnarray}
&& \Bigl[ (\mbox{4th term of (\ref{eq:6-2})}) ,~ \mathcal{K}^{K}_{\scriptscriptstyle{XP}} \Bigr] \sim  \Bigl[ (\mbox{4th term of (\ref{eq:6-2})}) ,~(\mbox{2nd line of (\ref{eq:6-4})}) \Bigr] \nonumber \\
&&~~ \sim  -\frac{D-2}{2p_{-}}\sum _{n=1}^{\infty } \Bigl( X^{I}_{-n}P^{K}_{n} -h.c. \Bigr) -\frac{D-6}{2p_{-}}\sum _{n=1}^{\infty }\Bigl( X^{K}_{-n}P^{I}_{n} - h.c. \Bigr) ~.
\label{eq:6-10}
\end{eqnarray}}
Similarly, the contribution from the 5th part in (\ref{eq:6-2}) to the anomalous part comes from commutators with the 3rd, 4th and 5th lines in (\ref{eq:6-4}). We collect them to obtain
{\setlength\arraycolsep{1pt}\begin{eqnarray}
&& \Bigl[ -\frac{\bar{\Lambda }^{I}_{\scriptscriptstyle{XP}}}{p_{-}} ,~\mathcal{K}^{K}_{\scriptscriptstyle{XP}} \Bigr] \sim  \Bigl[ -\frac{\bar{\Lambda }^{I}_{\scriptscriptstyle{XP}}}{p_{-}} ,~(\mbox{3rd, 4th and 5th line of (\ref{eq:6-4})}) \Bigr] \nonumber \\
&&~ \sim -\frac{2}{p_{-}}\sum _{n=1}^{\infty }(n-1)\Bigl( X^{I}_{-n}P^{K}_{n} -h.c. \Bigr) +\frac{D-1}{2p_{-}}\sum _{n=1}^{\infty }(n-1)\Bigl( X^{K}_{-n}P^{I}_{n} - h.c. \Bigr) \nonumber \\
&&~~~ +\frac{D-4}{2p_{-}}\sum _{n=1}^{\infty }(n-1)\Bigl( X^{I}_{-n}P^{K}_{n} -h.c. \Bigr) -\frac{1}{2p_{-}}\sum _{n=1}^{\infty }(n-1)\Bigl( X^{K}_{-n}P^{I}_{n} - h.c. \Bigr) \nonumber \\
&&~~~ +\frac{1}{p_{-}}\sum _{n=1}^{\infty }(n-1)\Bigl( X^{I}_{-n}P^{K}_{n} -h.c. \Bigr) \nonumber \\
&&~~~ -\frac{1}{p_{-}}\sum _{n=1}^{\infty }\Bigl( X^{K}_{-n}P^{I}_{n} - h.c. \Bigr) \left[ \Bigl( \frac{D-2}{6} -2 \Bigr) n +\Bigl( 2M_{0} -\frac{D-2}{6} \Bigr) \frac{1}{n} \right] \nonumber \\
&&~ \sim \frac{D-6}{2p_{-}}\sum _{n=1}^{\infty }(n-1)\Bigl( X^{I}_{-n}P^{K}_{n} -h.c. \Bigr) +\frac{D-2}{2p_{-}}\sum _{n=1}^{\infty }(n-1)\Bigl( X^{K}_{-n}P^{I}_{n} - h.c. \Bigr) \nonumber \\
&&~~~ -\frac{1}{p_{-}}\sum _{n=1}^{\infty }\Bigl( X^{K}_{-n}P^{I}_{n} - h.c. \Bigr) \left[ \Bigl( \frac{D-2}{6} -2 \Bigr) n +\Bigl( 2M_{0} -\frac{D-2}{6} \Bigr) \frac{1}{n} \right] ~.
\label{eq:6-11}
\end{eqnarray}}
\\
We collect (\ref{eq:6-7})-(\ref{eq:6-11}) to get 
\begin{eqnarray}\begin{split}
\Bigl[ \mathcal{J}^{-I}_{\scriptscriptstyle{XP}} ,~ \mathcal{K}^{K}_{\scriptscriptstyle{XP}} \Bigr] \sim & \frac{1}{p_{-}}\sum _{n=1}^{\infty }\Bigl[ (X^{I}_{-n}P^{K}_{n}+X^{K}_{-n}P^{I}_{n}) - h.c. \Bigr] \left[ \frac{D-6}{2}n -(D-4) \right] \\
& -\frac{1}{p_{-}}\sum _{n=1}^{\infty }\Bigl( X^{K}_{-n}P^{I}_{n} - h.c. \Bigr) \left[ \Bigl( \frac{D-2}{6} -4 \Bigr) n +\Bigl( 2M_{0} -\frac{D-2}{6} \Bigr) \frac{1}{n} \right] 
\end{split}\label{eq:6-12}
\end{eqnarray}
As expected from (\ref{eq:6-1}), the first line in the r.h.s. of (\ref{eq:6-12}) is symmetric in exchanging $I$ and $K$ and the second line vanishes when $D=26$ and $M_{0}=2$. Then the first line is anomalous.\\
Thus we have found explicitly the anomaly of the spacetime conformal symmetry for the tensionless string in the XP-normal order.\\

If we interpret anomalous terms in $[\mathcal{J}^{-I}_{\scriptscriptstyle{XP}}, \mathcal{K}^{K}_{\scriptscriptstyle{XP}}]$ as a new constraint condition to avoid the anomaly, commutators of such a constraint and generators produce other new constraints, as we see in \cite{bib:012, bib:013}. Under all constraint conditions, only the string ground state $|0\rangle _{\scriptscriptstyle{XP}}$ survives. This is consistent with the fact that the mass eigenstate with positive norm in the XP-normal order is only the string ground state.\\
Because another type of anomalous terms, $i\frac{\tilde{C}}{p_{-}^2}$, does not exist in the dangerous commutator, the theory restricted to the string ground state preserves the spacetime conformal symmetry. It means that the tensionless string theory with the spacetime conformal symmetry in the XP-normal order is point-like. 

\subsection{$\tilde{\mathcal{K}}^{-}_{\scriptscriptstyle{XP}}\equiv -i[\mathcal{J}^{-}_{\scriptscriptstyle{XP}}, \tilde{\mathcal{K}}^{-}_{\scriptscriptstyle{XP}}]$ and $[\mathcal{J}^{-}_{\scriptscriptstyle{XP}}, \tilde{\mathcal{K}}^{-}_{\scriptscriptstyle{XP}}]$ in $D=3$ }
$[\mathcal{J}^{-}_{\scriptscriptstyle{XP}}, \tilde{\mathcal{K}}^{-}_{\scriptscriptstyle{XP}}]$ in $D=3$ has $XP$-quadratic terms as the quantum effect term as well as (\ref{eq:4-16}). However, because there is only trace part, it is not anomalous. It defines $\tilde{\mathcal{K}}^{-}_{\scriptscriptstyle{XP}}$. In $D=3$, $[\mathcal{J}^{-}_{\scriptscriptstyle{XP}}, \tilde{\mathcal{K}}^{-}_{\scriptscriptstyle{XP}}]$ can become anomalous. \\
Because the dangerous commutator $[\mathcal{J}^{-}, \mathcal{J}^{-}]$ is trivially zero in $D=3$, the ordering constant in $M_{0}$, $a$, is not determined in $D=3$ only by the requirement of the Lorentz symmetry, unlike in $D>3$. Thanks to the freedom of $a$, the anomaly in $[\mathcal{J}^{-}_{\scriptscriptstyle{XP}}, \tilde{\mathcal{K}}^{-}_{\scriptscriptstyle{XP}}]$ may be avoided. From only the information about the structure of commutators, we can not know whether the anomaly is avoided or not. To know it, we must calculate explicitly $\tilde{\mathcal{K}}^{-}_{\scriptscriptstyle{XP}}\equiv -i[\mathcal{J}^{-}_{\scriptscriptstyle{XP}}, \tilde{\mathcal{K}}^{-}_{\scriptscriptstyle{XP}}]$ and then $[\mathcal{J}^{-}_{\scriptscriptstyle{XP}}, \tilde{\mathcal{K}}^{-}_{\scriptscriptstyle{XP}}]$.

\subsubsection{Preparation for explicit calculation}
Here we give the explicit representation of important generators and some useful commutation relations.

First, we give the representation of generators which we need for the calculation, $\mathcal{J}^{-}_{\scriptscriptstyle{XP}}$ and $\mathcal{K}_{\scriptscriptstyle{XP}}$ and $\mathcal{K}^{-}_{\scriptscriptstyle{XP}}$. $\mathcal{J}^{-}_{\scriptscriptstyle{XP}}$ is written as
\begin{eqnarray}
\mathcal{J}^{-}_{\scriptscriptstyle{XP}} = -x^{-}p -\frac{1}{2p_{-}}xL_{0} +\frac{i}{2p_{-}}p +\frac{\Lambda _{\scriptscriptstyle{XP}}}{p_{-}} ~,
\label{eq:6-13}
\end{eqnarray}
where $L_{0} =pp+\mathcal{M}^2 =pp+2\sum _{n>0}P_{-n}P_{n}$ and $\Lambda _{\scriptscriptstyle{XP}}$ is defined as
\begin{eqnarray}
\Lambda _{\scriptscriptstyle{XP}} \equiv \sum _{l=1}^{\infty } \Bigl[ -\frac{1}{2} (X_{-l}L_{l}+L_{-l}X_{l}) +\frac{i}{l}(P_{-l}M_{l}-M_{-l}P_{l}) \Bigr] ~.
\label{eq:6-14}
\end{eqnarray}
Here we note that $\Lambda _{\scriptscriptstyle{XP}}$ is hermitian and $p$ zero mode in $\Lambda _{\scriptscriptstyle{XP}}$ is canceled.\\
The important generators among the dilatation and the special conformal transformation are  
\begin{eqnarray}\begin{split}
\mathcal{K}_{\scriptscriptstyle{XP}} =& xx^{-}p_{-} +\frac{1}{2}xxp -ix +x\sum _{n=1}^{\infty }(X_{-n}P_{n}+P_{-n}X_{n}) +\sum _{n=1}^{\infty }X_{-n}X_{n}p \\
& +\frac{1}{2}\sum _{n=1}^{\infty }\sum _{m=1}^{\infty } \Bigl[ (X_{-n}X_{-m}P_{n+m}+2X_{-n}P_{-m}X_{n+m}) + h.c. \Bigr] \\
& -i\sum _{m=1}^{\infty }\frac{1}{m}(X_{-m}M_{m}-M_{-m}X_{m}) ~,\\
\end{split}\label{eq:6-15}
\end{eqnarray}
and 
{\setlength\arraycolsep{1pt}\begin{eqnarray}
\mathcal{K}^{-}_{\scriptscriptstyle{XP}} &=& x^{-}\mathcal{D}_{\scriptscriptstyle{XP}} -\frac{i}{2}x^{-} +\frac{1}{8p_{-}}(xxL_{0}+L_{0}xx) -\frac{1}{p_{-}}x\Lambda _{\scriptscriptstyle{XP}} \nonumber \\
&&+\frac{1}{2p_{-}}\sum _{n=1}^{\infty }X_{-n}L_{0}X_{n} +\frac{1}{4p_{-}}\sum _{n=1}^{\infty }\sum _{m=1}^{\infty } \Bigl[ (X_{-n}X_{-m}L_{n+m}+2X_{-n}L_{-m}X_{n+m}) + h.c. \Bigr] \\
&&+\frac{i}{p_{-}}\sum _{n=1}^{\infty }\sum _{m=1}^{\infty } \Bigl[ \Bigl( \frac{n+m}{m^2}X_{-n}M_{-m}P_{n+m}+\frac{n}{m^2}X_{-n-m}M_{m}P_{n}-\frac{m}{(n+m)^2}X_{-n}P_{-m}M_{n+m} \Bigr) + h.c. \Bigr] ~. \nonumber
\label{eq:6-16}
\end{eqnarray}}

Next we give some useful commutation relations below. 
\begin{eqnarray}\begin{split}
& [\Lambda _{\scriptscriptstyle{XP}}, x] = [\Lambda _{\scriptscriptstyle{XP}}, p] = 0 ~,\\
& [\Lambda _{\scriptscriptstyle{XP}}, X_{n}] = \frac{1}{n}\bar{M}_{n} +in \biggl[ \sum _{l=1}^{n-1}\frac{1}{l}X_{n-l}P_{l} +\sum _{l=1}^{\infty }\Bigl( \frac{1}{n+l}X_{-l}P_{n+l}-\frac{1}{l}P_{-l}X_{n+l} \Bigr) \biggr] ~,\\
& [\Lambda _{\scriptscriptstyle{XP}}, P_{n}] = -\frac{i}{2}\bar{L}_{n} +in \biggl[ \sum _{l=1}^{n-1}\frac{1}{l}P_{n-l}P_{l} +\sum _{l=1}^{\infty }\Bigl( \frac{1}{n+l}-\frac{1}{l} \Bigr) P_{-l}P_{n+l} \biggr] ~,\\
& [\Lambda _{\scriptscriptstyle{XP}}, \bar{M}_{n}] = iP_{n}\Bigl( 2M_{0} +\frac{1}{6}(n^2 -1) -n(n+1) \Bigr) +n\bar{L}_{0}X_{n} \\
&~~~~~~~~~~~~ +\frac{1}{2}n \biggl[ \sum _{l=1}^{n-1}\bar{L}_{n-l}X_{l} +\sum _{l=1}^{\infty }(X_{-l}\bar{L}_{n+l}+\bar{L}_{-l}X_{n+l}) \biggr] \\
&~~~~~~~~~~~~ +in\biggl[ \sum _{l=1}^{n-1}\frac{1}{l}\bar{M}_{n-l}P_{l} +\sum _{l=1}^{\infty }\Bigl( \frac{1}{n+l}\bar{M}_{-l}P_{n+l}-\frac{1}{l}P_{-l}\bar{M}_{n+l} \Bigr) \biggr] ~,\\
& [\Lambda _{\scriptscriptstyle{XP}}, X_{-n}] = -[\Lambda _{\scriptscriptstyle{XP}}, X_{n}]^{\dagger } ~,~ [\Lambda _{\scriptscriptstyle{XP}}, P_{-n}] = -[\Lambda _{\scriptscriptstyle{XP}}, P_{n}]^{\dagger } ~,~ [\Lambda _{\scriptscriptstyle{XP}}, M_{-n}] = -[\Lambda _{\scriptscriptstyle{XP}}, M_{n}]^{\dagger } ~,
\end{split}\label{eq:6-17}
\end{eqnarray}
where the barred operators are defined by removing to zero mode from unbarred ones.

\subsubsection{Calculation of $\tilde{\mathcal{K}}^{-}_{\scriptscriptstyle{XP}}$}
For the algebraic requirement, we must use $\tilde{\mathcal{K}}^{-}_{\scriptscriptstyle{XP}}$ defined below instead of $\mathcal{K}^{-}_{\scriptscriptstyle{XP}}$.
\begin{eqnarray}
\tilde{\mathcal{K}}^{-}_{\scriptscriptstyle{XP}} \equiv -i[\mathcal{J}^{-}_{\scriptscriptstyle{XP}},\mathcal{K}_{\scriptscriptstyle{XP}}] = \mathcal{K}^{-}_{\scriptscriptstyle{XP}} + \frac{\tilde{C}}{p_{-}} + \delta \mathcal{K}^{-}_{\scriptscriptstyle{XP}} ~.
\label{eq:6-18}
\end{eqnarray}
The first term in the r.h.s. of (\ref{eq:6-18}) is the original XP-normal ordered generator in (\ref{eq:6-16}). The second term is the quantum effect term arising only from zero modes. The third term consists of quantum effect terms of degree 2 and terms with $M_{0}$ at the right end, such as
\begin{eqnarray}
\delta \mathcal{K}^{-}_{\scriptscriptstyle{XP}} = \frac{i}{p_{-}}\sum _{n=1}^{\infty } (X_{-n}P_{n}-P_{-n}X_{n}) (g_{(n)}+h_{(n)}M_{0}) ~,
\label{eq:6-19}
\end{eqnarray}
where $g_{(n)}$ and $h_{(n)}$ is coefficients dependent on Fourier-mode index, $n$. 

We calculate $[\mathcal{J}^{-}_{\scriptscriptstyle{XP}},\mathcal{K}_{\scriptscriptstyle{XP}}]$ explicitly to determine $\tilde{C}$ and $\delta \mathcal{K}^{-}_{\scriptscriptstyle{XP}}$ in (\ref{eq:6-18}).
Because the calculation is lengthy, we omit the contribution to $\mathcal{K}^{-}_{\scriptscriptstyle{XP}}$ obtained by the classical calculation.

The contribution from the commutator of the first three terms in (\ref{eq:6-13}) and $\mathcal{K}_{\scriptscriptstyle{XP}}$ is
\begin{eqnarray}
\Bigl[ -x^{-}p -\frac{1}{2p_{-}}xL_{0} +\frac{i}{2p_{-}}p,~ \mathcal{K}_{\scriptscriptstyle{XP}} \Bigr] \sim \frac{i}{4p_{-}} -\frac{1}{2p_{-}}\sum _{n=1}^{\infty }(X_{-n}P_{n}-P_{-n}X_{n}) ~,
\label{eq:6-20}
\end{eqnarray}
where ``$\sim $" means the extraction of the contribution to $\tilde{C}$ and $\delta \mathcal{K}^{-}_{\scriptscriptstyle{XP}}$.

The contribution from the commutator of the last terms in (\ref{eq:6-13}) is lengthy. Therefore, firstly we calculate the contribution from each line of $\mathcal{K}_{\scriptscriptstyle{XP}}$ in (\ref{eq:6-15}) and next collect them to obtain $[\frac{\Lambda _{\scriptscriptstyle{XP}}}{p_{-}}, \mathcal{K}_{\scriptscriptstyle{XP}}]$.\\
In the calculation, we may use the help of (\ref{eq:6-17}). \\
After the calculation, we find that there is no contribution from the first and second lines of $\mathcal{K}_{\scriptscriptstyle{XP}}$ in (\ref{eq:6-15}).
\begin{eqnarray}\begin{split}
\Bigl[ \frac{\Lambda _{\scriptscriptstyle{XP}}}{p_{-}}, (\mbox{First line in } (\ref{eq:6-15})) \Bigr] \sim 0 ~,\\
\Bigl[ \frac{\Lambda _{\scriptscriptstyle{XP}}}{p_{-}}, (\mbox{Second line in } (\ref{eq:6-15})) \Bigr] \sim 0 ~.
\label{eq:6-21}\end{split}
\end{eqnarray}
Because $x$- or $p$-zero modes remain in the first commutator in (\ref{eq:6-21}), it does not contribute to $\frac{tilde{C}}{p_{-}}$ and $\delta \mathcal{K}^{-}_{\scriptscriptstyle{XP}}$. The contribution from the second commutator is canceled.\\
The contribution from the third line of $\mathcal{K}_{\scriptscriptstyle{XP}}$ in (\ref{eq:6-15}) is \footnote{We must not forget the contribution from terms arising in moving $M_{0}$ to the right end.}
\begin{eqnarray}
\Bigl[ \frac{\Lambda _{\scriptscriptstyle{XP}}}{p_{-}}, (\mbox{Third line in } (\ref{eq:6-15})) \Bigr] \sim \frac{1}{p_{-}}\sum _{n=1}^{\infty }(X_{-n}P_{n}-P_{-n}X_{n}) f_{(n)} ~,
\label{eq:6-22}
\end{eqnarray}
where 
\begin{eqnarray}
f_{(n)} = \bigl( \frac{1}{6}-3 \bigr) n +\frac{1}{2} +\bigl( 2M_{0} -\frac{1}{6} \bigr) \frac{1}{n} -2\sum_{m=1}^{n-1}\frac{n(n-m)}{m^2} ~.
\label{eq:6-23}
\end{eqnarray}
The sum of the last term in $f_{(n)}$ arises when we order $\frac{2i}{p_{-}}\sum _{m>0}\frac{1}{m^2}\bar{M}_{-m}\bar{M}_{m}$ correctly. 

Thus we collect (\ref{eq:6-20}) and (\ref{eq:6-21})-(\ref{eq:6-22}) to find
\begin{eqnarray}
\tilde{\mathcal{K}}^{-}_{\scriptscriptstyle{XP}} = \mathcal{K}^{-}_{\scriptscriptstyle{XP}} + \frac{1}{4p_{-}} -\frac{i}{p_{-}}\sum _{n=1}^{\infty }(X_{-n}P_{n}-P_{-n}X_{n}) f_{(n)} ~.
\label{eq:6-24}
\end{eqnarray}

\subsubsection{Anomalous terms in $[\mathcal{J}^{-}_{\scriptscriptstyle{XP}}, \tilde{\mathcal{K}}^{-}_{\scriptscriptstyle{XP}}]$}
In section 4, we have investigate the structure of $[\mathcal{J}^{-}_{\scriptscriptstyle{XP}}, \tilde{\mathcal{K}}^{-}_{\scriptscriptstyle{XP}}]$. By using the mode expansion in (\ref{eq:4-22}), terms with $M_{0}$ at the right end and quantum effect terms are unified into the following parts.
\begin{eqnarray}\begin{split}
& i\frac{p}{p_{-}^2}\times (\mbox{const.}) ~,\\
& \frac{p}{p_{-}^2}\sum _{n=1}^{\infty }(X_{-n}P_{n} -h.c. ) k_{(n)} ~,
\end{split}\label{eq:6-25}
\end{eqnarray}
\begin{eqnarray}\begin{split}
& \frac{1}{p_{-}^2}\sum _{n=1}^{\infty }\sum _{m=1}^{\infty }(X_{-n-m}P_{m}P_{n} -h.c. ) g_{(n,m)} ~,\\
& \frac{1}{p_{-}^2}\sum _{n=1}^{\infty }\sum _{m=1}^{\infty }(X_{-n}P_{-m}P_{n+m} -h.c. ) h_{(n,m)} ~,
\end{split}\label{eq:6-26}
\end{eqnarray}
where $g_{(n,m)}$ and $h_{(n,m)}$ are coefficients which depend of $n$ and $m$ and contain $M_{0}\approx a$. 

The first line in (\ref{eq:6-25}) has the same structure as in the Hermitian R-order and the second line in (\ref{eq:6-25}) is new one. Both structures in (\ref{eq:6-25}) contain zero mode, $p$. Therefore, after the similar calculation to the case of the Hermitian R-order in which we have considered the commutator of $x$ and $[\mathcal{J}^{-}, \tilde{\mathcal{K}}^{-}]$, we will find the absence of them.

Two structures in (\ref{eq:6-26}) are caused by the fact that the XP-normal order breaks the mode flipping symmetry of coefficients. 
The anomaly in $[\mathcal{J}^{-}_{\scriptscriptstyle{XP}}, \tilde{\mathcal{K}}^{-}_{\scriptscriptstyle{XP}}]$ vanishes if and only if all $g_{(n,m)}$ and $h_{(n,m)}$ equal to zero. Because there is only one freedom in the choice of $M_{0}$, this dangerous commutator is anomalous unless the profound reason exists. We extract the lowest part in two structures of (\ref{eq:6-26}), $(n,m)=(1,1)$, to verify the existence of the anomaly which can not be removed by the choice of $M_{0}$. \\

Because of the lengthy calculation, we separate $\tilde{\mathcal{K}}^{-}_{\scriptscriptstyle{XP}}$ into $\mathcal{K}^{-}_{\scriptscriptstyle{XP}}+\frac{1}{4p_{-}}$ and $\delta \mathcal{K}^{-}$.\\
First we separate the commutator of $\mathcal{J}^{-}_{\scriptscriptstyle{XP}}$ and $\mathcal{K}^{-}_{\scriptscriptstyle{XP}}+\frac{1}{4p_{-}}$ into two parts as 
\begin{eqnarray}
\Bigl[ \mathcal{J}^{-}_{\scriptscriptstyle{XP}}, \mathcal{K}^{-}_{\scriptscriptstyle{XP}}+\frac{1}{4p_{-}} \Bigr] = \Bigl[ -x^{-}p -\frac{1}{2p_{-}}xL_{0} +\frac{i}{2p_{-}}p ,~ \mathcal{K}^{-}_{\scriptscriptstyle{XP}}+\frac{1}{4p_{-}} \Bigr] +\Bigl[ \frac{\Lambda _{\scriptscriptstyle{XP}}}{p_{-}} ,~ \mathcal{K}^{-}_{\scriptscriptstyle{XP}}+\frac{1}{4p_{-}} \Bigr] ~.
\label{eq:6-27}
\end{eqnarray}
The contribution from the first part in the r.h.s. of (\ref{eq:6-27}) is obtained as \footnote{Note that $\mathcal{P}_{+N}=-\frac{L_{0}}{2p_{-}}$ and $[\mathcal{P}_{+N},\mathcal{K}^{-}_{\scriptscriptstyle{XP}}]=0$.}
\begin{eqnarray}
\Bigl[ -x^{-}p-\frac{1}{2p_{-}}xL_{0}+\frac{i}{2p_{-}}p ,~ \mathcal{K}_{\scriptscriptstyle{XP}} \Bigr] \sim \frac{3}{4p_{-}^2}(X_{-2}P_{1}P_{1}-h.c.) +\frac{9}{4p_{-}^2}(X_{-1}P_{-1}P_{2}-h.c.) ~,
\label{eq:6-28}
\end{eqnarray}
where $\sim $ means the extraction of the contribution to $(X_{-2}P_{1}P_{1}-h.c.)$ or $(X_{-1}P_{-1}P_{2}-h.c.)$.
The contribution from the second part in the r.h.s. of (\ref{eq:6-27}) is calculated for each line of (\ref{eq:6-15}) separately as follows.
\begin{eqnarray}\begin{split}
\Bigl[ \frac{\Lambda _{\scriptscriptstyle{XP}}}{p_{-}} ,~ (\mbox{First line in (\ref{eq:6-15})}) \Bigr] &\sim \frac{9}{4p_{-}^2}(X_{-2}P_{1}P_{1}-h.c.) -\frac{3}{4p_{-}^2}(X_{-1}P_{-1}P_{2}-h.c.) ~,\\
\Bigl[ \frac{\Lambda _{\scriptscriptstyle{XP}}}{p_{-}} ,~ (\mbox{Second line in (\ref{eq:6-15})}) \Bigr] &\sim -\frac{3}{4p_{-}^2}(X_{-2}P_{1}P_{1}-h.c.) +\frac{3}{2p_{-}^2}(X_{-1}P_{-1}P_{2}-h.c.) ~,\\
\Bigl[ \frac{\Lambda _{\scriptscriptstyle{XP}}}{p_{-}} ,~ (\mbox{Third line in (\ref{eq:6-15})}) \Bigr] &\sim \frac{1}{p_{-}^2}(X_{-2}P_{1}P_{1}-h.c.)[4-2M_{0}] \\
&~~~ +\frac{1}{p_{-}^2}(X_{-1}P_{-1}P_{2}-h.c.)\Bigl[ \frac{25}{4} -\frac{7}{2}M_{0} \Bigr] ~.
\end{split}\label{eq:6-29}
\end{eqnarray}
Then we collect them to obtain
\begin{eqnarray}\begin{split}
\Bigl[ \frac{\Lambda _{\scriptscriptstyle{XP}}}{p_{-}} ,~ \mathcal{K}^{-}_{\scriptscriptstyle{XP}}+\frac{1}{4p_{-}} \Bigr] \sim & \frac{1}{p_{-}^2}(X_{-2}P_{1}P_{1}-h.c.)\Bigl[ \frac{11}{2} -2M_{0} \Bigr] +\frac{1}{p_{-}^2}(X_{-1}P_{-1}P_{2}-h.c.)\Bigl[ 7 -\frac{7}{2}M_{0} \Bigr] ~.
\end{split}\label{eq:6-30}
\end{eqnarray}
From (\ref{eq:6-28}) and (\ref{eq:6-30}), we find 
\begin{eqnarray}\begin{split}
\Bigl[ \mathcal{J}^{-}_{\scriptscriptstyle{XP}} ,~ \mathcal{K}^{-}_{\scriptscriptstyle{XP}}+\frac{1}{4p_{-}} \Bigr] \sim & \frac{1}{p_{-}^2}(X_{-2}P_{1}P_{1}-h.c.)\Bigl[ \frac{25}{4}-2M_{0} \Bigr] +\frac{1}{p_{-}^2}(X_{-1}P_{-1}P_{2}-h.c.)\Bigl[ \frac{37}{4} -\frac{7}{2}M_{0} \Bigr] ~.
\end{split}\label{eq:6-31}
\end{eqnarray}

Next we calculate the contribution from the commutator of $\mathcal{J}^{-}_{\scriptscriptstyle{XP}}$ and $\delta \mathcal{K}^{-}_{\scriptscriptstyle{XP}}$. The result is
\begin{eqnarray}\begin{split}
\Bigl[ \mathcal{J}^{-}_{\scriptscriptstyle{XP}} ,~ \delta \mathcal{K}^{-}_{\scriptscriptstyle{XP}} \Bigr] \sim & \frac{1}{p_{-}^2}\bigl( (X_{-2}P_{1}P_{1}-h.c.) +(X_{-1}P_{-1}P_{2}-h.c.) \bigr) \Bigl[ \frac{3}{2}f_{(2)}-3f_{(1)} \Bigr] \\
=& \frac{1}{p_{-}^2}\bigl( (X_{-2}P_{1}P_{1}-h.c.) +(X_{-1}P_{-1}P_{2}-h.c.) \bigr) \Bigl[ -\frac{51}{8}-\frac{9}{2}M_{0} \Bigr] ~.
\end{split}\label{eq:6-32}
\end{eqnarray}
Note that the coincidence of coefficients of $(X_{-2}P_{1}P_{1}-h.c.)$ and $(X_{-1}P_{-1}P_{2}-h.c.)$ is the specialty in $n=m=1$. \footnote{$\frac{m}{n+m}+\frac{n}{m} = \frac{1}{2}\left( 1+\frac{n}{m}+\frac{m}{n} \right)$ for $n=m=1$.}\\

Thus we collect (\ref{eq:6-31}) and (\ref{eq:6-32}) to obtain 
\begin{eqnarray}\begin{split}
\bigl[ \mathcal{J}^{-}_{\scriptscriptstyle{XP}} ,~ \tilde{\mathcal{K}}^{-}_{\scriptscriptstyle{XP}} \bigr] \sim &  \frac{1}{p_{-}^2}(X_{-2}P_{1}P_{1}-h.c.)\Bigl[ -\frac{1}{8}-\frac{13}{2}M_{0} \Bigr] +\frac{1}{p_{-}^2}(X_{-1}P_{-1}P_{2}-h.c.)\Bigl[ \frac{23}{8} -8M_{0} \Bigr] ~.
\end{split}\label{eq:6-33}
\end{eqnarray}
From this, we find that there is no choice of $M_{0}$ such that two coefficients vanish. Therefore the dangerous commutator, $[ \mathcal{J}^{-}_{\scriptscriptstyle{XP}},~ \tilde{\mathcal{K}}^{-}_{\scriptscriptstyle{XP}} ] $, does not vanish. Then there is the spacetime conformal anomaly in the 3-dim. tensionless string theory in the XP-normal order. \\

As well as the case of $D>3$, if we interpret anomalous terms in $[\mathcal{J}^{-}_{\scriptscriptstyle{XP}}, \tilde{\mathcal{K}}^{-}_{\scriptscriptstyle{XP}}]$ as a new constraint condition, commutators of such a constraint and generators produce other new constraints. Under all constraint conditions, only the string ground state $|0\rangle _{\scriptscriptstyle{XP}}$ survives. This is consistent with the fact that the mass eigenstate with positive norm in the XP-normal order is only the string ground state.\\
Because another type of anomalous terms, $i\tilde{C}\frac{p}{p_{-}^2}$, does not exist in the dangerous commutator, the theory restricted to the string ground state preserves the spacetime conformal symmetry. It means that the tensionless string theory with the spacetime conformal symmetry in the XP-normal order is point-like.

\section{Summary and Outlook}
In this paper, we have considered the tensionless closed bosonic string theory in the light-cone gauge and calculated dangerous commutators in various dimensions and in two types of operator orders. The main products of our study in this paper are the avoidance of the spacetime conformal anomaly in the Hermitian R-order (and the Weyl order) and the explicit calculation of the spacetime conformal anomaly in the XP-normal order. Both products are independent of the spacetime dimension.

The first product has been obtained mainly in section 4 and 5. In section 4, we have considered the structure of (dangerous) commutators in the Hermitian R-order, where the mode flipping symmetry corresponding to the world sheet parity is preserved. By using the information of it, we have calculated the dangerous commutators of the spacetime conformal symmetry in the Hermitian R-order. Against the critical dimension of the conformal string in the BRST formalism or the calculation in the light-cone gauge without the mode expansion \cite{bib:012, bib:047, bib:013}, we have found that there is no spacetime conformal anomaly unrelated to the regularization in the Hermitian R-order (and the Weyl order) by the calculation with the Fourier mode expansion. Furthermore, in appendix, we have calculated the dangerous commutators in the Hermitian R-order in three dimensions with the cut-off regularization to obtain the set of the expected commutation relations for the spacetime conformal symmetry in the limit of the cut-off scale, $N\rightarrow \infty $. \\
The results obtained is a new example for the differences between the results in the BRST formalism and in the light-cone gauge quantization, such as the critical dimension of the tensionful string theory in the BRST quantization formalism and the avoidance of the Lorentz anomaly in the 3-dim. light-cone gauge quantization. Such differences between the results in the BRST formalism and in the light-cone gauge quantization are strange and interesting and then need the further studies.

The second product has been obtained from section 4 and 6. By using the information about the structure of dangerous commutators considered in section 4, we have calculated concretely the dangerous commutators in the spacetime conformal symmetry of $D>3$ and $D=3$
in the XP-normal order, where the string ground state breaks the mode flipping symmetry. Then we have verified the existence of the spacetime conformal anomaly for the tensionless string theory in the XP-normal order. Or, if we have interpreted the anomalous commutators as additional constraints to avoid the spacetime conformal anomaly, we have found that the tensionless string theory in the XP-normal order becomes point-like. It is similar to other works by the discussion of the norm of mass eigenstates\cite{bib:012, bib:013, bib:011}.\\

The outlooks for the future works are roughly divided as follows:
\begin{eqnarray*}
&\bullet & \mbox{Other types of tensionless string in light-cone gauge}\\
&\bullet & \mbox{Reproduction of results in 3-dim. theories in light-cone gauge by a covariant method} \\
&~& \mbox{ or Reason for differences between results in BRST and in light-cone gauge}\\
&\bullet & \mbox{Further study of mass spectrum in (Hermitian) R-order}\\
&~& \mbox{ and Tensionless string theory with interaction} \\
&\bullet & \mbox{Application of duality} \\
&\bullet & \mbox{Relation between tensionless string theory and higher spin gauge theory} 
\end{eqnarray*} 
Firstly, other types of the tensionless string in the light-cone gauge are interesting. The spacetime conformal anomalies of these theories in the light-cone gauge are not satisfyingly investigated yet, against many studies of the Lorentz anomaly\cite{bib:008, bib:010, bib:006}.\\
In the supersymmetric case, we have a fermionic field. We choose naturally the normal order as its operator order. Because the string ground state for the normal order breaks the mode flipping symmetry, the normal order for the fermionic modes and the Hermitian R-order in this paper are incompatible. To avoid the spacetime conformal anomaly, we may need the choice of the normal order also for bosonic modes and a miracle.\\
In the case of open tensionless string, the commutation relation of $X_{n}=X_{-n}$ and $P_{m}=P_{-m}$ is different from the case of the closed tensionless string as seen in this paper: $[X_{n}, P_{m}]=\frac{i}{2}(\delta _{n,-m}+\delta _{n,m})$. Therefore the dangerous commutator $[\mathcal{J}^{-}, \tilde{\mathcal{K}}^{-}]$ can have terms consisting of modes with even integer index, $\frac{i}{p_{-}}\sum _{n>0}C_{n}P_{2n}$ even in the Hermitian R-order and without the explicit regularization. In fact, from the explicit calculation, we will find such terms.\\
Thus both cases is difficult and then we probably need some other method.

Next, it is important to reproduce the results obtained in the light-cone gauge by some covariant method. At present, the special results in three dimensional theories comes only from the light-cone gauge quantization. It is not understood well whether the BRST formalism and the light-cone gauge quantization give really the same physics. Therefore it is important to reproduce a result obtained in one method by other method(s). If we cannot reproduce, we must consider its reason.

In this paper, we have not considered the mass spectrum. However we must study more the mass spectrum of the tensionless string in the (Hermitian) R-order, especially the massless spectrum. In other studies such as \cite{bib:010, bib:999}, the eigenfunctions only of the mass square operator, $\mathcal{M}^2$, are considered. However there is another Poincar\'{e} invariance corresponding to spin, $\Lambda $ in three dimensions. Although we use it to construct the non-separable mass eigenfunctions in \cite{bib:999}, diagonalizing simultaneously $\mathcal{M}^2$ and $\Lambda $ is possible in principle. The investigation of states with various helicities or spins, such as the case of the 3-dim. tensionful string in the light-cone gauge\cite{bib:003, bib:004, bib:005, bib:020, bib:032,bib:038, bib:022}, is interesting and important for the construction of a interacting tensionless string theory. Although there are abundant spectrum even in a single string, there will be new interesting feature or some restriction in the interacting theory. Then the interacting theory probably tell us the relation between tensionless string theories and higher spin gauge theories. 

Furthermore the application of dualities to the tensionless string theory and the understanding of the relation between tensionless string theories and higher spin gauge theories are interesting and important.

\section*{Acknowledgements}
The work of K.M. was supported by the Japan Society for the Promotion of Science (JSPS). I would like to thank to it.

\appendix

\section{Commutation Relations of Spacetime Conformal Symmetry}\label{app:com-dan}
In this appendix, we give the commutation relations of the spacetime conformal symmetry in the light-cone base. Then we indicate dangerous commutators in the tensionless string theory in light-cone gauge. Because dangerous commutators are different between the case of $D>3$ and the case of $D=3$, we consider two cases separately.

\subsection{Commutation relations and dangerous commutators in $D>3$}
The generators of the spacetime conformal symmetry are the translation $\mathcal{P}_{\mu }$, the Lorentz transformation $\mathcal{J^{\mu \nu }}$, the dilatation $\mathcal{D}$ and the special conformal transformation $\mathcal{K}^{\mu }$.\\  
The commutation relations of the Poincar\'{e} symmetry in the light-cone base\footnote{The spacetime index in the light-cone base: $\{ \mu \} = \{ +, -, I \}$ and $I=2, \cdots , D-1 $. \par The metric in the light-cone base: $\eta _{+-}=\eta ^{+-}=1$ and $\eta _{IJ}=\eta ^{IJ}=\delta ^{IJ}$. } and in $D>3$ are
\begin{eqnarray}\begin{split}
& [\mathcal{P}^{\pm }, \mathcal{J}^{+-}] =\pm i\mathcal{P}^{\pm } ~,~~ [\mathcal{P}^{\pm }, \mathcal{J}^{\mp I}]=-i\mathcal{P}_{I} ~,\\
& [\mathcal{P}^{I}, \mathcal{J}^{\pm J}]=i\delta ^{IJ}\mathcal{P}^{\pm } ~,~~ [\mathcal{P}^{I}, \mathcal{J}^{JK}]=i\left( \delta ^{IK}\mathcal{P}^{J} -\delta ^{IJ}\mathcal{P}^{K} \right) ~,\\
& [\mathcal{J} ^{+-} , \mathcal{J} ^{\pm I} ] =\mp i\mathcal{J}^{\pm I} ~,~~ [\mathcal{J} ^{+I} , \mathcal{J} ^{-K} ] =i\left( \mathcal{J}^{IK}+\delta ^{IK}\mathcal{J}^{+-} \right) ~,\\
& [\mathcal{J} ^{\pm I} ,\mathcal{J} ^{KL} ] =i\left( \delta ^{IL}\mathcal{J}^{\pm K}-\delta ^{IK}\mathcal{J}^{\pm L} \right) ~,\\
& [\mathcal{J} ^{IJ} , \mathcal{J} ^{KL} ] = i\left( \delta ^{JL}\mathcal{J}^{IK} -\delta ^{JK}\mathcal{J}^{IL} -\delta ^{IL}\mathcal{J}^{JK} +\delta ^{IK}\mathcal{J}^{JL} \right) ~,
\end{split}\label{eq:a-1}
\end{eqnarray}
and other commutators vanish.\\
The rest commutation relations of the spacetime conformal symmetry in the light-cone base are  
\begin{eqnarray}\begin{split}
& [\mathcal{D},\mathcal{P}^{\pm }]=i\mathcal{P}^{\pm } ~,~~ [\mathcal{D},\mathcal{P}^{I}]=i\mathcal{P}^{I} ~,~~  [\mathcal{D},\mathcal{K}^{\pm }]=-i\mathcal{K}^{\pm } ~,~~  [\mathcal{D},\mathcal{K}^{I}]=-i\mathcal{K}^{I} , \\
& [\mathcal{K}^{\pm } , \mathcal{P}^{\mp } ] =i\left( \mathcal{D} \pm \mathcal{J}^{+-} \right) ~,~~ [\mathcal{K}^{\pm } , \mathcal{P} ^{I} ] = -[\mathcal{K}^{I} , \mathcal{P} ^{\pm } ] =i\mathcal{J}^{\pm I} , \\
& [\mathcal{K}^{I} , \mathcal{P} ^{J} ] =i \left( \delta ^{IJ} D+\mathcal{J}^{IJ} \right) , \\
& [\mathcal{K}^{\pm } , \mathcal{J} ^{+-} ] = \pm i \mathcal{K}^{\pm } ~,~~ [\mathcal{K}^{\pm }, \mathcal{J}^{\mp I}]=-i\mathcal{K}^{I} ~,~~ [\mathcal{K}^{I}, \mathcal{J}^{\pm J}]=i\delta ^{IJ}\mathcal{K}^{\pm } , \\
& [\mathcal{K}^{I}, \mathcal{J}^{JK}]=i\left( \delta ^{IL}\mathcal{K}^{K}-\delta ^{IK}\mathcal{K}^{L} \right) , 
\end{split}\label{eq:a-2}
\end{eqnarray}
and other commutators vanish.\\

In string theories in the light-cone gauge,\footnote{The light-cone gauge is defined by $X^{+}=\tau $ and $P_{-}=p_{-}(\tau )\not= 0$.} $X^{-}(\sigma )$ is obtained by integrating other fields, $X^{I}(\sigma )$ and $P_{I}(\sigma )$, and then it is non-local. Generally, commutators of two generators which contain the part in $X^{-}(\sigma )$ dependent on $\sigma $, $\bar{X}^{-}(\sigma )$,\footnote{$\bar{X}^{-}(\sigma )=X^{-}(\sigma )-x^{-}$ where $x^{-}\equiv \oint \frac{d\sigma }{2\pi } X^{-}(\sigma )$.} are dangerous unless they are trivially zero. Generators which contain $\bar{X}^{-}(\sigma )$ are $\mathcal{J}^{-I}$, $\mathcal{K}^{J}$ and $\mathcal{K}^{-}$. Therefore dangerous commutators of the spacetime conformal symmetry are the infamous one, $[\mathcal{J}^{-I},\mathcal{J}^{-J}]$, and new ones. The new dangerous commutators are $[\mathcal{J}^{-I},\mathcal{K}^{K}]$, $[\mathcal{J}^{-I},\mathcal{K}^{-}]$, $[\mathcal{K}^{I},\mathcal{K}^{J}]$ and $[\mathcal{K}^{I},\mathcal{K}^{-}]$.\\
The correct relation of the first of new dangerous commutators is 
\begin{eqnarray}
[\mathcal{J}^{-I},\mathcal{K}^{K}]=-i\delta ^{IK}\mathcal{K}^{-} ~.
\label{eq:a-3}
\end{eqnarray}
However, the possibility that the traceless part of the transverse indexes, $I$ and $K$, exists as anomaly is shown in some operator order\cite{bib:012, bib:013}.\\
If (\ref{eq:a-3}) is correct, by using it and non-dangerous commutation relations and the Jacobi identity, we find that the rest of the new dangerous commutators are related to the infamous dangerous commutator and the first of new ones: 
\begin{eqnarray}\begin{split}
[\mathcal{J}^{-I}, \mathcal{K}^{-}] =& \left[ \mathcal{J}^{-I}, i[\mathcal{J}^{-J}, \mathcal{K}^{J}] \right] = -i\left[ \mathcal{K}^{J}, [\mathcal{J}^{-I}, \mathcal{J}^{-J}]\right] -i\left[ \mathcal{J}^{-J}, [\mathcal{K}^{J}, \mathcal{J}^{-I}] \right] ~,\\
[\mathcal{K}^{I}, \mathcal{K}^{J}] =& \left[ i[\mathcal{K}^{+}, \mathcal{J}^{-I}], \mathcal{K}^{J}\right] = -i\left[ [\mathcal{K}^{J}, \mathcal{K}^{+}], \mathcal{J}^{-I}\right] -i\left[ [\mathcal{J}^{-I}, \mathcal{K}^{J}], \mathcal{K}^{+}\right] ~,\\
[\mathcal{K}^{I}, \mathcal{K}^{-}] =& \left[ i[\mathcal{K}^{+}, \mathcal{J}^{-I}], \mathcal{K}^{-}\right] = -i\left[ [\mathcal{K}^{-}, \mathcal{K}^{+}], \mathcal{J}^{-I}\right] -i\left[ [\mathcal{J}^{-I}, \mathcal{K}^{-}], \mathcal{K}^{+}\right] ~,
\end{split}\label{eq:a-4}
\end{eqnarray}
where $I \not= J$. Therefore, to check the spacetime conformal symmetry of the tensionless string theory in the light-cone gauge and in $D>3$, we must calculate at least $[\mathcal{J}^{-I},\mathcal{J}^{-J}]$ and $[\mathcal{J}^{-I},\mathcal{K}^{K}]$. In the operator order in which we need some regularization, we may need to investigate other dangerous commutators to check that commutators with dropping terms in the limit of the regulator factor do not cause any problem. The necessity depends on the choice of the regularization. Here we don't discuss more.

\subsection{Commutation relations of and dangerous commutators in $D=3$}
In $D=3$, by $\mathcal{J}^{\mu } =\frac{1}{2}\epsilon ^{\mu \nu \rho }\mathcal{J}_{\nu \rho }$, the Lorentz generators are rewritten as vector, $\mathcal{J}^{\pm }= \pm \mathcal{J}^{\pm 2} ~,~ \mathcal{J} =-\mathcal{J}^{+-}$. By using the representations as vector and omitting the transverse index, the commutation relations of the spacetime symmetry become simple. \\ 
The commutation relations of the Poincar\'{e} symmetry in the light-cone base and $D=3$ are
\begin{eqnarray}\begin{split}
& [\mathcal{J}^{\pm }, \mathcal{P}^{\mp }]=\pm i \mathcal{P} ~,~~ [\mathcal{J}, \mathcal{P}^{\pm }]=\pm i \mathcal{P}^{\pm } ~,~~ [\mathcal{J}^{\pm }, \mathcal{P}]=\mp i \mathcal{P}^{\pm } ~, \\
& [\mathcal{J} , \mathcal{J} ^{\pm } ] = \pm i\mathcal{J}^{\pm } ~,~~  [\mathcal{J}^{+} , \mathcal{J} ^{-} ] = i\mathcal{J} ~,
\end{split}\label{eq:a-5}
\end{eqnarray}
and other commutators vanish. $[\mathcal{J}^{-}, \mathcal{J}^{-}]$ corresponding to the infamous dangerous commutator in the Lorentz symmetry is trivially zero because there is only one transverse direction in $D=3$ \cite{bib:003, bib:004, bib:005, bib032}. Therefore the 3-dimensional (string) theory in the light-cone gauge have no Lorentz anomaly and then preserves the Poincar\'{e} symmetry.\\
The rest commutation relations of the spacetime conformal symmetry in the light-cone base are  
\begin{eqnarray}\begin{split}
& [\mathcal{D},\mathcal{P}^{\pm }]=i\mathcal{P}^{\pm } ~,~~ [\mathcal{D},\mathcal{P}]=i\mathcal{P} ~,~~ [\mathcal{D},\mathcal{K}^{\pm }]=-i\mathcal{K}^{\pm } ~,~~ [\mathcal{D},\mathcal{K}]=-i\mathcal{K} ~,\\
& [\mathcal{K}^{\pm } , \mathcal{P}^{\mp } ] =i\left( \mathcal{D} \mp \mathcal{J} \right) ~,~~ [\mathcal{K}^{\pm } , \mathcal{P} ] =-[\mathcal{K} , \mathcal{P}^{\pm } ] =\pm i\mathcal{J}^{\pm } ~,~~ [\mathcal{K} , \mathcal{P} ] =i\mathcal{D} ~,\\
& [\mathcal{K}^{\pm } , \mathcal{J} ^{\mp } ] = \pm i \mathcal{K} ~,~~ [\mathcal{K}^{\pm } , \mathcal{J} ] = \mp i \mathcal{K}^{\pm } ~,~~ [\mathcal{K} , \mathcal{J} ^{\pm } ] = \pm i \mathcal{K}^{\pm } ~,
\end{split}\label{eq:a-6}
\end{eqnarray}
and other commutators vanish. \\

The non-trivial commutators of two generators which contain $\bar{X}^{-}$ are $[\mathcal{J}^{-}, \mathcal{K}]$, $[\mathcal{J}^{-}, \mathcal{K}^{-}]$ and $[\mathcal{K}, \mathcal{K}^{-}]$. Because the first of them has no traceless part unlike the case of $D>3$, it is not dangerous. Regardless of the existence of the quantum effect, it gives the definition: 
\begin{eqnarray}
\tilde{\mathcal{K}}^{-}\equiv -i[\mathcal{J}^{-}, \mathcal{K}] ~.
\label{eq:a-7}
\end{eqnarray}
From the algebraic requirement, we must use $\tilde{\mathcal{K}}^{-}$ in the calculation of commutators instead of $\mathcal{K}^{-}$. Most of commutators with $\tilde{\mathcal{K}}^{-}$ are easily calculated and then we find that they are correct. However, $[\mathcal{J}^{-}, \tilde{\mathcal{K}}^{-}]$ and $[\mathcal{K}, \tilde{\mathcal{K}}^{-}]$ are dangerous. Unlike the case of $D>3$, by the Jacobi identity we can not find whether $[\mathcal{J}^{-}, \tilde{\mathcal{K}}^{-}]=0$ is correct. On the other hand, $[\mathcal{K}, \tilde{\mathcal{K}}^{-}]$ is related to other commutators by the Jacobi identity:
\begin{eqnarray}
[\mathcal{K}, \tilde{\mathcal{K}}^{-}] = \bigl[ i[\mathcal{J}^{-}, \mathcal{K}^{+}], \tilde{\mathcal{K}}^{-}\bigr] = -i\bigl[ [\tilde{\mathcal{K}}^{-}, \mathcal{J}^{-}],~ \mathcal{K}^{+}\bigr] -i\bigl[ [\mathcal{K}^{+}, \tilde{\mathcal{K}}^{-}],~ \mathcal{J}^{-}\bigr] = -i\bigl[ [\tilde{\mathcal{K}}^{-}, \mathcal{J}^{-}],~ \mathcal{K}^{+}\bigr] ~.
\label{eq:a-8}
\end{eqnarray}
Therefore, to check the spacetime conformal symmetry of the tensionless string theory in the light-cone gauge and in $D=3$, we must calculate at least $\tilde{\mathcal{K}}^{-}\equiv -i[\mathcal{J}^{-}, \mathcal{K}] $ and $[\mathcal{J}^{-},\tilde{\mathcal{K}}^{-}]$.

\subsection{Dangerous commutators}
Here we summarize dangerous commutators in $D=3$ and $D>3$ which we must calculate at least on the next table.

\begin{table}[htb]\begin{center}
Dangerous commutators which we must calculate
\begin{tabular}{|c|c|c|}
\hline
Spacetime dimension & Lorentz anomaly & Spacetime conformal anomaly \\
\hline
\hline
$D>3$ & $[\mathcal{J}^{-I}, \mathcal{J}^{-J}]$ & $[\mathcal{J}^{-I}, \mathcal{K}^{K}]$ \\ \hline
$D=3$ & nothing & $\tilde{\mathcal{K}}^{-} \equiv -i [\mathcal{J}^{-}, \mathcal{K}]$, $[\mathcal{J}^{-}, \hat{\mathcal{K}}^{-}]$ \\ \hline
\end{tabular}\end{center}\end{table}

\section{Cut-off Regularization in R-order and Hermitian R-order}\label{app:reg}
In some operator orders such as the R-order and the Hermitian R-order, some Hermitian generators of the spacetime symmetry have divergent terms\cite{bib:012, bib:013}. Therefore we have to regularize them for explicit calculations. For example, the dilatation in the R-order is
\begin{eqnarray}
\mathcal{D}_{R} = x^{-}p_{-} +\sum _{n}X_{n}\cdot P_{-n} .
\label{eq:r-1}
\end{eqnarray}
Because $\mathcal{D}_{R}$ is not hermitian, we must {\it hermitianize} it. If we reorder the hermitian generator into the R-order, we find the divergent term as follows:
\begin{eqnarray}
\mathcal{D} = \frac{1}{2}\Bigl( \mathcal{D}_{R}+(\mathcal{D}_{R})^{\dagger }\Bigr) = \mathcal{D}_{R} -\frac{i}{2}\Bigl[1 +\sum _{n}(D-2) \Bigr] .
\label{eq:r-2}
\end{eqnarray}
In the Hermitian R-order where the ordered result is the second formula, the regularization is not necessary in the dilatation. However, as seen in $\frac{C}{p_{-}}$ of (\ref{eq:4-11}) or (\ref{eq:5-7}), generators of higher degree such as $\tilde{\mathcal{K}}^{-}$ defined by the commutator can contain a divergent term from the quantum effect. Therefore we need some regularization also in the Hermitian R-order. \\

In this appendix, we regularize divergences with the cut-off regularization, which drops higher Fourier modes of the transverse fields. The advantages of the cut-off regularization are that infinite series become the sum of finite terms and that all commutators of generators except for $\tilde{\mathcal{K}}^{-}$ satisfy the relations expected in the spacetime conformal group.\footnote{Of course, other regularizations exist. A example is using the approximate delta function. This regularization smear the delta function of the commutation relation $[X(\sigma ), P(\sigma ')] =2\pi i \delta (\sigma -\sigma ')$. Explicit example is $[X_{n}, P_{m}]_{\epsilon } = \frac{i}{|n|^{\epsilon }} \delta _{n,-m}$ in Fourier mode, where the regulator $\epsilon $ measures the scale of smearing. In this regularization, the divergent series may be represented by zeta functions. However all commutation relations have differences of $\mathcal{O}(\epsilon )$ from the expected relations and multiple commutators are too complicated. }

For the simplicity, we consider the case of three dimensions below. The higher dimensional case is similarly discussed.

\subsection{Cut-off regularization}
In the cut-off regularization, we drop $X$- and $P$-modes with larger index than $N$ as follows:
\begin{eqnarray}
X_{n} = P_{n} = 0 ~~ \mbox{for} ~ |n|>N ~,
\label{eq:r-3}
\end{eqnarray}
where the regulator $N$ is some large positive integer. From (\ref{eq:r-3}), modes of $P_{+}$ and $X^{-}$ are\footnote{There is no difference in $L^{\scriptscriptstyle{(N)}}_{n}$ and $M^{\scriptscriptstyle{(N)}}_{n}$ between the R-order, the Hermitian R-order and the Weyl order, as well as $L_{n}$ and $M_{n}$. Differences arise in operators of higher degree, as seen in the next subsection.}
\begin{eqnarray}\begin{split}
& L^{\scriptscriptstyle{(N)}}_{n} \equiv  \sum _{\substack{\scriptscriptstyle{|m|\leq N} \\ \scriptscriptstyle{|n-m|\leq N}}} P_{m}P_{n-m} ~,~ M^{\scriptscriptstyle{(N)}}_{n} \equiv  -i \sum _{\substack{\scriptscriptstyle{|m|\leq N} \\ \scriptscriptstyle{|n-m|\leq N}}} m X_{x}P_{n-m} ~,\\
& L^{\scriptscriptstyle{(N)}}_{n} = M^{\scriptscriptstyle{(N)}}_{n} = 0 ~~ \mbox{for} ~ |n|>2N ~.
\end{split}\label{eq:r-4}
\end{eqnarray}
For $|n|\leq N$ and $|n+m|\leq N$, the fundamental commutation relations are 
\begin{eqnarray}\begin{split}
& [X_{n}, L_{m}^{\scriptscriptstyle{(N)}}] = 2iP_{n+m} ~,~ [P_{n}, L_{m}^{\scriptscriptstyle{(N)}}] = 0 ~,\\
& [X_{n}, M_{m}^{\scriptscriptstyle{(N)}}] = (n+m)X_{n+m} ~,~ [P_{n}, M_{m}^{\scriptscriptstyle{(N)}}] = nP_{n+m} ~,
\end{split}\label{eq:r-5}
\end{eqnarray}
and otherwise vanish. ``$|n+m|\leq N$" is new restriction for the summation. Therefore $L^{\scriptscriptstyle{(N)}}_{n}$ and $M^{\scriptscriptstyle{(N)}}_{m}$ with finite $N$ do not satisfy 2D Galilean conformal algebra (GCA)\footnote{The commutation relations of 2D GCA are $[L_{n},L_{m}]=0,~ [L_{n},M_{m}]=(n-m)L_{n+m},~ [M_{n},M_{m}]=(n-m)M_{n+m}$. The detail of 2D GCA is referred to \cite{bib:007, bib:029, bib:014}.}~ and the algebra of these is not even closed: 
\begin{eqnarray}\begin{split}
\bigl[ L_{n}^{\scriptscriptstyle{(N)}}, L_{m}^{\scriptscriptstyle{(N)}} \bigr] =& 0 ~,\\
\bigl[ L_{n}^{\scriptscriptstyle{(N)}}, M_{m}^{\scriptscriptstyle{(N)}} \bigr] =& (n-m) L_{n+m}^{\scriptscriptstyle{(N)}} -2\sum _{\substack{\scriptscriptstyle{|l|\leq N}, \scriptscriptstyle{|l-n| > N} \\ \scriptscriptstyle{|n+m-l|\leq N}}} (n-l) P_{l}P_{n+m-l} ~,\\
\bigl[ M_{n}^{\scriptscriptstyle{(N)}}, M_{m}^{\scriptscriptstyle{(N)}} \bigr] =& (n-m) M_{n+m}^{\scriptscriptstyle{(N)}} -i\sum _{\substack{\scriptscriptstyle{|l|\leq N}, \scriptscriptstyle{|l-n| > N} \\ \scriptscriptstyle{|n+m-l|\leq N}}} l(l-n) X_{l}P_{n+m-l} +i\sum _{\substack{\scriptscriptstyle{|l|\leq N}, \scriptscriptstyle{|l-m| > N} \\ \scriptscriptstyle{|n+m-l|\leq N}}} l(l-m) X_{l}P_{n+m-l} ~.
\end{split}\label{eq:r-6}
\end{eqnarray}
The first commutator is good, but the 2nd term of the second commutator and the 2nd and 3rd terms of the third commutator are extra.
However, because we expect that taking the limit of $N\rightarrow \infty $ recovers the original algebra, such extra terms should be dropped in the limit if they are correctly ordered. Dropping the correctly ordered extra terms in the limit corresponds to the shift of the index of summation in series of ordered operators. The shift of the index of summation in series of ordered operators is done when we derive the Virasoro algebra for the usual tensionful string theory. Therefore, we assume that the correctly ordered extra terms have a good convergence property and are dropped in the limit of $N\rightarrow \infty $.

The regularized generators are defined by the restriction in (\ref{eq:r-3}) and indicated by $\scriptscriptstyle{(N)}$ in the subscript. For example, 
{\setlength\arraycolsep{1pt}\begin{eqnarray}
\mathcal{J}^{-} &=& -x^{-}p -\frac{1}{2p_{-}}\sum _{l}X_{l}L_{-l} +\frac{i}{p_{-}}\sum _{l \not= 0} \frac{1}{l}M_{l}P_{-l} +\frac{i}{2p_{-}}p \nonumber \\
\rightarrow \mathcal{J}^{-}_{\scriptscriptstyle{(N)}} &=& -x^{-}p -\frac{1}{2p_{-}}\sum _{|l|\leq N}X_{l}L^{\scriptscriptstyle{(N)}}_{-l} +\frac{i}{p_{-}}\sum _{0<|l|\leq N} \frac{1}{l}M^{\scriptscriptstyle{(N)}}_{l}P_{-l} +\frac{i}{2p_{-}}p ~.
\label{eq:r-7}
\end{eqnarray}}

\subsection{Problem of R-order in limit of $N\rightarrow \infty $}
Here we show that the above dropping in the R-order breaks the Hermitian property. For example, we consider the commutator of the following two hermitian operators in the R-order case and the Hermitian R-order case,
{\setlength\arraycolsep{1pt}\begin{eqnarray}
\frac{1}{2} \mathop{\sum _{\scriptscriptstyle{|n|\leq N}} \sum _{\scriptscriptstyle{|m|\leq N}} }_{\scriptscriptstyle{|n+m|\leq N}} ( X_{n}X_{m}P_{-n-m}+h.c. ) &=& \mathop{\sum _{\scriptscriptstyle{|n|\leq N}} \sum _{\scriptscriptstyle{|m|\leq N}} }_{\scriptscriptstyle{|n+m|\leq N}} X_{n}X_{m}P_{-n-m} -i(2N+1)x ~,\\
\frac{1}{2} \mathop{\sum _{\scriptscriptstyle{|k|\leq N}} \sum _{\scriptscriptstyle{|l|\leq N}} }_{\scriptscriptstyle{|k+l|\leq N}} ( X_{k+l}P_{-k}P_{-l}+h.c. ) &=& \mathop{\sum _{\scriptscriptstyle{|k|\leq N}} \sum _{\scriptscriptstyle{|l|\leq N}} }_{\scriptscriptstyle{|k+l|\leq N}} X_{k+l}P_{-k}P_{-l} -i(2N+1)p ~.
\label{eq:r-8}
\end{eqnarray}}
In order to obtain terms of lower degree correctly, we write all the range of summation explicitly as well as extra ranges. 

First the commutator of above two hermitian operators in the R-order is 
{\setlength\arraycolsep{1pt}\begin{eqnarray}
&\biggl[ &\mathop{\sum _{\scriptscriptstyle{|n|\leq N}} \sum _{\scriptscriptstyle{|m|\leq N}} }_{\scriptscriptstyle{|n+m|\leq N}} X_{n}X_{m}P_{-n-m} -i(2N+1)x ~,~ \mathop{\sum _{\scriptscriptstyle{|k|\leq N}} \sum _{\scriptscriptstyle{|l|\leq N}} }_{\scriptscriptstyle{|k+l|\leq N}} X_{k+l}P_{-k}P_{-l} -i(2N+1)p \biggr] \nonumber \\
&=& 3i \mathop{\sum _{\scriptscriptstyle{|n|\leq N}} \sum _{\scriptscriptstyle{|m|\leq N}} \sum _{\scriptscriptstyle{|l|\leq N}}}_{\scriptscriptstyle{|n+m+l|\leq N}} X_{n}X_{m}P_{l}P_{-n-m-l} +2\sum _{|l|\leq N} \bigl[ 3(2N+1)-|l| \bigr] X_{l}P_{-l} -i(2N+1)^2 \nonumber \\
&& -4i \mathop{\sum _{\scriptscriptstyle{|n|\leq N}} \sum _{\scriptscriptstyle{|m|\leq N}} \sum _{\scriptscriptstyle{|l|\leq N}}}_{\scriptscriptstyle{|n+m+l|\leq N,~ |m+l|>N}} X_{n}X_{m}P_{l}P_{-n-m-l} +i \mathop{\sum _{\scriptscriptstyle{|n|\leq N}} \sum _{\scriptscriptstyle{|m|\leq N}} \sum _{\scriptscriptstyle{|l|\leq N}}}_{\scriptscriptstyle{|n+m+l|\leq N,~ |n+m|>N}} X_{n}X_{m}P_{l}P_{-n-m-l} ~,
\label{eq:r-9}
\end{eqnarray}}
where the right hand side is anti-hermitian. The last line consists of the R-ordered extra terms. However it is not anti-hermitian. Therefore, according to the assumption in the last subsection, dropping the correctly ordered extra terms in the limit of $N\rightarrow \infty $ breaks the anti-hermitian property in the equation. Such a breaking causes the anomaly unrelated to one which we would like to know. If such breaking is accepted, anomalous quadratic terms such as the second part in the r.h.s. of (\ref{eq:r-9}) arise in $[\mathcal{J}^{-I}, \mathcal{K}^{K}]$. This is probably the cause of the anomaly in the traceless part of $[\mathcal{J}^{-I}, \mathcal{K}^{K}]$ shown in \cite{bib:012, bib:013}. 

Next we reorder the commutator of (\ref{eq:r-9}) in the Hermitian R-order, instead of the R-order.
{\setlength\arraycolsep{1pt}\begin{eqnarray}
&\biggl[ & \mathop{\sum _{\scriptscriptstyle{|n|\leq N}} \sum _{\scriptscriptstyle{|m|\leq N}} }_{\scriptscriptstyle{|n+m|\leq N}} X_{n}X_{m}P_{-n-m} -i(2N+1)x ~,~ \mathop{\sum _{\scriptscriptstyle{|k|\leq N}} \sum _{\scriptscriptstyle{|l|\leq N}} }_{\scriptscriptstyle{|k+l|\leq N}} X_{k+l}P_{-k}P_{-l} -i(2N+1)p \biggr] \nonumber \\
&=& \frac{3i}{2} \mathop{\sum _{\scriptscriptstyle{|n|\leq N}} \sum _{\scriptscriptstyle{|m|\leq N}} \sum _{\scriptscriptstyle{|l|\leq N}}}_{\scriptscriptstyle{|n+m+l|\leq N}} (X_{n}X_{m}P_{l}P_{-n-m-l}+h.c.) +i(7N^2+7N+1) \\
&& -2i \mathop{\sum _{\scriptscriptstyle{|n|\leq N}} \sum _{\scriptscriptstyle{|m|\leq N}} \sum _{\scriptscriptstyle{|l|\leq N}}}_{\scriptscriptstyle{|n+m+l|\leq N, |m+l|>N}} (X_{n}X_{m}P_{l}P_{-n-m-l} +h.c. ) +\frac{i}{2} \mathop{\sum _{\scriptscriptstyle{|n|\leq N}} \sum _{\scriptscriptstyle{|m|\leq N}} \sum _{\scriptscriptstyle{|l|\leq N}}}_{\scriptscriptstyle{|n+m+l|\leq N, |n+m|>N}} (X_{n}X_{m}P_{l}P_{-n-m-l}+h.c.) \nonumber ~.
\label{eq:r-10}
\end{eqnarray}}
Because of hermitian combinations, dropping the extra terms in the last line of (\ref{eq:r-10}) does not break the anti-hermitian property in the equation. After taking the limit, the quantum effect appears only in constant, which is 4 lower degree than the highest. Though, because such a quantum effect may cause the anomaly in commutators of the spacetime conformal symmetry, we must calculate concretely such dangerous commutators. \\

In this way, when we consider operators of higher degree such as the special conformal transformation and their commutators, the Hermitian R-order is supposed to be appropriate. Therefore, we should choose the Hermitian R-order instead of the R-order. In the Hermitian R-order, all generators and commutators are ordered in the form with the hermitian partner as follows.
\begin{eqnarray}
\mathcal{G}_{\scriptscriptstyle{HR}} \equiv \frac{1}{2} \bigl( \mathcal{G}_{\scriptscriptstyle{R}} + (\mathcal{G}_{\scriptscriptstyle{R}})^{\dagger } \bigr) ~.
\label{eq:r-11}
\end{eqnarray}
The Weyl order is possible. However the calculation in the Weyl order is more difficult.

\section{Calculations of $\tilde{\mathcal{K}}^{-}_{\scriptscriptstyle{HR (N)}}$ and $[\mathcal{J}^{-}_{\scriptscriptstyle{HR (N)}}, \tilde{\mathcal{K}}^{-}_{\scriptscriptstyle{HR (N)}}]$ in $D=3$ with Cut-off Regularization}\label{app:cal}
In this appendix, we calculate $\tilde{\mathcal{K}}^{-}_{\scriptscriptstyle{HR (N)}} \equiv -i[\mathcal{J}^{-}_{\scriptscriptstyle{HR (N)}}, \mathcal{K}_{\scriptscriptstyle{HR (N)}}]$ and $[\mathcal{J}^{-}_{\scriptscriptstyle{HR (N)}}, \tilde{\mathcal{K}}^{-}_{\scriptscriptstyle{HR (N)}}]$ in $D=3$ with the cut-off regularization explicitly to investigate whether the anomaly exists or not. Note that some commutators have extra restrictions in the range of sum. For the simplicity, we omit the range of sum which we know from the mode indexes of $X_{n}$ and $P_{n}$. 

\subsection{Definition of $\tilde{\mathcal{K}}^{-}_{\scriptscriptstyle{HR (N)}}$}
First we calculate $\tilde{\mathcal{K}}^{-}_{\scriptscriptstyle{HR (N)}} \equiv -i[\mathcal{J}^{-}_{\scriptscriptstyle{HR (N)}}, \mathcal{K}_{\scriptscriptstyle{HR (N)}}]$ and then see the effect of the cut-off regularization. \\
Because  
\begin{eqnarray}
[\mathcal{J}^{-}_{\scriptscriptstyle{HR (N)}}, \mathcal{K}_{\scriptscriptstyle{HR (N)}}] = \frac{1}{2} \left( [\mathcal{J}^{-}_{\scriptscriptstyle{HR (N)}}, \mathcal{K}_{\scriptscriptstyle{R (N)}}] -h.c. \right) ~,
\label{eq:c-1}\end{eqnarray}
it is enough to calculate the commutator of the following two generators:
\begin{eqnarray}\begin{split}
\mathcal{J}^{-}_{\scriptscriptstyle{HR (N)}} =& -x^{-}p +\Bigl( -\frac{1}{2p_{-}}xL^{\scriptscriptstyle{(N)}}_{0}+i\frac{p}{2p_{-}} \Bigr) +\frac{\Lambda ^{\scriptscriptstyle{(N)}}_{\scriptscriptstyle{HR}} }{p_{-}} ~,~~ \Lambda ^{\scriptscriptstyle{(N)}}_{\scriptscriptstyle{HR}} \equiv  \sum _{l \not= 0} \Bigl[-\frac{1}{2}X_{l}L^{\scriptscriptstyle{(N)}}_{-l}+\frac{i}{l}M^{\scriptscriptstyle{(N)}}_{l}P_{-l} \Bigr] 
\end{split}\label{eq:c-2}
\end{eqnarray}
and
\begin{eqnarray}
\mathcal{K}_{\scriptscriptstyle{(N)},R} = xx^{-}p_{-} +\frac{1}{2}\sum _{n}\sum _{m}X_{n}X_{m}P_{-n-m} +i\sum _{m \not= 0} \frac{1}{m}X_{m}M^{\scriptscriptstyle{(N)}}_{-m} ~.
\label{eq:c-3}
\end{eqnarray}
Note that, in the Hermitian R-order, all terms always have their hermitian conjugate partners.\\

Because the calculation is lengthy, we divide the commutator $[\mathcal{J}^{-}_{\scriptscriptstyle{HR (N)}}, \mathcal{K}_{\scriptscriptstyle{HR (N)}}]$ into three hermitian parts of $\mathcal{J}^{-}_{\scriptscriptstyle{HR (N)}}$ in (\ref{eq:c-2}).
The first part is 
\begin{eqnarray}
\Bigl[ -x^{-}p,~ \mathcal{K}_{\scriptscriptstyle{HR (N)}} \Bigr] = \frac{i}{2}\biggl( \Bigl( x^{-}x^{-}p_{-} +x^{-}\sum _{l\not= 0}X_{l}P_{-l} \Bigr) +h.c. \biggr) ~.
\label{eq:c-4}
\end{eqnarray}
The second part is 
{\setlength\arraycolsep{1pt}\begin{eqnarray}
\Bigl[ -\frac{1}{2p_{-}}x L^{\scriptscriptstyle{(N)}}_{0} +i\frac{p}{2p_{-}},~ \mathcal{K}_{\scriptscriptstyle{HR (N)}} \Bigr] = \frac{i}{2}\biggl( \Bigl( x^{-}xp +\frac{1}{4p_{-}} \sum _{n}X_{n}X_{-n} L^{\scriptscriptstyle{(N)}}_{0} -x\frac{\Lambda ^{\scriptscriptstyle{(N)}}_{\scriptscriptstyle{HR}}}{p_{-}} \Bigr) +h.c. \biggr) +\frac{i}{4p_{-}} 
\label{eq:c-5}
\end{eqnarray}}
and the third part is 
{\setlength\arraycolsep{1pt}\begin{eqnarray}
&& \biggl[ \frac{\Lambda ^{\scriptscriptstyle{(N)}}_{\scriptscriptstyle{HR}}}{p_{-}},~ \mathcal{K}_{\scriptscriptstyle{HR (N)}} \biggr] \nonumber \\
&& = \frac{i}{2} \Biggl( \biggl( \frac{1}{4p_{-}}\mathop{\sum _{n}\sum_{m}}_{n+m \not= 0}X_{n}X_{m}L^{\scriptscriptstyle{(N)}}_{-n-m} -\frac{i}{p_{-}}\sum _{n}\sum _{m \not= 0} \frac{n+m}{m^2}X_{n}M^{\scriptscriptstyle{(N)}}_{m}P_{-n-m} +x\frac{\Lambda ^{\scriptscriptstyle{(N)}}_{\scriptscriptstyle{HR}}}{p_{-}} \biggr) +h.c. \Biggr) \nonumber \\
&& ~ +\frac{i}{p_{-}}\biggl( -\frac{3}{4}N^2 -\frac{1}{4}N -\frac{1}{2}\mathop{\sum _{\scriptscriptstyle{|n|\leq N}}\sum _{\scriptscriptstyle{0<|m|\leq N}}}_{\scriptscriptstyle{|n+m|\leq N}}\frac{n^2}{m^2} \biggr) -\frac{2}{p_{-}}\sum _{n \not= 0} \frac{1}{n}X_{n}P_{-n} M^{\scriptscriptstyle{(N)}}_{0} +\delta \mathcal{K}^{-}_{\scriptscriptstyle{HR (N)}} ~,
\label{eq:c-6}
\end{eqnarray}}
where $\delta \mathcal{K}^{-}_{\scriptscriptstyle{HR (N)}}$ is the effect of the cut-off regularization and its concrete form is
\begin{eqnarray}\begin{split}
\delta \mathcal{K}^{-}_{\scriptscriptstyle{HR (N)}} =& \frac{1}{2p_{-}}\mathop{\sum _{n}\sum _{m}\sum _{l}}_{\scriptscriptstyle{|n+m|>N}}\Bigl( \frac{1}{4}+\frac{n}{2m}+\frac{m^2}{nl}+\frac{n+m}{l} \Bigr) (X_{n}X_{m}P_{l}P_{-n-m-l}+h.c.) \\
& -\frac{1}{2p_{-}}\mathop{\sum _{n}\sum _{m}\sum _{l}}_{\scriptscriptstyle{|m+l|>N}}\Bigl( \frac{l}{m+l}+\frac{m^2}{nl}+\frac{n+m}{l}-\frac{nm}{(m+l)^2} \Bigr) (X_{n}X_{m}P_{l}P_{-n-m-l} +h.c. ) ~.
\end{split}\label{eq:c-7}
\end{eqnarray}

We collect (\ref{eq:c-4})-(\ref{eq:c-6}) to get
\begin{eqnarray}
[\mathcal{J}^{-}_{\scriptscriptstyle{HR (N)}}, \mathcal{K}_{\scriptscriptstyle{HR (N)}}] = i\biggl( \mathcal{K}^{-}_{\scriptscriptstyle{HR (N)}} +\frac{C_{\scriptscriptstyle{(N)}}}{p_{-}} +\delta \mathcal{K}^{-}_{\scriptscriptstyle{HR (N)}} +\frac{2i}{p_{-}}\sum _{n \not= 0} \frac{1}{n}X_{n}P_{-n} M_{0}^{\scriptscriptstyle{(N)}} \biggr) \equiv i\tilde{\mathcal{K}}^{-}_{\scriptscriptstyle{HR (N)}} ~,
\label{eq:c-8}
\end{eqnarray}
where $\mathcal{K}^{-}_{\scriptscriptstyle{HR (N)}}$ is the ordered generator without extra terms,
\begin{eqnarray}\begin{split}
& \mathcal{K}^{-}_{\scriptscriptstyle{HR (N)}} = \frac{1}{2}\Bigl( \mathcal{K}^{-}_{\scriptscriptstyle{R (N)}} +h.c. \Bigr) ~,\\
& \mathcal{K}^{-}_{\scriptscriptstyle{R (N)}} = x^{-}\mathcal{D}_{\scriptscriptstyle{R (N)}} +\frac{1}{4p_{-}}\sum _{n}\sum_{m}X_{n}X_{m}L^{\scriptscriptstyle{(N)}}_{-n-m} -\frac{i}{p_{-}}\sum _{n}\sum _{m \not= 0} \frac{n+m}{m^2}X_{n}M^{\scriptscriptstyle{(N)}}_{m}P_{-n-m} 
\end{split}\label{eq:c-9}
\end{eqnarray}
and 
\begin{eqnarray}
C_{\scriptscriptstyle{(N)}} = -\frac{3}{4}N^2 -\frac{1}{4}N +\frac{1}{4} -\frac{1}{2}\mathop{\sum _{\scriptscriptstyle{|n|\leq N}}\sum _{\scriptscriptstyle{0<|m|\leq N}}}_{\scriptscriptstyle{|n+m|\leq N}}\frac{n^2}{m^2} ~.
\label{eq:c-10}
\end{eqnarray}
Here we emphasize that $\delta \mathcal{K}^{-}_{\scriptscriptstyle{HR (N)}}$ is obviously hermitian and does not contain any zero mode, $x$ and $p$. Although $\delta \mathcal{K}^{-}_{\scriptscriptstyle{HR (N)}}$, which is ordered correctly in the Hermitian R-order, is dropped in the limit of $N\rightarrow \infty $, we must verify whether the commutators of it and other generators becomes consist of only terms dropped in the limit.

\subsection{Dangerous part in $[\mathcal{J}^{-}_{\scriptscriptstyle{HR (N)}}, \tilde{\mathcal{K}}^{-}_{\scriptscriptstyle{HR (N)}}]$}
From the consideration in section 4, the structure of $[\mathcal{J}^{-}_{\scriptscriptstyle{HR (N)}}, \tilde{\mathcal{K}}^{-}_{\scriptscriptstyle{HR (N)}}]$ in the Hermitian R-order is 
\begin{eqnarray}
[\mathcal{J}^{-}_{\scriptscriptstyle{HR (N)}}, \tilde{\mathcal{K}}^{-}_{\scriptscriptstyle{HR (N)}}] = i(\mbox{ordered classical}) + i\frac{1}{p_{-}}(\mbox{XPP-cubic}) \times M_{0}^{\scriptscriptstyle{(N)}} + i\frac{p}{p_{-}^2}\times (\mbox{const.}) ~.
\label{eq:c-11}
\end{eqnarray}
The first term of (\ref{eq:c-11}) corresponds to the part which is zero classically in the calculation without the regularization. Therefore it consists of the terms which are dropped in the limit of $N\rightarrow \infty $.\footnote{Because of the classical part, we do not have to mind the divergence.}~ The second part\footnote{We can know the second part from the classical calculation as well as quantum one.} is related to the choice of the ordering constant, $M_{0}^{\scriptscriptstyle{(N)}}\rightarrow M_{0}\approx a$. The third part is the quantum effect term of the lowest degree. The anomalous part consists of the second and third parts. 

Here we calculate the anomalous part directly without considering the commutator with $x$ unlike the case of the main text. Then we verify that there is no contribution to the anomalous part from the commutator of $\delta \mathcal{K}^{-}_{\scriptscriptstyle{HR (N)}}$. \\
Because of the lengthy results, we divide $[\mathcal{J}^{-}_{\scriptscriptstyle{HR (N)}}, \tilde{\mathcal{K}}^{-}_{\scriptscriptstyle{HR (N)}}]$ into the commutators of each part of $\tilde{\mathcal{K}}^{-}_{\scriptscriptstyle{HR (N)}}$ defined by (\ref{eq:c-8}) and then extract only the anomalous part, $\propto i\frac{p}{p_{-}^2}$.\\

First, we consider
\begin{eqnarray}
[\mathcal{J}^{-}_{\scriptscriptstyle{HR (N)}}, \mathcal{K}^{-}_{\scriptscriptstyle{HR (N)}}] = \frac{1}{2}\left( [\mathcal{J}^{-}_{\scriptscriptstyle{HR (N)}}, \mathcal{K}^{-}_{\scriptscriptstyle{R (N)}}] -h.c. \right) ~.
\label{eq:c-12}
\end{eqnarray}
and then see the contribution to the anomalous part.\footnote{In the calculation of (\ref{eq:c-12}), to obtain the Hermitian  R-ordered version of terms proportional to $x^{-}\frac{1}{p_{-}}$, we deform such terms by using $x^{-}\frac{1}{p_{-}}= \frac{1}{2}\bigl( x^{-}\frac{1}{p_{-}}+\frac{1}{p_{-}} \bigr) -\frac{i}{2p_{-}}$.}\\
Because the calculation of (\ref{eq:c-12}) is lengthy, we divide $\mathcal{J}^{-}_{\scriptscriptstyle{HR (N)}}$ into three hermitian parts in (\ref{eq:c-2}). We calculate the contribution from the commutator of the first part of $\mathcal{J}^{-}_{\scriptscriptstyle{HR (N)}}$ in detail.
\begin{eqnarray}\begin{split}
[-x^{-}p, \mathcal{K}^{-}_{\scriptscriptstyle{R (N)}}] =& \frac{i}{4p_{-}^2}\sum _{n}\sum _{m}X_{n}X_{m}L^{\scriptscriptstyle{(N)}}_{-n-m} p -\frac{i}{p_{-}^2}\sum _{n}\sum _{m \not= 0} \frac{n+m}{m^2}X_{n}M^{\scriptscriptstyle{(N)}}_{m}P_{-n-m} p \\
& +ix^{-}\frac{1}{p_{-}} \Bigl( \frac{1}{2}xL^{\scriptscriptstyle{(N)}}_{0} -\Lambda ^{\scriptscriptstyle{(N)}}_{\scriptscriptstyle{HR}} \Bigr) \\
=& \frac{i}{4p_{-}^2}\sum _{n}\sum _{m}X_{n}X_{m}L^{\scriptscriptstyle{(N)}}_{-n-m} p -\frac{i}{p_{-}^2}\sum _{n}\sum _{m \not= 0} \frac{n+m}{m^2}X_{n}M^{\scriptscriptstyle{(N)}}_{m}P_{-n-m} p \\
& +\frac{i}{2}\bigl( x^{-}\frac{1}{p_{-}} + \frac{1}{p_{-}}x^{-} \bigr) \Bigl( \frac{1}{2}xL^{\scriptscriptstyle{(N)}}_{0} -\Lambda ^{\scriptscriptstyle{(N)}}_{\scriptscriptstyle{HR}} \Bigr) +\frac{1}{2p_{-}^2}\Bigl( \frac{1}{2}xL^{\scriptscriptstyle{(N)}}_{0} -\Lambda ^{\scriptscriptstyle{(N)}}_{\scriptscriptstyle{HR}} \Bigr) ~.
\end{split}\label{eq:c-13}
\end{eqnarray}
From this, we obtain 
\begin{eqnarray}
\frac{1}{2}\left( [-x^{-}p, \mathcal{K}^{-}_{\scriptscriptstyle{R (N)}}] -h.c. \right) \sim \frac{1}{2} \left( \frac{1}{2p_{-}^2}\Bigl( \frac{1}{2}xL^{\scriptscriptstyle{(N)}}_{0} -\Lambda ^{\scriptscriptstyle{(N)}}_{\scriptscriptstyle{HR}} \Bigr) -h.c. \right) \sim i\frac{p}{4p_{-}^2} ~,
\label{eq:c-14}
\end{eqnarray}
where ``$\sim $" indicates extracting of the contribution to the anomalous part, $\propto i\frac{p}{p_{-}^2}$. \\
Similarly, the contributions of other parts are calculated. The contribution from the commutator of the second part of $\mathcal{J}^{-}_{\scriptscriptstyle{HR (N)}}$ is
\begin{eqnarray}
\frac{1}{2}\Bigl( \Bigl[ -\frac{1}{2p_{-}}xL^{\scriptscriptstyle{(N)}}_{0} +i\frac{p}{2p_{-}} ,~ \mathcal{K}^{-}_{\scriptscriptstyle{R (N)}} \Bigr] -h.c. \Bigr) \sim -i\frac{p}{2p_{-}^2} ~.
\label{eq:c-15}
\end{eqnarray}
The contribution from the commutator of the third part of $\mathcal{J}^{-}_{\scriptscriptstyle{HR (N)}}$ is
\begin{eqnarray}\begin{split}
\frac{1}{2}\Bigl( \Bigl[ \frac{\Lambda ^{\scriptscriptstyle{(N)}}_{\scriptscriptstyle{HR}}}{p_{-}} ,~ \mathcal{K}^{-}_{\scriptscriptstyle{R (N)}} \Bigr] -h.c. \Bigr) &\sim \frac{2i}{p_{-}^2} \sum _{n \not= 0} \frac{1}{n^2} M_{n}^{\scriptscriptstyle{(N)}}P_{-n} M_{0}^{\scriptscriptstyle{(N)}} +i\Biggl( \frac{3}{4}N^2 +\frac{1}{4}N +\frac{1}{2}\mathop{\sum _{\scriptscriptstyle{|n|\leq N}}\sum _{\scriptscriptstyle{0<|m|\leq N}}}_{\scriptscriptstyle{|n+m|\leq N}}\frac{n^2}{m^2} \Biggr) \frac{p}{p_{-}^2} ~,
\end{split}\label{eq:c-16}
\end{eqnarray}
where we added the contribution from the term arisen when we move $M_{0}^{\scriptscriptstyle{(N)}}$ to the right end. \\
We collect (\ref{eq:c-14})-(\ref{eq:c-16}) to obtain 
\begin{eqnarray}\begin{split}
[\mathcal{J}^{-}_{\scriptscriptstyle{(N)}}, \mathcal{K}^{-}_{\scriptscriptstyle{(N)}}] &\sim \frac{2i}{p_{-}^2} \sum _{n \not= 0} \frac{1}{n^2} M_{n}^{\scriptscriptstyle{(N)}}P_{-n} M_{0}^{\scriptscriptstyle{(N)}} +i\Biggl( \frac{3}{4}N^2 +\frac{1}{4}N -\frac{1}{4} +\frac{1}{2}\mathop{\sum _{\scriptscriptstyle{|n|\leq N}}\sum _{\scriptscriptstyle{0<|m|\leq N}}}_{\scriptscriptstyle{|n+m|\leq N}}\frac{n^2}{m^2} \Biggr) \frac{p}{p_{-}^2} \\
& = \frac{2i}{p_{-}^2} \sum _{n \not= 0} \frac{1}{n^2} M_{n}^{\scriptscriptstyle{(N)}}P_{-n} M_{0}^{\scriptscriptstyle{(N)}} -iC_{\scriptscriptstyle{(N)}}\frac{p}{p_{-}^2} ~.
\end{split}\label{eq:c-17}
\end{eqnarray}

Next the commutator of $\frac{C_{\scriptscriptstyle{(N)}}}{p_{-}}$ is
\begin{eqnarray}
\Bigl[ \mathcal{J}^{-}_{\scriptscriptstyle{HR (N)}}, \frac{C_{\scriptscriptstyle{(N)}}}{p_{-}} \Bigr] = \Bigl[ -x^{-}p, \frac{C_{\scriptscriptstyle{(N)}}}{p_{-}} \Bigr]  = iC_{\scriptscriptstyle{(N)}}\frac{p}{p_{-}^2} ~.
\label{eq:c-18}
\end{eqnarray}
This is canceled by (\ref{eq:c-17}).

Next we consider the contribution from the commutator with $\delta \mathcal{K}^{-}_{\scriptscriptstyle{HR (N)}}$ to the anomalous part. Because of the extra restriction on the range of the summation in (\ref{eq:c-7}), $\delta \mathcal{K}^{-}_{\scriptscriptstyle{HR (N)}}$ does not contain any zero mode. Therefore the contribution to $i\frac{p}{p_{-}^2}$ comes from the commutators of terms with $p$ in $\mathcal{J}^{-}_{\scriptscriptstyle{HR (N)}}$ and $\delta \mathcal{K}^{-}_{\scriptscriptstyle{(N)}}$. \\
The terms with $p$ in $\mathcal{J}^{-}_{\scriptscriptstyle{HR (N)}}$ are
\begin{eqnarray}
\mathcal{J}^{-}_{\scriptscriptstyle{HR (N)}}|_{p} = -x^{-}p -\frac{1}{2p_{-}}xpp +\frac{i}{2p_{-}}p ~.
\label{eq:c-19}
\end{eqnarray}
Then we obtain
\begin{eqnarray}
\Bigl[ \mathcal{J}^{-}_{\scriptscriptstyle{HR (N)}}, \delta \mathcal{K}^{-}_{\scriptscriptstyle{HR (N)}} \Bigr] \sim \Bigl[ \mathcal{J}^{-}_{\scriptscriptstyle{HR (N)}}|_{p}, \delta \mathcal{K}^{-}_{\scriptscriptstyle{HR (N)}} \Bigr] = -\Bigl[ x^{-}, \delta \mathcal{K}^{-}_{\scriptscriptstyle{HR (N)}} \Bigr] p = i\delta \mathcal{K}^{-}_{\scriptscriptstyle{HR (N)}} \frac{p}{p_{-}} ~.
\label{eq:c-20}
\end{eqnarray}
This is dropped in the limit of $N\rightarrow \infty $ and has no contribution to the anomalous part.

Finally the commutator of the rest term in $\tilde{\mathcal{K}}^{-}$ of (\ref{eq:c-8}) is again the term with $M_{0}^{\scriptscriptstyle{(N)}}$ at the right end because $M_{0}^{\scriptscriptstyle{(N)}}$ commutes with all regularized generators. 
\begin{eqnarray}\begin{split}
& \Bigl[ \mathcal{J}^{-}_{\scriptscriptstyle{HR (N)}},~ \frac{2i}{p_{-}}\sum _{n \not= 0}\frac{1}{n}X_{n}P_{-n} M_{0}^{\scriptscriptstyle{(N)}} \Bigr] \\
& = \Bigl( -\frac{6}{p_{-}^2}\sum _{n \not= 0}\frac{1}{n}X_{n}P_{-n}p +\frac{1}{p_{-}^2}\sum _{n \not= 0}\frac{1}{n}X_{n}L_{-n} +\frac{2i}{p_{-}^2}\sum _{n \not= 0}\frac{1}{n^2}M_{n}P_{-n} \Bigr) M_{0}^{\scriptscriptstyle{(N)}} ~.
\end{split}\label{eq:c-21}
\end{eqnarray}
All the terms with $M_{0}^{\scriptscriptstyle{(N)}}$ at the right end in $[\mathcal{J}^{-}_{\scriptscriptstyle{HR (N)}}, \tilde{\mathcal{K}}^{-}_{\scriptscriptstyle{HR (N)}}]$ vanishes if and only if we choose $a=0$.\footnote{In the cut-off regularization, the constraint is $M_{0}^{\scriptscriptstyle{(N)}}\rightarrow M_{0}\approx a=0$. }

Thus the anomalous part in $[\mathcal{J}^{-}_{\scriptscriptstyle{HR (N)}}, \tilde{\mathcal{K}}^{-}_{\scriptscriptstyle{HR (N)}}]$ vanishes if and only if $a=0$.

\subsection{The rest commutators with $\tilde{\mathcal{K}}^{-}_{\scriptscriptstyle{HR (N)}}$}
Here we consider the rest of commutators with $\tilde{\mathcal{K}}^{-}_{\scriptscriptstyle{HR (N)}}$ and other generators. We can easily verify that most of them give the expected results even for finite $N$, 
\begin{eqnarray}\begin{split}
& [\mathcal{P}_{- \scriptscriptstyle{HR (N)}}, \tilde{\mathcal{K}}^{-}_{\scriptscriptstyle{HR (N)}}] = -i(\mathcal{D}_{\scriptscriptstyle{HR (N)}}+\mathcal{J}_{\scriptscriptstyle{HR (N)}}) ~,~ [\mathcal{P}_{\scriptscriptstyle{HR (N)}}, \tilde{\mathcal{K}}^{-}_{\scriptscriptstyle{HR (N)}}] = i\mathcal{J}^{-}_{\scriptscriptstyle{HR (N)}} ~,~ [\mathcal{P}_{+ \scriptscriptstyle{HR (N)}}, \tilde{\mathcal{K}}^{-}_{\scriptscriptstyle{HR (N)}}] = 0 \\
& [\mathcal{J}^{+}_{\scriptscriptstyle{HR (N)}}, \tilde{\mathcal{K}}^{-}_{\scriptscriptstyle{HR (N)}}] = i\mathcal{K}_{\scriptscriptstyle{HR (N)}} ~,~ [\mathcal{J}_{\scriptscriptstyle{HR (N)}}, \tilde{\mathcal{K}}^{-}_{\scriptscriptstyle{HR (N)}}] = -i\tilde{\mathcal{K}}^{-}_{\scriptscriptstyle{HR (N)}} \\
& [\mathcal{D}_{\scriptscriptstyle{HR (N)}}, \tilde{\mathcal{K}}^{-}_{\scriptscriptstyle{HR (N)}}] = -i\tilde{\mathcal{K}}^{-}_{\scriptscriptstyle{HR (N)}} ~,~ [\mathcal{K}^{+}_{\scriptscriptstyle{HR (N)}}, \tilde{\mathcal{K}}^{-}_{\scriptscriptstyle{HR (N)}}] = 0 ,
\end{split}\label{eq:c-22}
\end{eqnarray}
Although the rest commutator which should be calculated, $[\mathcal{K}_{\scriptscriptstyle{HR (N)}}, \tilde{\mathcal{K}}^{-}_{\scriptscriptstyle{HR (N)}}]$, is complicated, the calculation of this is parallel to that of $[\mathcal{J}^{-}_{\scriptscriptstyle{HR (N)}}, \tilde{\mathcal{K}}^{-}_{\scriptscriptstyle{HR (N)}}]$. It consists of the classical ordered part, the terms with $M_{0}^{\scriptscriptstyle{(N)}}$ at the right end and the term proportional to $i\frac{x}{p_{-}}$ such as (\ref{eq:c-11}). The commutator of $\delta \mathcal{K}^{-}_{\scriptscriptstyle{HR (N)}}$ does not contribute to the anomalous part in the same way as the last subsection. Therefore we get $[\mathcal{K}_{\scriptscriptstyle{HR (N)}}, \tilde{\mathcal{K}}^{-}_{\scriptscriptstyle{HR (N)}}]\rightarrow 0$ in the limit of $N\rightarrow \infty $ and $a=0$.\\

Thus we obtain the set of commutators which give the expected result for the spacetime conformal symmetry in the limit of $N\rightarrow \infty $.


\begin{thebibliography}{999}
\bibitem{bib:080} I.R. Klebanov, A.M. Polyakov, 
``AdS Dual of the Critical O(N) Vector Model," Phys. Lett. B {\bf 550} (2002) 213-219 [arXiv:hep-th/0210114]. 
\bibitem{bib:071} Simone Giombi, Xi Yin, 
``Higher Spin Gauge Theory and Holography: The Three-Point Functions," JHEP {\bf 1009} (2010) 115 [arXiv:0912.3462[hep-th]].
\bibitem{bib:078} Simone Giombi and Xi Yin, 
``On Higher Spin Gauge Theory and the Critical $O(N)$ Model," Phys. Rev. D {\bf 85} (2012) 086005 [arXiv:1105.4011 [hep-th]].
\bibitem{bib:074} Chi-Ming Chang and Xi Yin, 
``Higher Spin Gravity with Matter in $AdS_3$ and Its $CFT$ Dual," JHEP {\bf 1210} (2012) 024 [arXiv:1106.2580[hep-th].
\bibitem{bib:077} Simone Giombi, Shiraz Minwalla, Shiroman Prakash, Sandip P. Trivedi, Spenta R. Wadia and Xi Yin, 
``Chern-Simons Theory with Vector Fermion Matter," Eur. Phys. J. C {\bf 72} (2012) 2112 [arXiv:1110.4386[hep-th]].
\bibitem{bib:052} Chi-Ming Chang, Shiraz Minwalla, Tarun Sharma and Xi Yin, 
``ABJ Triality: from Higher Spin Fields to Strings," J. Phys. A {\bf 46} (2013) 214009 [arXiv:1207.4485 [hep-th]].
\bibitem{bib:018} A. Campoleoni, S. Fredenhagen, S. Pfenninger, and S. Theisen, 
``Asymptotic symmetries of three-dimensional gravity coupled to higher-spin fields," JHEP {\bf 1011} (2010) 007 [arXiv:1008.4744[hep-th]].
\bibitem{bib:054} Andrea Campoleoni, Stefan Fredenhagen and Stefan Pfenninger, 
``Asymptotic W-symmetries in three-dimensional higher-spin gauge theories," JHEP {\bf 1109} (2011) 113 [arXiv:1107.0290 [hep-th]].
\bibitem{bib:073} Matthias R Gaberdiel and Rajesh Gopakumar, 
``An $AdS_3$ Dual for Minimal Model CFTs," Phys. Rev. D {\bf 83} (2011) 066007 [arXiv:1011.2986 [hep-th]].
\bibitem{bib:076} Matthias R. Gaberdiel and Thomas Hartman, 
``Symmetries of Holographic Minimal Models," JHEP {\bf 1105} (2011) 031 [arXiv:1101.2910 [hep-th]].
\bibitem{bib:075} Matthias R. Gaberdiel and Rajesh Gopakumar, Thomas Hartman and Suvrat Raju, 
``Partition Functions of Holographic Minimal Models," JHEP {\bf 1108} (2011) 077 [arXiv:1106.1897 [hep-th]].
\bibitem{bib:066} Martin Ammon, Michael Gutperle, Per Kraus and Eric Perlmutter, 
``Spacetime Geometry in Higher Spin Gravity," JHEP {\bf 1110} (2011) 053 [arXiv:1106.4788 [hep-th]].
\bibitem{bib:057} Alfredo Pereza, David Tempoa and Ricardo Troncosoa, 
``Higher spin gravity in 3D: Black holes, global charges and thermodynamics," Phys. Lett. B {\bf 726} (2013) 444-449 [arXiv:1207.2844 [hep-th]].
\bibitem{bib:053} Ippei Fujisawa, Kenta Nakagawa and Ryuichi Nakayama, 
``AdS/CFT for 3D Higher-Spin Gravity Coupled to Matter Fields," Class. Quant. Grav. {\bf 31} (2014) 065006 [arXiv:1311.4714 [hep-th]].
\bibitem{bib:055} Andrea Campoleoni, Marc Henneaux, 
``Asymptotic symmetries of three-dimensional higher-spin gravity: the metric approach," arXiv:1412.6774 [hep-th].
\bibitem{bib:035} Hong Lu, C.N. Pope, K. Thielemans, X.J. Wang and K.W. Xu, 
``Quantizing higher spin string theories," Int. J. Mod. Phys. A {\bf 10} (1995) 2123-2142 [arXiv:hep-th/9410005].
\bibitem{bib:061} Amihay Hanany and Igor R. Klebanov, 
``On tensionless strings in (3+1)-dimensions," Nucl. Phys. B {\bf 482} (1996) 105-118 [arXiv:hep-th/9606136].
\bibitem{bib:031} Giulio Bonelli,
``On the tensionless limit of bosonic strings, infinite symmetries and higher spins," Nucl.Phys. B {\bf 669} (2003) 159-172 [arXiv:hep-th/0305155].
\bibitem{bib:045} G.K. Savvidy, 
``Tensionless strings: Physical Fock space and higher spin fields," Int. J. Mod. Phys. A {\bf 19} (2004) 3171-3194 [arXiv:hep-th/0310085].
\bibitem{bib:059} A. Sagnotti, 
``Notes on Strings and Higher Spins," J. Phys. A {\bf 46} (2013) 214006 [arXiv:1112.4285 [hep-th]].
\bibitem{bib:041} K. Surya Kiran, Chethan Krishnan, Ayush Saurabh and Joan Simon, 
``Strings vs Spins on the Null Orbifold," JHEP {\bf 1412} (2014) 002 [arXiv:1408.3296 [hep-th]].
\bibitem{bib:072} C. Fronsdal, 
``Massless fields with integer spin," Phys. Rev. D {\bf 18} (1978) 3624.
\bibitem{bib:069} Mikhail Vasiliev, 
``Higher-Spin Gauge Theories in Four, Three and Two Dimensions," Int. J. Mod. Phys. D {\bf 5} (1996) 763-797 [arXiv:hep-th/9611024].
\bibitem{bib:017} Mikhail Vasiliev, 
``Higher Spin Gauge Theories: Star-Product and AdS Space," In *Shifman, M.A. (ed.): The many faces of the superworld* 533-610 [arXiv:hep-th/9910096].
\bibitem{bib:060} M.A. Vasiliev, 
``Higher-Spin Theory and Space-Time Metamorphoses," Lect. Notes Phys. 892 (2015) 227-264 [arXiv:1404.1948 [hep-th]].
\bibitem{bib:001} M. B. Green, J. H. Schwarz and E. Witten, 
``Superstring Theory. Vol. 1 : Introduction," Cambridge, UK: Univ. Pr. (1987).
\bibitem{bib:002} J. Polchinski, 
``String theory. Vol. 1: An introduction to bosonic string," Cambridge, UK: Univ. Pr. (1988).
\bibitem{bib:081} Barton Zwiebach, 
``A First Course in STRING THEORY," Cambridge Univ. Pr. (2004).
\bibitem{bib:025} Mitsuhiro Kato and Kaku Ogawa, 
``Covariant Quantization of String Based on BRS Invariance," Nucl.Phys. B {\bf 212} (1983) 443.
\bibitem{bib:026} Nobuyoshi Ohta,
``Covariant quantization of superstrings based on Becchi-Rouet-Stora invariance," Phys.Rev. D {\bf 33} (1986) 1681.
\bibitem{bib:027} Stephen Hwang, 
``Covariant Quantization of the String in Dimensions D <= 26 Using a BRS Formulation," Phys.Rev. D {\bf 28} (1983) 2614.
\bibitem{bib:046} A. Karlhede and U. Lindstrom, 
``The Classical Bosonic String in the Zero Tension Limit," Class. Quant. Grav. {\bf 3} (1986) L73-L75.
\bibitem{bib:006} F. Lizzi, B. Rai, G. Sparano and A. Srivastava, 
``Quantization Of The Null String And Absence Of Critical Dimensions," Phys.Lett. B {\bf 182} (1986) 326. 
\bibitem{bib:037} J. Gamboa, C. Ramirez and M. Ruiz-Altaba, 
``Quantum Null (super)strings," Phys.Lett. B {\bf 225} (1989) 335-339.
\bibitem{bib:010} J. Gamboa, Cupatitzio Ramirez and M. Ruiz-Altaba, 
``Null spinning strings," Nuc. Phys. B {\bf 338} (1990) 143-187.
\bibitem{bib:008} H.Gustafsson, U.Lindstrom, P.Saltsidis, B.Sundborg and R.v.Unge, 
``Hamiltonian BRST Quantization of the Conformal String," Nucl.Phys. B {\bf 440} (1995) 495-520 [arXiv:hep-th/9410143].
\bibitem{bib:012} J. Isberg, U. Lindstrom and B. Sundborg, 
``Space-Time Symmetries of Quantized Tensionless Strings," Phys. Lett. B {\bf 293} (1992) 321-326 [arXiv:hep-th/9207005].
\bibitem{bib:047} Ulf Lindstrom, 
``The Zero tension limit of strings and superstrings," arXiv:hep-th/9303173.
\bibitem{bib:013} J. Isberg and U. Lindstrom, B. Sundborg and G. Theodoridis, 
``Classical and Quantized Tensionless Strings," Nucl. Phys. B {\bf 411} (1994) 122-156 [arXiv:hep-th/9307108].
\bibitem{bib:015} J. Barcelos-Neto1 and M. Ruiz-Altaba, 
``Superstrings with zero tension," Phys. Let. B {\bf 228} (1989) 193-199.
\bibitem{bib:016} U. Lindstrom, B. Sundborg and G. Theodoridis, 
``The zero tension limit of the superstring," Phys. Let. B {\bf 253} (1991) 319-323.
\bibitem{bib:048} U. Lindstrom, B. Sundborg and G. Theodoridis, 
``The zero tension limit of the spinning string," Phys. Lett. B {\bf 258} (1991) 331-334.
\bibitem{bib:030} P. Saltsidis, 
``Hamiltonian BRST quantization of the conformal spinning string," Nucl.Phys. B {\bf 446} (1995) 286-298 [arXiv:hep-th/9503062].
\bibitem{bib:009} P. Saltsidis, 
``Tensionless p-branes with manifest conformal invariance," Phys. Lett. B {\bf 401} (1997) 21-29 [arXiv:hep-th/9702081].
\bibitem{bib:011} P. Saltsidis, 
``The Mass Spectrum of the 2-dimensional Conformal String," Phys.Lett. B {\bf 396} (1997) 107-114 [arXiv:hep-th/9609169].
\bibitem{bib:003} Luca Mezincescu and Paul K. Townsend, 
``Anyons from Strings,"  Phys. Rev. Lett. {\bf 105}, 191601,2010 [arXiv:1008.2334 [hep-th]].
\bibitem{bib:004} L. Mezincescu and P. K. Townsend, 
``Quantum 3D Superstrings," Phys.Rev. D {\bf 84} (2011) 106006 [arXiv:1106.1374 [hep-th]].
\bibitem{bib:005} L. Mezincescu and P. K. Townsend, 
``3D strings and other anyonic things," Fortsch.Phys. {\bf 60} (2012) 1076-1079 [arXiv:1111.3384 [hep-th]].
\bibitem{bib:020} L. Mezincescu, A. J. Routh and P. K. Townsend, 
``Equivalence of 3D Spinning String and Superstring," JHEP {\bf 1307} (2013) 024 [arXiv:1305.5049 [hep-th]].
\bibitem{bib:999} K. M. , 
``Quantum 3D Tensionless String in Light-cone Gauge," JHEP {\bf 1402} (2014) 038 [arXiv:1303.7202 [hep-th]].
\bibitem{bib:007} F. Lizzi, 
``The Zero Tension Limit of The Virasoro Algebra and the Central Extension," Mod.Phys.Lett. A {\bf 9} (1994) 1495-1500 [arXiv:hep-th/9404148].
\bibitem{bib:029} Arjun Bagchi, Rajesh Gopakumar, Ipsita Mandal, and Akitsugu Miwa,
``GCA in 2d," JHEP {\bf 1008} (2010) 004 [arXiv:0912.1090 [hep-th]].
\bibitem{bib:014} Arjun Bagchi, 
``Tensionless Strings and Galilean Conformal Algebra," JHEP {\bf 1305} (2013) 141 [arXiv:1303.0291[hep-th]]. 
\bibitem{bib:032} Luca Mezincescu and Paul K. Townsend, 
``The Quantum 3D Superparticle," SIGMA {\bf 7} (2011) 005 [arXiv:1011.5049 [hep-th]].
\bibitem{bib:038} Luca Mezincescu and Paul K. Townsend,
``Semionic Supersymmetric Solitons," J. Phys. A {\bf 43} (2010) 465401 [arXiv:1008.2775 [hep-th]].
\bibitem{bib:022} Luca Mezincescu, Alasdair J. Routh and Paul K. Townsend, 
``All Superparticles are BPS," J. Phys. A {\bf 47} (2014) 175401 [arXiv:1401.5116 [hep-th]].
\end{thebibliography}
\end{document}